\documentclass[aps,pra,showkeys,preprintnumbers,twocolumn,nobalancelastpage,longbibliography,superscriptaddress]{revtex4-1}
\usepackage{lmodern}
\usepackage[utf8]{inputenc}
\usepackage[T1]{fontenc}
\usepackage[english]{babel}
\usepackage{url}
\usepackage{graphicx}
\usepackage{dsfont}
\usepackage{amsmath,amssymb}
\usepackage{hyperref}
\usepackage{braket}
\usepackage{enumerate}
\usepackage{float}
\usepackage{bm}
\usepackage{textcomp}
\usepackage{siunitx}
\usepackage{todonotes}
\begin{document}

\newcommand{\MR}[1]{{\textcolor{red}{MR:#1}}}
\newcommand{\pr}[0]{\mathrm{pr}}

\title{The random coupled-plaquette gauge model and the surface code under circuit-level noise}
\author{Manuel Rispler}
\affiliation{Institute for Theoretical Nanoelectronics (PGI-2), Forschungszentrum Jülich, 52428 Jülich, Germany}
\affiliation{JARA-Institute for Quantum Information, RWTH Aachen University, 52056 Aachen,
Germany}

\author{Davide Vodola}
\affiliation{Dipartimento di Fisica e Astronomia ``Augusto Righi'' dell'Universit\`a di Bologna, I-40127 Bologna, Italy}

\author{Markus Müller}
\affiliation{Institute for Theoretical Nanoelectronics (PGI-2), Forschungszentrum Jülich, 52428 Jülich, Germany}
\affiliation{JARA-Institute for Quantum Information, RWTH Aachen University, 52056 Aachen,
Germany}
\author{Seyong Kim}
\affiliation{Department of Physics, Sejong University, 05006 Seoul, Republic of Korea}

\begin{abstract}
We map the decoding problem of the surface code under depolarizing and syndrome noise to a disordered spin model, which we call the random coupled-plaquette gauge model (RCPGM). By coupling X- and Z-syndrome volumes, this model allows us to optimally account for genuine Y-errors in the surface code in a setting with noisy measurements. Using Parallel Tempering Monte Carlo simulations, we determine the code's fundamental error threshold. Firstly, for the phenomenological noise setting we determine a threshold of $6\%$ under uniform depolarizing and syndrome noise. This is a substantial improvement compared to results obtained via the previously known "uncoupled" random plaquette gauge model (RPGM) in the identical setting, where marginalizing Y-errors leads to a threshold of $4.3\%$. Secondly, we tackle the circuit-level noise scenario, where we use a reduction technique to find effective asymmetric depolarizing and syndrome noise rates to feed into the RCPGM mapping. Despite this reduction technique breaking up some of the correlations contained in the intricacies of circuit-level noise, we find an improvement exceeding that for the phenomenological case. We report a threshold of up to $1.4\%$, to be compared to $0.7\%$ under the identical noise model when marginalizing the Y-errors and mapping to the anisotropic RPGM. These results enlarge the landscape of statistical mechanical mappings for quantum error correction. In particular they provide an underpinning for the broadly held belief that accounting for Y-errors is a major bottleneck in improving surface code decoders. This is highly encouraging for leading efficient practical decoder development, where heuristically accounting for Y-error correlations has seen recent developments such as belief-matching. This suggests that there is further room for improvement of the surface code for fault-tolerant quantum computation.

\end{abstract}
\maketitle

\section{Introduction}

In the light of two decades of research, Kitaev's surface (toric) code stands strong as the archetypal candidate for realizing fault-tolerant quantum computation~\cite{Kitaev2003, Dennis2002, Terhal2015}. It provides attractive features such as a two-dimensional planar layout with local low weight stabilizers, which make it viable for experimental realization. One of the guiding figures of merit of a quantum error correcting (QEC) code is its threshold value, i.e. the value of noise strength below which one can suppress logical errors arbitrarily well by (moderately) increasing the size of the QEC code~\cite{Aharonov1997}. One of the main features which make the surface code a leading candidate for fault-tolerant quantum computation is its very high threshold value under realistic error models. This threshold value, reported in the vicinity of $1\%$ error rate under realistic error models called \emph{circuit-level noise}~\cite{Raussendorf2007, Fowler2009}, is occasionally almost synonymously used with the notion of \emph{the fault-tolerance threshold}. With recent substantial leaps in quantum information hardware, we are currently witnessing the implementation of surface codes ushering in the new era of fault-tolerant quantum processing~\cite{Krinner2022, Zhao2022, Acharya2023, Bluvstein2024, Acharya2024}.\\

Importantly, this threshold value is tied to the decoding strategy, i.e. the classical processing of the syndrome information. A paradigmatic decoding strategy for the surface code is the minimum-weight perfect matching (MWPM). Given an observed X- (Z-) syndrome, this decoder finds the most likely Z- (X-) error configuration and is computationally efficient thanks to Edmond's discovery of the blossom algorithm~\cite{Edmonds1965,Higgott2023}. The performance of MWPM is known to possess two major shortcomings: firstly it does not account for code degeneracy and secondly it does not handle (Pauli-)$Y$-errors well. Let us briefly review these two shortcomings here. Firstly, for degenerate codes such as the surface code many distinct microscopic configurations are logically equivalent. This leads to the observation that maximizing the probability of successful decoding is not necessarily the same as finding the most likely error. The optimal strategy to take degeneracy fully into account is known as maximum likelihood decoding (MLD). Here, one accounts for all possible \emph{logically equivalent} error configurations and chooses the most likely error \emph{class} as the optimal recovery operation. While MLD is generally a provably computationally hard problem~\cite{Berlekamp1978,Iyer2015}, comparisons can be drawn in special cases. Notably, for the surface code under independent bit- and phase-flip noise with perfect measurements, Bravyi et al. found an exact method for MLD and observe that MWPM performs sub-optimally in comparison~\cite{Bravyi2014}. This trend is even more visible when moving to depolarizing data noise, where genuine Pauli-$Y$ errors pose a challenge for MWPM. The latter inherently rests on the assumption that syndromes are produced in pairs, which while true for $X-$ as well as $Z-$errors (ignoring the boundary), breaks down for depolarizing noise. Here, $Y$-errors can no longer be viewed as independent $X$ and $Z$-errors (the probability of a $Y$-error does not factorize: $\pr(Y)\neq \pr(X)\pr(Z)$). MWPM has no inherent ability to reflect this departure from independence and as such generally fails to pick up correlations between both syndromes~\cite{Bombin2012,Wootton2012}. The picture of sub-optimality of MWPM generalizes to the case of phenomenological noise. Here one models the syndrome extraction itself as unreliable by introducing a flip probability for the syndrome bit alongside the data qubit bit-flip probability. It has been shown that also here, thresholds under MWPM decoding are strictly smaller compared to thresholds using MLD~\cite{Wang2003}. 

\begin{figure}
    \centering
    \includegraphics[width=\columnwidth]{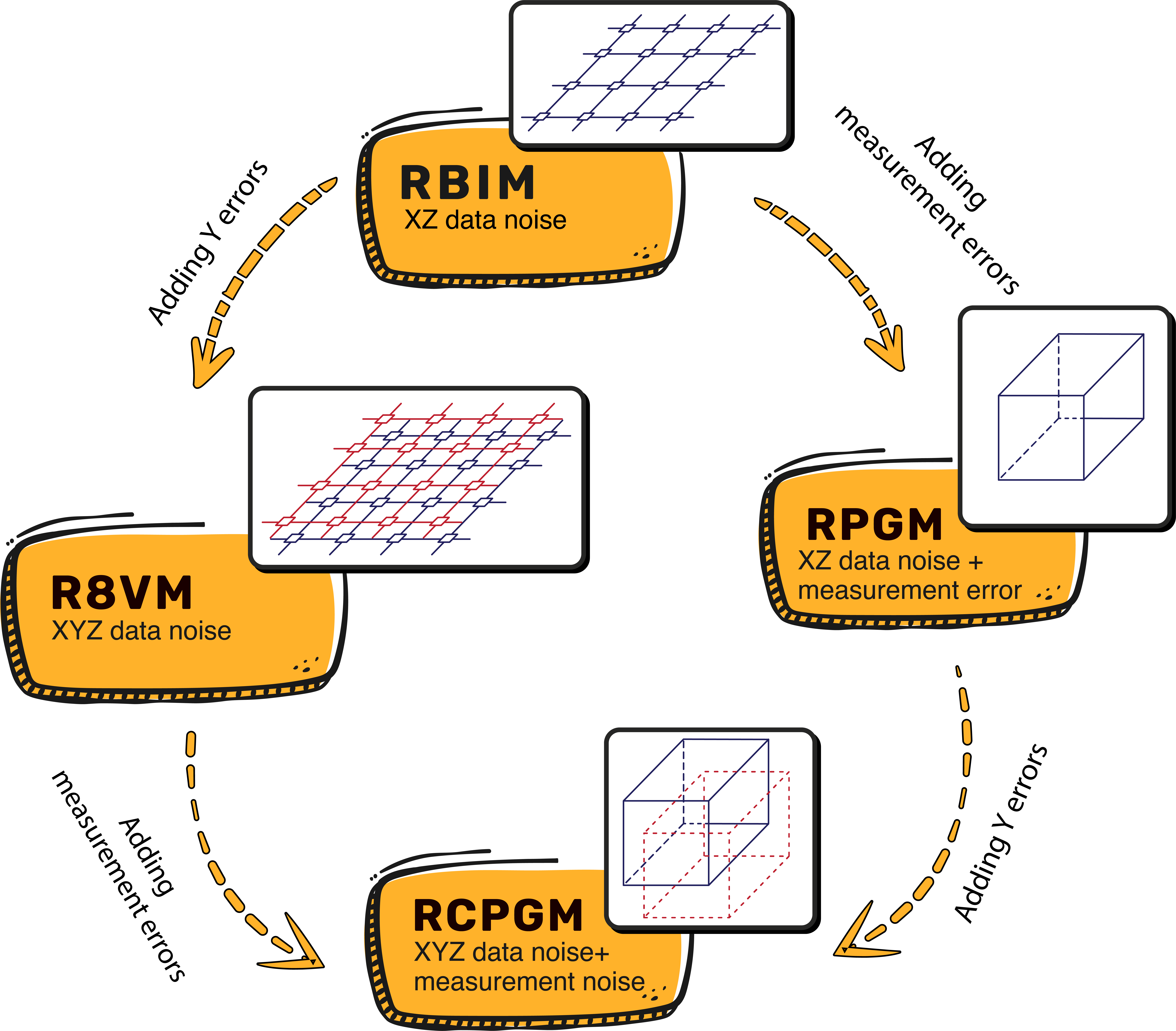}
    \caption{Overview of statistical-mechanical models and their application to different QEC scenarios in the toric/surface code. The original model was the random bond Ising model (RBIM), which describes decoding of data qubit independent bit-flip and phase-flip noise (two uncoupled lattices, only one shown). This model can on the one hand be furnished either by a coupling of the two lattices to describe depolarizing noise (genuine Y errors), which leads to the random eight-vertex model (R8VM). On the other hand, one can stick to independent XZ noise but introduce measurement (syndrome) errors, which lead to time-like equivalences similar to stabilizer equivalences in the time-direction. This leads to the random plaquette gauge model (RPGM). Both phenomena can be fused together into the random coupled-plaquette gauge model (RCPGM), which describes depolarizing data noise and syndrome noise simultaneously.}
    \label{fig:models-overview}
\end{figure}

Strikingly, in all these cases, this observation can be elucidated and unified by mapping the decoding problem of the surface code to quenched disorder many-body spin systems, occasionally simply known as statistical mechanics mappings in the QEC literature~\cite{Dennis2002, Chubb2021}. Broadly speaking in this language, the error rate can be translated to a temperature and quenching probability. This leads to the emergence of a regime, where the spin system orders, corresponding to errors being arbitrarily well suppressed. The critical point at which the spin system ceases to order is then identified with the threshold of the QEC code beyond which error correction is rendered useless (for a sketch see Fig.~\ref{fig:random_bond_phase_diagram}). The original construction focused on MLD decoding, which can be identified with a finite temperature in the phase diagram known as the Nishimori temperature~\cite{Nishimori2002} and the corresponding threshold coinciding with the so-called Nishimori point. This method has been used to compute thresholds under optimal decoding (MLD) for various noise models. These models and the relationship between them is illustrated in Fig.~\ref{fig:models-overview}. The corresponding numerical values are summarized in Tab.~\ref{tab:models_and_thresholds}). We mention in passing that this mapping has also been extended beyond the surface code e.g.to color codes~\cite{Bombin2012,Katzgraber2009,Kubica2018}, hypergraph product codes~\cite{Kovalev2018} or post-selected QEC~\cite{English2024}. The MLD task of finding the most likely successful recovery operation can be cast as the problem of minimizing the statistical mechanical free energy~\cite{Chubb2021}. Interestingly, this offers a somewhat intuitive connection between MLD and MWPM: the task of minimizing the internal energy corresponds to finding the most likely error, precisely the task MWPM is designed to find (for independent XZ noise models). The difference between MLD and MWPM lies in ignoring the degeneracy, in other words the entropy, which corresponds to setting the temperature to zero, such that MWPM thresholds can be identified with the phase-boundary at zero temperature. A key challenge then is to generalize this picture to more realistic noise models. An established realistic noise model is that of circuit-level noise, where one models all components in a quantum circuit realizing the measurement of the code stabilizers as noisy. This can be seen as an extension of the simpler noise models of code capacity and phenomenological noise. All previously mentioned noise models can be seen as containing proper subsets of circuit locations contained in circuit-leve noise. It is thus reasonable to expect that MWPM also performs sub-optimally for the case of circuit level noise. 
Here the statistical-mechanical mapping becomes significantly more challenging. Errors at locations in the circuit propagate and turn into correlated error events. Given the large number of circuit locations this translates to a Hamiltonian with an intractable number of variables and interactions. A benchmark from such mappings to compare MWPM or other suboptimal decoding strategies to is thus far missing beyond simpler cases such as repetition codes~\cite{Vodola2022} .\\
\begin{table*}
\begin{tabular}{c|c|c}
noise model & stat-mech model & threshold\\
\hline
data bit-flip $p$ & random-bond Ising (RBIM) &  10.9\%~\cite{Merz2002}\\
\hline
data depolarizing  $p$ & random-bond Ising & 16.3\%~\cite{Merz2002}\\
& random eight vertex (R8VM) & 18.9\%~\cite{Bombin2012}\\
\hline
data bit-flip $p$ + measurement bit-flip $p$ & random plaquette gauge (RPGM) & 3.3\%~\cite{Ohno2004,Kubica2018}\\
\hline
data depolarizing $p$ + measurement flip $p$ &  \textbf{(anisotropic) random plaquette gauge} & 4.3\%~\cite{Harrington2004}\\
data depolarizing $p$ + measurement flip $p$ &  \textbf{random coupled-plaquette gauge (RCPGM)} & $6\%$\\
\hline
circuit-level noise & random plaquette gauge & $\approx 0.7\%$\\
 & \textbf{random coupled-plaquette gauge} & $\approx 1.4\%$
\end{tabular}
\caption{Table listing noise models for the surface code alongside statistical mechanics models and resultant thresholds. This contains the central message of the present work: fully accounting for $Y$-errors allows for substantial improvements in threshold value. This was known for code capacity noise, where refining the RBIM to a R8VM allows a $15\%$ relative increase of the threshold value (second row block). We find that refining the RPGM to the RCPGM allows for a relative increase of threshold of $40\%$ for uniform depolarizing data and syndrome noise (penultimate row block), which is even surpassed by the observation for an effective error model capturing circuit level noise in the toric code (see Sec.~\ref{sec:circuit-level-noise-toric-code}), where the RCPGM outperforms the RPGM by a relative improvement of $100\%$ (last row block). }
\label{tab:models_and_thresholds}
\end{table*}

In this work, we introduce the random coupled-plaquette gauge model (RCPGM). This model unifies the random plaquette gauge model and the random eight vertex model (illustrated in Fig.~\ref{fig:models-overview}). It emerges by mapping the decoding problem of maximum likelihood decoding of the surface code under depolarizing and measurement noise, for which it enables the discovery of the optimal threshold value. We give a careful derivation how it arises and illustrate how it can be seen both as either a generalization of the random eight-vertex model to which one adds the statistical-mechanical equivalent of syndrome noise or as an extension of the ("uncoupled") random plaquette gauge model where one introduces Pauli-$Y$ errors by coupling the syndrome lattices. This quenched bond disordered spin model model undergoes a Higgs-deconfinement transition with increasing quenching probability and/or temperature, corresponding to an error suppression to error enhancement transition of the surface code with increasing noise strength on the Nishimori line. This model is significant in particular because it allows to shed light onto the two main shortcomings of the leading practical decoding strategy of minimum weight perfect matching, which are to neglect correlations of $Y-$errors as well as code degeneracy. We use a Parallel Tempering Monte Carlo (PTMC) approach to map out the relevant phase diagram of the model. This leads us to establishing the threshold under uniform depolarizing and measurement noise of $6\%$. 
In the second part of this manuscript, we work towards casting circuit-level depolarizing noise into an effective noise model that feeds into the RCPGM. This enables us to shed light onto the threshold error rate of the toric code under circuit-level noise, i.e. the realistic noise scenario for which the toric code is believed to be among the best known QEC code(s). Here we observe that this threshold value could be significantly improved towards up to $1.4\%$. We discuss the scope and implications of our results.

\section{Kitaev's toric and surface codes}
\begin{figure}
    \centering
    \includegraphics[width=\columnwidth]{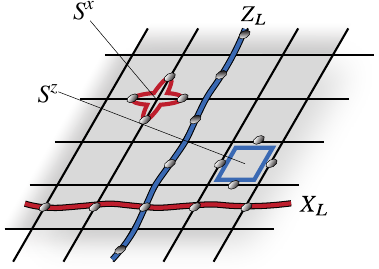}
    \caption{The stabilizer generators and logical operators defining the surface (toric) code on a square lattice. Faces (blue plaquettes) are associated with $Z$ operators on their edges, red stars are associated with $X$ operators on the edges emanating from a vertex, such that all operators commute by design. Logical operators are the non-trivial string operators on the lattice (logical $Z_L$ operators) and the dual lattice respectively (logical $X_L$ operators). In case of periodic boundary conditions these are the two non-contractible loops of the torus and similarly for open boundaries the string operators extending from one boundary to the other.
    }
    \label{fig:toric-code-lattice}
\end{figure}

\begin{figure}
    \centering
    \includegraphics[width=\columnwidth]{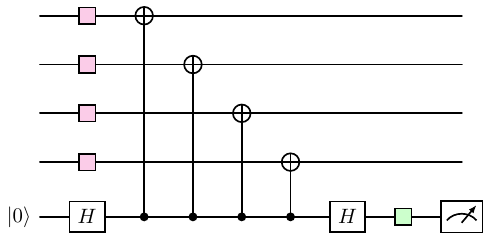}
    \includegraphics[width=0.3\columnwidth]{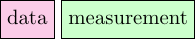}
    \caption{Quantum circuit of an $X$-stabilizer measurement. The eigenvalue of the operator $XXXX$ with support on the top four qubits is measured via coupling with four CNOT gates to an ancilla qubit (bottom qubit). The location of data qubit errors is indicated by purple boxes, here bit/phase flip errors or depolarizing errors happen with probability $p$. Phenomenological syndrome noise is indicated by the green box: the measurement outcome flips with some probability $q$. A $Z$-stabilizer measurement circuit is analogous, except we have to change the basis, which is equivalent to reversing the CNOT direction and removing the Hadamard gates.}
    \label{fig:stabilizer_measurement_circuit}
\end{figure}

The toric code~\cite{Kitaev2003} is a CSS stabilizer code, where the data qubits reside on the edges of a square lattice shown pictorially in Fig.~\ref{fig:toric-code-lattice}. The stabilizer checks consist of weight four $Z$ plaquette operators, which are the four qubits touching a face of the lattice and the weight four $X$ star operators, which are the four edges emanating from a vertex of the lattice (denoted as $S^z$ and $S^x$ in Fig.~\ref{fig:toric-code-lattice}). Given this definition of stabilizer check operators and a finite square lattice, one can either periodically close at the boundary by identifying opposing qubits, leading to the toric code. Here, two logical operators (denoted as $Z_L$ and $X_L$ in Fig.~\ref{fig:toric-code-lattice}) can be uncovered in each principal direction of the (dual) lattice, implementing two logical qubits via the non-contractible loops of the torus on the (dual) lattice. This picture can be adapted when desiring open boundary conditions leading to the surface code~\cite{Dennis2002}. In this case the number of logical qubits is reduced to one with the corresponding logical operators connecting the opposite boundaries of the lattice. Both versions in particular behave identically in the bulk and share the main properties such as threshold values. In our work we use periodic boundary conditions for convenience.
In order to detect errors, measurement qubits are needed and they are coupled to the four data qubits defining the $X$ and $Z$ check operators. As an example, Fig.~\ref{fig:stabilizer_measurement_circuit} shows the circuit for measuring the $X$ stabilizer operators. Once erroneous events happen on the data qubits, the state of the check operators will change and errors will be indicated by changes in the measurement outcomes.

\section{Noise models}
\label{sec:noise-models}
In this section we will present the noise models we are going to use throughout this work. With increasing complexity, these are two data qubit noise models, two phenomenological noise models and a circuit-level noise model. The first four will directly correspond to their dedicated statistical-mechanical model (cf. Fig.~\ref{fig:models-overview}), whereas circuit-level noise requires more intricate treatment. We choose to present all noise models here as an overview, we will revisit circuit-level noise in further depth in Sec.~\ref{sec:circuit-level-noise-toric-code}.
\subsection{Data qubit $XZ$ and depolarizing noise}

Generally, we model every operation in a circuit as an ideal operation $U_{\mathrm{ideal}}$ followed by an error $E$ drawn from an error set with a given probability $\pr(E)$. Here, we consider noise channels of the form
\begin{equation}
    \mathcal{E}_1(\rho) = (1 - \sum_{j=1}^3 p_j)\rho + \sum_{j= 1}^3 p_j E^{j}_1 \rho E^{j}_1 \\
     \label{eq:depol_single_qubit}
\end{equation}
with errors in the error set 
 $E^j \in \{ X, Y, Z \}$ (the Pauli matrices) occurring with probability $p_j = \pr(E^j)$. 

The simplest noise model called data-qubit noise, also known as code-capacity noise, assumes that the qubits comprising the code are noisy but the code can be operated ideally, i.e.~the stabilizer eigenvalues can be extracted perfectly. The data qubits in this setting are typically modeled with independent XZ noise (due to being a CSS code, $X$ and $Z$ syndrome can then be treated separately). While this is not a realistic model for a quantum memory setting, it serves as a first benchmark for stabilizer codes and an indicator for more realistic noise models. In cases where $X$ and $Z$ syndrome behave identically, one can focus on a sole Pauli type noise, as is the case for the toric code since stars and plaquettes are dual to each other (the dual lattice of the square lattice is again a square lattice). The data qubit noise model can be extended  to data qubit depolarizing noise (Pauli error $X$, $Y$, or $Z$ with e.g. probability $p/3$). Note that the difference to the former lies solely in the probability distribution ($\pr(Y)\neq \pr(X)\pr(Z)$), i.e. correlations between $X$ and $Z$, which can substantially affect code performance.
\subsection{Phenomenological syndrome noise}
In order to diagnose and remove errors, we will have to measure the stabilizer operators defining the QEC code. The measurement outcomes cannot be always trusted in practice. The first step towards modeling this situation is to flip the bit-value of the Pauli stabilizer measurement with some flip probability $q$. While still ignoring how to actually implement the stabilizer measurements, this adds a degree of realisticness to the noise model. Here we can no longer trust an individual syndrome outcome, since it may have flipped by itself instead of hinting at a closeby 'real' error on a data qubit. To overcome this problem, we will repeat the stabilizer measurements a sufficient number of times to gain confidence in distinguishing data qubit errors from measurement errors since only the former should be corrected for at the end. One such period is then called a QEC cycle -  the fact that repetition helps can be seen e.g.~by noting that an individual data qubit error will persist for all subsequent measurement rounds, whereas an individual measurement error will disappear in the subsequent round.
\subsection{Circuit-level noise}
\begin{figure}
    \centering
    \includegraphics[width=\columnwidth]{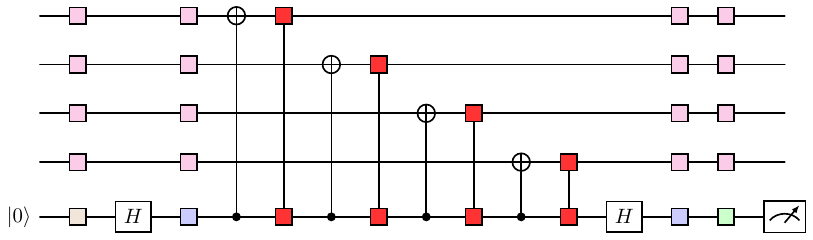}
    \includegraphics[width=\columnwidth]{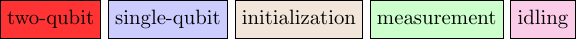}
    \caption{Quantum circuit of an $X$-stabilizer measurement in the circuit-level noise model. The eigenvalue of the operator $XXXX$ with support on the top four qubits is measured via coupling with four CNOT gates to an ancilla qubit (bottom qubit). The location of possible errors is indicated with colored squares, where a single box indicates that one of the three Paulis is applied with probability $p_{loc}/3$ (see Eq.~\ref{eq:depol_single_qubit}). CNOT errors are indicated by boxes connected with a vertical line, where one of the 15 nontrivial two-qubit Paulis is stochastically applied with probability $p/15$ (see Eq.~\ref{eq:depol_two_qubit}). Note that we suppress CNOT gates to other ancillas, every data qubit participates in a CNOT at each of the four steps.  A $Z$-stabilizer measurement circuit is analogous, except we have to change the basis, which is equivalent to reversing the CNOT direction and removing the Hadamard gates.}
    \label{fig:stabilizer_measurement_circuit_circuit_level}
\end{figure}

To actually implement the measurements of the stabilizer operators, we will typically have to resort to using ancilla qubits which we couple to the data qubits in a quantum circuit (Fig.~\ref{fig:stabilizer_measurement_circuit_circuit_level}), such that the measurement of the ancilla implements the measurement of the respective stabilizer operator. Given that the stabilizers are Pauli product operators, a single ancilla per stabilizer suffices in principle. We will use a separate ancilla for each stabilizer operator. Since we have to perform sequences of gates to implement the stabilizer, we will model those gates and the other operations of qubit initialization and qubit measurements as noisy (the collection of those are known as the circuit \emph{locations}). Note that we add the idling (identity) gate to the gate set, which we assign to qubit locations where the respective qubit has to wait for other qubits to finish their operation. We will model all errors as depolarizing errors according to the following definition: imperfect single qubit gates (and idling qubits) are modeled as the perfect gate followed by a Pauli error with probability $p$, where we choose from the three Pauli errors equally, i.e.~with probability $p/3$ (cf. Eq.~\ref{eq:depol_single_qubit}). Imperfect qubit initialization is modeled as perfect initialization followed by a depolarizing channel and imperfect measurement is modeled as a depolarizing channel followed by a perfect measurement. Imperfect two-qubit gates are modeled as perfect two-qubit gates followed by a two-qubit depolarizing channel of the form
\begin{align}
    \mathcal{E}_2(\rho) &= (1 - p_{2q})\rho + \frac{p_{2q}}{15} \sum_{j= 1}^{15}   E_2^{j} \, \rho\, E_2^{j}.\label{eq:depol_two_qubit}
\end{align}
with the error set
\begin{align}
	E_2 &\in \{\sigma_k \otimes \sigma_l, \forall k, l \in \{0, 1, 2, 3 \}  \} \backslash  \{\sigma_0 \otimes \sigma_0 \}, \nonumber
\end{align}
where $\sigma_k$ are the Pauli operators now also including the identity operator $\sigma_k = \{I, X, Y, Z \}$ with $k=0, 1, 2, 3$ and the two-qubit gate error rate $p_{2q}$. 
Let us remark here that there are slightly varying versions of circuit-level noise in the literature, in particular one could choose bit-flip noise over depolarizing noise for initialization and measurement since these operations are only sensitive in one basis. Another possible modification is to increase the two-qubit depolarizing probability relative to the single-qubit rate with the intention of equalizing the marginal probability of a single qubit error in both cases (the marginal probability of a single qubit having an error under two-qubit depolarizing probability $p$ is $4p/5$)~\cite{Stephens2014}. We discuss the case of circuit-level noise in the toric code in much further detail in Sec.~\ref{sec:circuit-level-noise-toric-code}.

\subsection{Decoding of syndrome information: maximum likelihood decoding}
The collection of syndrome information has to be interpreted by a decoder, whose task is to find a recovery operation, which ideally removes the errors that have accumulated. By construction of stabilizer codes, this means that the decoder must find a recovery operation that removes the syndrome (such that all stabilizers are fulfilled). Under this premise, the essential question is whether accumulated error plus recovery operation are a trivial error (a product of stabilizers) or a non-trivial error (a non-contractible loop), i.e.~logical error. By this token, recovery operations fall into logical cosets consisting of errors $E$ related by elements of the stabilizer group $S$: 
    \begin{equation}
\overline E := \{ E' | \exists s\in S \rightarrow E' = sE \}
\end{equation}
The optimal decoder computes the probability of all logical cosets, upon which it is trivial to choose the optimal recovery by simply choosing a representative of the most probable coset. This is known as \emph{maximum likelihood decoding} (MLD).  MLD is generally computationally hard ~\cite{Berlekamp1978,Iyer2015} and hence typically not a practical decoding strategy. However it serves as a guideline for all more practical decoders and can be computationally feasible either by approximation and/or by exploiting the structure of the QEC code. The goal of statistical-mechanical mappings will be to relate the computation of the coset probabilities into computing the partition function of a spin-model Hamiltonian.

\section{Statistical mechanics mappings}

The basic idea of statistical mechanical mappings is to construct a Hamiltonian whose Boltzmann statistics reproduce the probabilities of the logical conjugacy classes in the decoding task. For clarity of presentation we will present this in a self-contained fashion starting from the simplest case of the Random Bond Ising model, which will allow us to introduce terminology. Readers familiar with known mappings can skip to Sec.~\ref{sec:RCPGM}.

\begin{figure}
\centering
\includegraphics[width=\columnwidth]{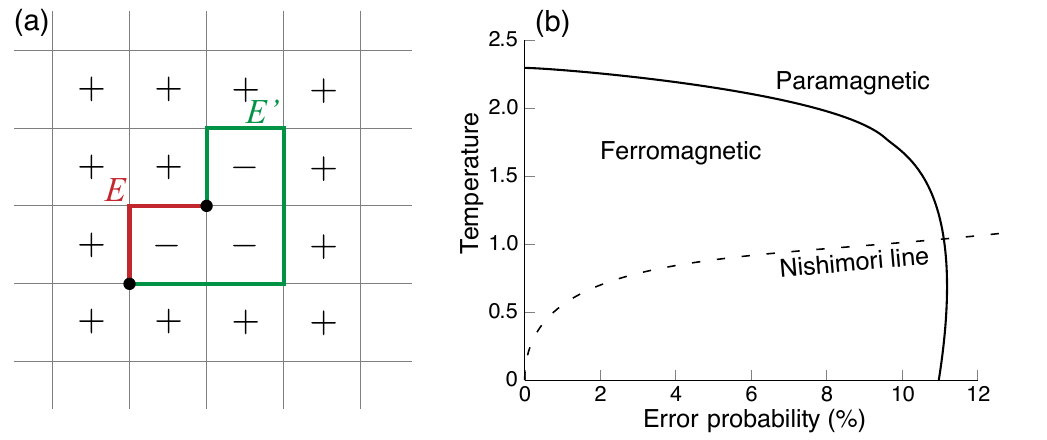}
\caption{(a) Random bond Ising model. The Ising variables $\sigma=\{\pm 1\}$ reside on the faces of a square lattice. The couplings in the error chain $E$ are antiferromagnetic (AFM) between neighboring variables. The faces with $-1$ correspond to thermal excitations of the spin variables and generate stabilizer-equivalent error chains $E'$. (b) Sketch of the phase diagram of the random bond Ising model. The solid line represents the boundary between the ordered (ferromagnetic) and disordered (paramagnetic) phase. The dashed line is the Nishimori condition. The point where the two lines cross (for $p_c = 10.9\%$) corresponds to the threshold of the toric code with perfect syndrome measurements.}\label{fig:random_bond_phase_diagram}
\end{figure}

\subsection{Random Bond Ising model}
When subjecting the toric code purely to bit-flip noise on the data qubits, the statistical model for the decoding of the syndrome is given by the quenched bond disorder two-body Hamiltonian
\begin{equation}
  H = - \sum_{\langle ij\rangle } J_{ij} \sigma_i \sigma_j \label{RBIM}
  \end{equation}
  with Ising variables $\sigma \in {\pm 1}$ on the faces of a square lattice, which are interacting with their four neighboring faces through the shared edges (see Fig.~\ref{fig:random_bond_phase_diagram}). These edges are subject to bimodal quenched bond disorder, i.e. they are made ferromagnetic with probability $1-p$  ($p_X=p$) and antiferromagnetic (AFM) with probability $p$:
  \begin{align}
    J_{ij} = \begin{cases} +J \quad \mathrm{with}\quad 1-p \\ -J \quad \mathrm{with}\quad p \end{cases}
  \end{align}
  Note that the latter (AFM bonds) are frequently also referred to as "wrong-sign" bonds.
  The condition that the Boltzmann factor of thermal excitations is consistent with the QEC noise model probabilities of flipping qubits is expressed as the Nishimori condition
  \begin{equation}
  e^{-2J} = \frac{p}{1-p}. \label{Nishimori_RBIM}
  \end{equation}
  The original derivation can be found in~\cite{Dennis2002}.

  The interpretation of the model is that by drawing the quenched bond disorder we draw a reference error configuration of the toric code, which is generated with its respective probability according to the disorder probability prescription. This pins down the syndrome configuration, which is revealed by the syndrome measurements. Given one such quench-disordered Hamiltonian, we then study thermal excitations of the spin variables in that model, whose role is to generate stabilizer-equivalent configurations, i.e. flipping a single $\sigma$ variable corresponds to applying the stabilizer generator sitting at that variable on the lattice. (The correct statistics being enforced by the Nishimori condition.) The picture is that error chains correspond to domain walls whose endpoints, called Ising vortices, correspond to the syndrome. Transitioning between equivalent configurations corresponds to fluctuations of domain walls in the model. As long as these domain walls remain localized, essentially all configurations belong to the same conjugacy class, such that we can recover from the error with probability approaching unity in the large system limit. As we increase the error probability beyond a critical value, domain walls start to delocalize, such that we no longer can be sure that all configurations correspond to the same conjugacy class and lose the ability to reliably recover from the error. In the RBIM, the transition coincides with a ferromagnetic to paramagnetic phase transition, such that one can use on-site magnetization as an order parameter to find the well-known critical value $p_c = 10.9\%$ corresponding to the threshold of the toric code with perfect syndrome measurements (see Tab.~\ref{tab:models_and_thresholds}).

  \subsection{Random eight vertex model}
  \begin{figure}
    \centering
    \includegraphics[width=\columnwidth]{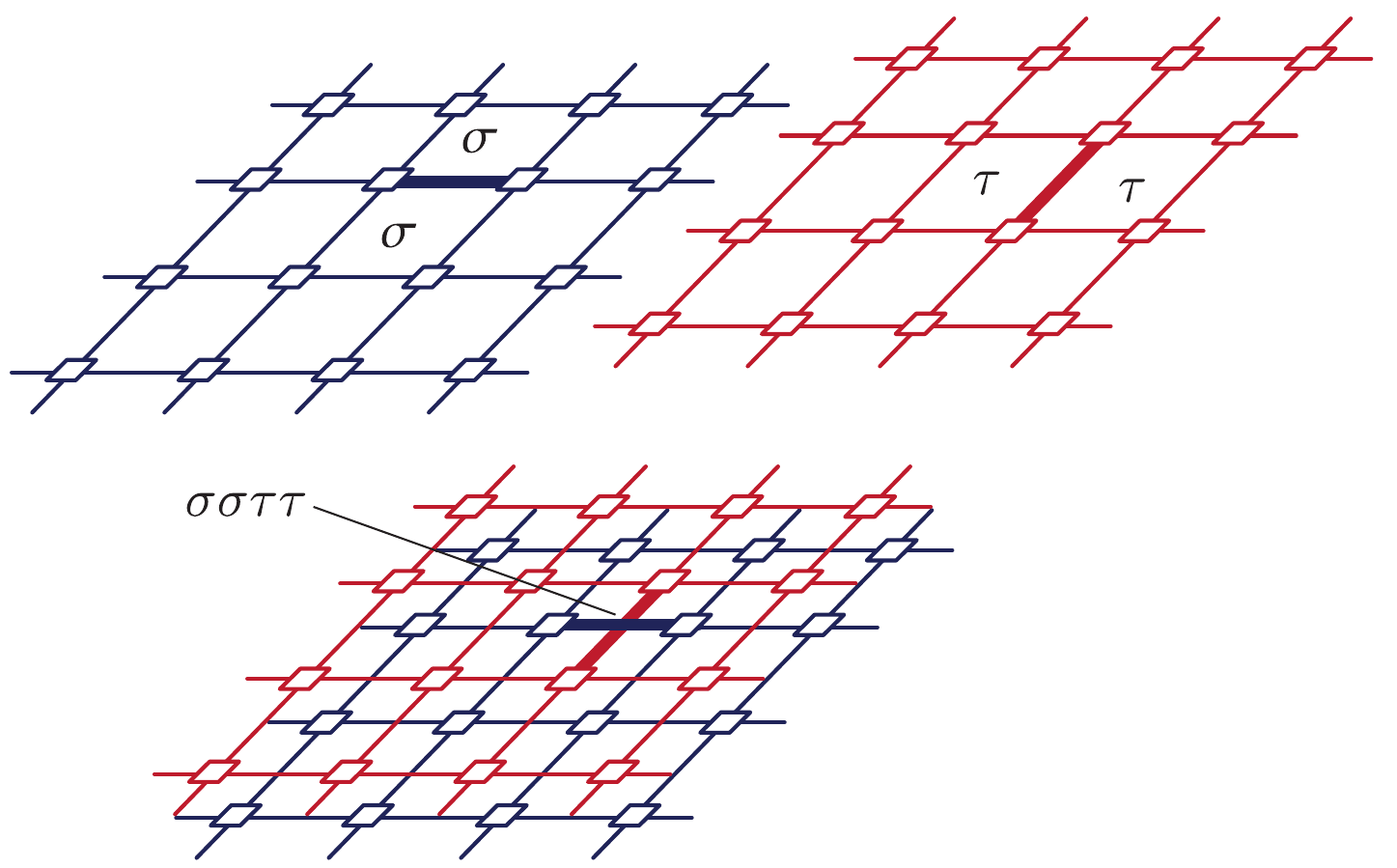}    \caption{The lattice of the Random Eight Vertex Model can be viewed as two juxtaposed square lattices accounting for $X$ and $Z$ syndrome. The edges of the lattice correspond to the qubits of the surface code. As in the RBIM, the variables on the faces of the lattice interact with neighboring faces via shared edges (with Hamiltonian terms of the form $\sigma \sigma$ or $\tau \tau$ similarly to Eq.~\eqref{RBIM}).  In the Random Eight vertex model Eq.~\eqref{random-eight-vertex}, a new interaction (of the form $\sigma\sigma\tau\tau$) term arises which is given by a cross involving one edge of each sublattice simultaneously, mediating an interaction between four spins (two of each sublattice) and thus coupling the two sublattices. Introducing this interaction allows to account for $Y$-errors.}
    \label{fig:R8VM}
\end{figure}

  Let us now illustrate how to incorporate Pauli-$Y$ errors into the model, as was done in~\cite{Bombin2012}. First of all, given that $Y$ affects both $X$- and $Z$- syndrome, we now have to consider both by considering two sets of spin variables. They each form a square lattice among themselves and moreover the spins of the two lattices will get coupled with an interaction term. Qubits are subjected to noise of all three Pauli types, which each come with a certain probability we write as $\pr(X), \pr(Y)$ and $\pr(Z)$. The Hamiltonian is

  \begin{equation}
    H = - \sum_{\sigma, \tau} J(X) \sigma \sigma + J(Y)\sigma\sigma \tau \tau + J(Z) \tau \tau, \label{random-eight-vertex}
  \end{equation}
where for each term, the two $\sigma$ variables and/or the two $\tau$ variables are the respective plaquette variables incident on the respective qubit, as shown in Fig.~\ref{fig:R8VM}. The interaction $J(W)$ ($W\in X,Y,Z$) is again bi-modally disordered: it has absolute value given by the following Nishimori conditions ($W \in X,Y,Z$):
  \begin{equation}
    \exp\left( -4 |J(W)| \right) = \frac{\pr(X)\pr(Y)\pr(Z)}{\left(\pr(W)\right)^2 \pr(\mathds{1})}\label{Nishimori-eight-vertex}
    \end{equation}
    and we flip the sign of the two terms conjugate with the error, i.e. with probability $\pr(X)$, we flip the sign of $J(Y)\rightarrow -J(Y)$ and $J(Z)\rightarrow -J(Z)$ making them AFM "wrong-sign bonds" and cyclically for $\pr(Y)$ and $\pr(Z)$.

        For uniform depolarizing noise we set all three probabilities equal: $\pr(W) = p/3$. This leads to
    \begin{equation}
      \exp(-4J) = \frac{p}{3(1-p)},
      \end{equation}
      such that all three interaction strengths are the same in magnitude. In this case, we can make contact with the eight-vertex model, which is given by switching off the disorder completely. In this edge case, which is the $p=0$ axis in the $(p,T)$ phase diagram and thus away from the Nishimori condition, the model is rendered analytically solvable and among other things the critical temperature of the phase transition is known to be $T_c = \frac{4}{\log(3)}$. Of course we are interested in the case with disorder according to the Nishimori condition, the corresponding numerical investigation was performed in Ref.~\cite{Bombin2012}, which reports a threshold value of
      \begin{equation}
        p_c = 18.9\%.
      \end{equation}

To interpret the role of the coupling term $J(Y)$, it is enlightening to note that for the case of independent XZ noise, we recover the random bond Ising model as follows. Due to
    \begin{align}
      \pr(X) &= p_X\cdot (1-p_Z)\\
      \pr(Y) &= p_X\cdot p_Z\\
      \pr(Z) &= (1-p_X)\cdot p_Z,
    \end{align}
    it follows that $J(Y) = 0 $, i.e. the interaction between the two lattices vanishes and we can study the $\sigma $ variables independently from the $\tau$ variables. Accordingly, the Nishimori conditions also reduce to Eq.~\eqref{Nishimori_RBIM}. A qualitative interpretation of the role of the coupling term $J(Y)$ thus would be that it represents how "genuine" Y errors are, i.e. how much the error distribution deviates from independent $X$ and $Z$ errors. Let us point out here that with the departure from independent XZ noise, the correspondence of disorder bonds and error chains becomes more subtle: whereas for the RBIM, chains of AFM bonds directly correspond to the reference error configuration, now a certain Pauli error on a particular qubit flips the sign of the two coupling terms of the Paulis conjugate to the drawn error. \\ 

\subsection{Comparing bit-flip noise against uniform depolarizing noise}
It can be a bit subtle to compare noise rates when looking at different noise models. The threshold under bit-flip noise is $0.109$, but how does this compare to $0.189$ under depolarizing noise? The situation to imagine is that we are presented a syndrome and we think it comes from a situation best described by XZ noise, however the data was actually generated by a uniform depolarizing noise model. This leads us to decode effectively ignoring the existence of $Y$ errors, which means the bit-flip probability of every qubit would amount to $p_{bf} = 2p_{depol.}/3$ (same for its phase-flip probability). This entails that we would find a threshold of $3/2\cdot 0.109 = 0.1635 = 16.35\%$. This observation can be interpreted in the way that by taking the real noise model into account, i.e. the correlations between X and Z syndrome, we managed to increase the threshold value from $16.35\%$ to $18.9\%$, i.e. by a relative factor of $15\%$.

\section{Random plaquette gauge model}
\begin{figure}
    \centering
\includegraphics[width=\columnwidth]{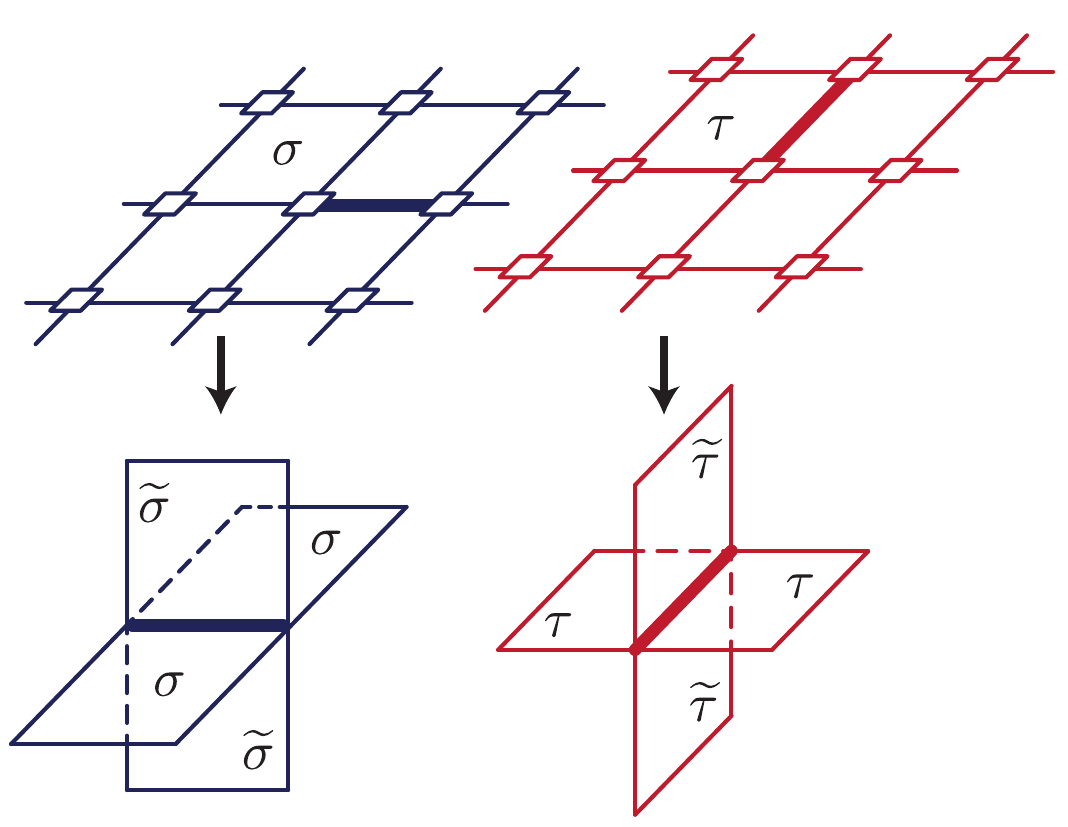}    
    \caption{Repeating the syndrome measurements to deal with syndrome noise adds a third dimension to the lattice. In this picture, qubits are identified with edges. Whereas previously in the perfect syndrome case, equivalences were generated by the spin variables $\sigma$, we now also get time-like equivalences generated by the variables $\tilde\sigma$. The corresponding interaction term mediated by the edge (corresponding to a qubit at a particular moment in time) is now lifted to a four body interaction, where the time-like equivalences incident on a qubit take part in the interaction alongside the two spatial equivalences.}
    \label{fig:rbim_rpgm_bonds}
\end{figure}

When the syndrome information is not reliable, we model this by a probability $q$ that the syndrome bit has been flipped. To deal with this type of noise, we repeat syndrome measurements, adding a discrete time dimension to the model. This entails that ``stabilizer-equivalences'', i.e.~error configurations that differ by the application of a code stabilizer, now generalize to so called space-time equivalences. Here, distinct error configurations of qubit and syndrome errors are now possibly equivalent if they produce the same syndrome volume. Remarkably, also the space-time equivalences can be generated systematically and for the toric code they behave similarly to stabilizer-equivalences. The generator for this is the event configuration, where a single data qubit error happens at a given round, but its signal on the two adjacent syndrome bits happens to be suppressed due to two measurement errors sitting on those two syndrome bits in the same round, which is then followed by another data qubit error in the subsequent round, such that in total, there is no data qubit error remaining and we did not see any signal on the syndrome information. By extending the square lattice into a cubic lattice, as shown in Fig.~\ref{fig:rbim_rpgm_bonds}, we can generate all possible space-time equivalences by assigning Ising variables to the time-like plaquettes (on the space-like edges the original spin variables generating the stabilizer-equivalences remain unchanged). We again introduce Ising variables, which receive the same Greek letter as their counterparts from before but we add a tilde to signify that they are time-like plaquettes. Firstly, focusing on the $\sigma$ lattice, the interaction term in the Hamiltonian again involves all plaquettes adjacent to a given qubit, which now increases from two to four $\sigma$-like variables:
\begin{equation}
h_{s.l.} = J_z \sigma\sigma \tilde \sigma \tilde \sigma,
\end{equation}
which applies to all space-like edges (i.e. all qubits at a given time-step). This Hamiltonian is a generalization of Eq.~\ref{RBIM}, where two time-like $\tilde\sigma$ now take part in the interaction. On top of that we furthermore get interactions around the time-like edges (not shown in Fig.~\ref{fig:rbim_rpgm_bonds})
\begin{equation}
h_{t.l.} = J_q \tilde\sigma\tilde\sigma \tilde \sigma \tilde \sigma
\end{equation}
with the accompanying Nishimori condition
  \begin{equation}
  e^{-2|J_q|} = \frac{q}{1-q}. \label{Nishimori-q}
  \end{equation}
To build (or confirm) some intuition, we can wonder what the effect of flipping a $\sigma$ ($\tilde\sigma$) variable is. For simplicity, we start in the all plus configuration. If we flip a $\sigma$, we pay an energy $2J_z$ for every edge around that plaquette, which under the Nishimori condition (Eq.~\eqref{Nishimori_RBIM}) corresponds to the relative probability $e^{-8J_z} = \left(\frac{p}{1-p}\right)^4$, the probability of activating one stabilizer. If we flip a $\tilde \sigma$, we note that this involves two space-like edges and two time-like edges, i.e.
\begin{equation}
  e^{-4J_q-4J_z} = \frac{p^2q^2}{(1-q)^2 (1-p)^2},
\end{equation}
which corresponds to the probability of having two data errors and two syndrome errors. Analogously to the situation with perfect syndrome, we also add quenched randomness to the interaction by choosing
  \begin{align}
    J_{q} = \begin{cases} +|J_q|  \quad \mathrm{with}\quad 1-q \\ -|J_q| \quad \mathrm{with}\quad q \end{cases}.
  \end{align}
  As a consistency check, we can recover the Random Bond Ising model by sending the syndrome noise parameter to zero $q\rightarrow 0$. In this case, the interaction strength diverges as $|J_q|\rightarrow +\infty $, such that all $\tilde\sigma$ variables freeze in. In effect we can essentially just ignore the $\tilde\sigma$ parts of the Hamiltonian, which leaves us with the Random Bond Ising model.
This model is frequently presented on the lattice dual to the one presented above. Under duality, vertices become cells, edges become faces, faces become edges and cells become vertices. This entails that the interaction, which on the original lattice acts between the four faces incident on an edge is now embodied by the plaquette consisting of the four edges (i.e. now the spin variables) that are interacting (this plaquette of the dual lattice is the plaquette pierced by the edge on the original lattice). Note that the notion of time-like and space-like also gets reversed under the duality mapping, i.e. from now on a time-like plaquette is the interaction term involving a qubit and a space-like plaquette describes the interaction stemming from a measurement (syndrome) error. Let us remark here that this is the picture lending the name random plaquette gauge model: the interaction is mediated by plaquette terms, which are furthermore randomly quenched with the concentration of "wrong-sign plaquettes" given by the error probability $p$ ($q$). For the noise model of bit-flip data noise and syndrome noise, where we furthermore set the noise rates equal, i.e. $p=q$, the two interaction strengths become the same and the model becomes uniform. The Hamiltonian then reads
\begin{equation}
  H = - \sum_{ ijkl \in \square } J_{\square} \sigma_i \sigma_j\sigma_k \sigma_l.
\end{equation}
This model (also the non-uniform generalizations) contains a non-trivial transformation of the states, which leaves the Hamiltonian invariant and is therefore a gauge symmetry. This transformation is given by flipping all variables incident on a vertex of the lattice (i.e. all faces around a cube on the original lattice). The model is related to Wegner's $\mathbb{Z}_2$ (Ising) lattice gauge theory, at least in the case of switching off disorder ($p=0$) - since the interaction is mediated via the plaquettes and they contain quenched randomness, this model is called the Random Plaquette Gauge Model (RPGM). This model was numerically studied in~\cite{Ohno2004,Kubica2018} and the critical point was determined to be
\begin{equation}
p_c = 3.3\%.
\end{equation}

\section{Random coupled-plaquette gauge model}
\label{sec:RCPGM}
\begin{figure}
    \centering
    \includegraphics[width=0.9\columnwidth]{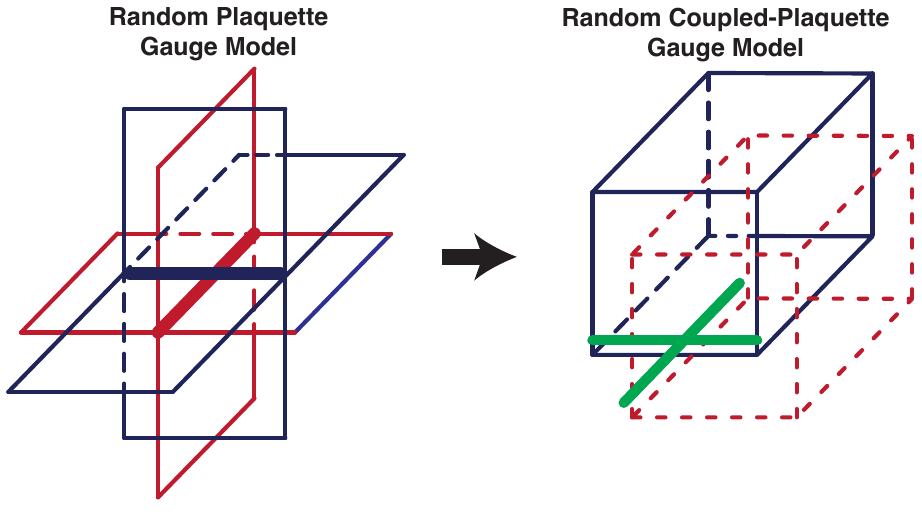}
\caption{Similarly to going from the RBIM to the R8VM (Fig.~\ref{fig:R8VM}), we can go from the random plaquette gauge model (LHS)  to the random coupled-plaquette gauge model (RHS) by introducing an interaction term coupling both cubic sublattices. The LHS shows $X$ and $Z$ syndrome lattices juxtaposed. LHS: An edge corresponding to $J(X)$ and the corresponding $J(Z)$ edge on the conjugate syndrome lattice coexist independently. RHS: a new interaction term J(Y) is introduced, which is a combination of the edges on both sublattices and thus a coupling between both syndrome sublattices, leading to a unit cell consisting of two cubes. Note that the figures are showing the lattice, where interactions are mediated via edges on a cubic lattice. On the dual lattice, these interactions turn into plaquettes, or \emph{coupled plaquettes} for the case of $J(Y)$. We abstain from attempting to draw the dual lattice for readability. }
    \label{fig:RCPGM}
\end{figure}

The extensive presentation of known statistical mechanical mappings in the previous section has shown the basics of mapping syndrome decoding problems to quenched disorder Hamiltonians, in particular we were able to see how to either include $Y-$errors or how to include noisy syndrome measurements. The goal of this section is to present a new statistical mechanical model that unifies these two directions into a new model we call the random coupled-plaquette gauge model (RCPGM). Starting from the perspective of the random eight vertex model, we have to generalize the Hamiltonian from Eq.~\ref{random-eight-vertex} to three dimensions. In analogy to the random plaquette gauge model, the interaction terms generalize as $\sigma\sigma \rightarrow \sigma\sigma\tilde\sigma\tilde\sigma$ and $\tau\tau \rightarrow \tau\tau\tilde\tau\tilde\tau$ for the time-like interaction terms. There are also the space-like plaquette interaction terms $J_q \tilde\sigma\tilde\sigma\tilde\sigma\tilde\sigma$ and analogously for $\tilde\tau$. In total, this leads to the Hamiltonian

\begin{align}
  \begin{split}
  H = \\
  - &\sum_{\mathrm{time-like}\square}J(X) \sigma \sigma \tilde \sigma \tilde\sigma + J(Y)\sigma\sigma \tilde\sigma\tilde\sigma \tau \tau \tilde\tau\tilde\tau+ J(Z) \tau \tau \tilde\tau\tilde\tau \\
  + &\sum_{\mathrm{space-like}\square} J_q \tilde\sigma\tilde\sigma\tilde\sigma\tilde\sigma + J_q \tilde\tau\tilde\tau\tilde\tau\tilde\tau,  \label{random-coupled-plaquette}
  \end{split}
\end{align}
where the eight-body interaction applied to a ``double-plaquette'', i.e. the plaquette of the $\sigma$-lattice together with the accompanying plaquette on the $\tau$ lattice around the same qubit. The Nishimori conditions are the same as above, i.e. Eq.~\eqref{Nishimori-eight-vertex} for $J(W), W \in X,Y,Z$ and Eq.~\eqref{Nishimori-q} for $J_q$. 
  \begin{align}
    \exp\left( -4 |J(W)| \right) = \frac{\pr(X)\pr(Y)\pr(Z)}{\left(\pr(W)\right)^2 p(\mathds{1})} \\
    \exp\left( -2 |J_q| \right) = \frac{q}{1-q}
    \label{Nishimori-RCPGM}
    \end{align}
Again analogously to the random eight vertex model above, we draw the disorder distribution according to the respective error probabilities, i.e. with $\pr(X)$ we flip the conjugate terms $J(Y)$ and $J(Z)$ (and cyclically for $Y$ and $Z$), whereas $J_q$ flips with probability $q$.

\section{Monte Carlo simulations of Random Coupled-Plaquette Models}

The statistical physics models in this work are variants of quenched disorder spin systems. These models are known to possess energy landscapes with many local minima, notably related to the field of spin glasses. Thus, we perform Parallel Tempering Monte Carlo simulations, where instead of performing Metropolis updates on a single configuration, we take an ensemble of configurations at different temperatures and interleave the Metropolis updates on each of those configurations with updates that swap configurations close in temperature, known as Parallel Tempering updates~\cite{Earl2005}. Parallel tempering steps range from high temperature to low temperature to facilitate transitions from local minima.

First, wrong-sign bond or wrong-sign plaquette interaction configurations for a given Hamiltonian is drawn with a given quenching probability. For the given configuration of interactions, we perform a fixed number of Metropolis steps for the thermalization at each temperature. Then for the measurement of observables, we take a certain number of runs for Metropolis updates followed by a parallel tempering step \cite{Earl2005} between neighboring temperatures starting from high temperature. Measurements of the observables are binned in a regular interval between these combined Monte Carlo updates. The whole process is repeated over different quenched interaction configurations. Thus, in general, there are two different averages: one is over the thermal ensemble and the other over random configurations of wrong-signs. 

Below a critical concentration of wrong-sign plaquettes, elementary plaquettes dominantly have a $+$ sign at low temperature and have non-vanishing average plaquette value ("Higgs phase"). At high temperature, plaquettes have both signs equally and have small average value ("confining phase"). To distinguish these two different phases, we can employ an order parameter. For gauge theories, any order parameter for the phase diagram has to be a gauge invariant quantity due to Elitzur's theorem~\cite{Elitzur1975}. In  Wang et al.~\cite{Wang2003} and Andrist et al.~\cite{Andrist2010}, the Wilson loop, 
\begin{equation}
\left< W_C \right> = \left< \prod_{i\in C} \sigma_i \right>,
\end{equation}
where $C$ denotes any closed curve on the lattice, is used as the order parameter for the studies of the RPGM. Wang et al. considered whether the Wilson loop follows the area law or the perimeter law to distinguish the phase and studied the transition at $T = 0$ in detail using the homology of error chains. Ohno et al.~\cite{Ohno2004} and Kubica et al.~\cite{Kubica2018} investigated the specific heat, in addition to the Wilson loop behavior. Andrist et al. studied the cumulant of the elementary (i.e. smallest area) Wilson loop to locate the thermal transition temperature.

In this work, we use the Polyakov line $P(\mathbf{x})=\prod_{t} \sigma_{\mathbf{x},t}$ as the order parameter, which is routinely used in studies of Yang-Mills theory (e.g., ~\cite{Borsanyi2022}) and is closely related to the Wilson loop. We consider the third order cumulant together with the susceptibility of the Polyakov line,
\begin{equation}
\langle |\overline{P}| \rangle, \;\; \overline{P} = \frac{1}{L^2}\sum_{\mathbf{x}} P (\mathbf{x}) = \frac{1}{L^2} \sum_{\mathbf{x}} \prod_{t} \sigma_{\mathbf{x},t} , \label{eq:polyakov}
\end{equation}
where $\mathbf{x}$ denotes the space-like sites and $\prod_t$ means taking a product along the time-direction at a given $\mathbf{x}$. Due to the periodic boundary condition, the Polyakov line is gauge-invariant. Since the Polyakov line in our model is a product of Ising spin variables, the Polyakov line at $\mathbf x$ itself has $\pm$-sign and $\langle \overline{P} \rangle$ serves as the "average magnetization" over the lattice volume and is less susceptible to short distance fluctuation since the product in Eq. \ref{eq:polyakov} is over the entire time-direction. Then, the susceptibility for the average Polyakov line and the third order cumulant are defined respectively as
\begin{equation}
\chi = \langle \tilde{P}^2 \rangle , \;\;\;\;\; B_3 = \langle \tilde{P}^3 \rangle / \langle \tilde{P}^2 \rangle^{3/2} 
\end{equation}
with $\tilde{P} = |\overline{P}| - \langle |\overline{P}|\rangle$.

Andrist et al.~\cite{Andrist2010} observe that the transition in the RPGM is generally of first order, signified by a double peak structure in the histogram of the smallest area Wilson loop expectation value distribution. The double peak structure can be analyzed more specifically by measuring the skewness of this distribution, which is related to the third order Binder cumulant. Thus, we also adopt the third-order cumulant and the susceptibility of the Polyakov line in our study of RCPGM in contrast to those of the smallest Wilson loop in ~\cite{Andrist2010}. Note that for a first order phase transition, the correlation length is finite and finite size scaling of the susceptibility of the order parameter shows $\sim L^{-d}$ ~\cite{binder1984finite} in contrast to a divergent behavior of the susceptibility peak of a critical transition.

Fig.~\ref{fig:order-parameter_RPM} shows typical behaviors of the average Polyakov line across different noise probabilities. Without wrong sign plaquettes (i.e., $p = 0$), behavior of the average Polyakov line shows a well-defined transition temperature in the infinite volume limit. Well above the threshold probability (an example at $p = 0.852\%$ is shown in said figure), the average Polyakov line does not show a transition as the lattice volume increases. Below and near the threshold probability ($p = 0.682\%$), the order parameter still shows a transition. The third-order cumulant and the susceptibility corroborate this observation. In Fig.~\ref{fig:B3_RPM} and Fig.~\ref{fig:susc_RPM}, $B_3$ for crosses zero at the temperature where $\chi$ reaches a peak. Well above the threshold probability ($p = 0.852\%$), even at low temperature, $B_3$ does not cross the zero and $\chi$ does not reach a peak. Between these two extreme noise cases, $B_3$ crosses zero and $\chi$ still shows a peak at a similar temperature. We note that finite volume effects are important near the threshold probability. For example, for $p = 7\%$ in Fig. \ref{fig:B3} and Fig. \ref{fig:susc}, the transition temperature does not give a limiting value as the lattice volume increases.

Details of the parameters chosen for Monte Carlo study of various statistical mechanics models reported in this work are given in Appendix~\ref{app:MonteCarlo}.

\section{Threshold of the Random Coupled-Plaquette Gauge model -  uniform noise}
The natural setting to explore the RCPGM is a uniform noise model with depolarizing data error rate set equal to the syndrome error rate $p=q$. This leads to
\begin{align}
    \exp\left( -4 |J_x| \right) = \frac{p}{3(1-p)}\\
    \exp\left( -2 |J_q| \right) = \frac{p}{1-p},\\
    |J_x|=|J_y|=|J_z|
    \label{Nishimori-RCPGM-uniform}
    \end{align}
which we can translate to a ratio between spatial and temporal couplings of
\begin{align}
\exp\left( -4 |J_x| \right) = \frac{1}{3}\exp\left( -2 |J_q| \right)\\
\Leftrightarrow |J_q| = 2|J_x|-\frac{1}{2}\log_e(3)
\end{align}
and subsequently explore the model as a function of $J_x$ and $p$ alone. The threshold will correspond to the crossing of the Nishimori condition of $J_x$ with phase-boundary in the $p-T_c$ diagram where $T_c$ is to be understood as $\frac{1}{\beta_c J_x}$.

\begin{figure}
    \centering
    \includegraphics[width=0.8\columnwidth]{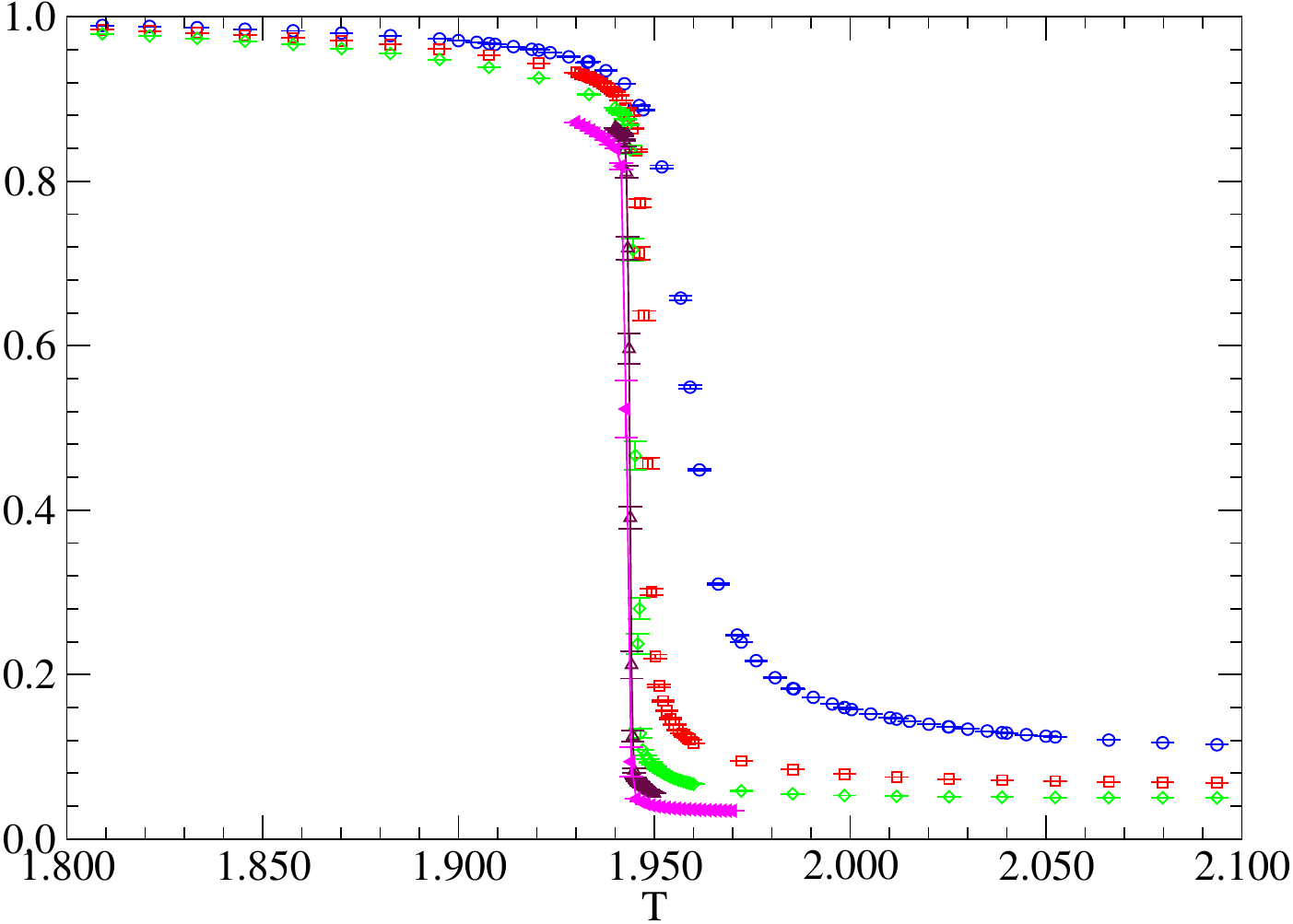}
   \includegraphics[width=0.8\columnwidth]{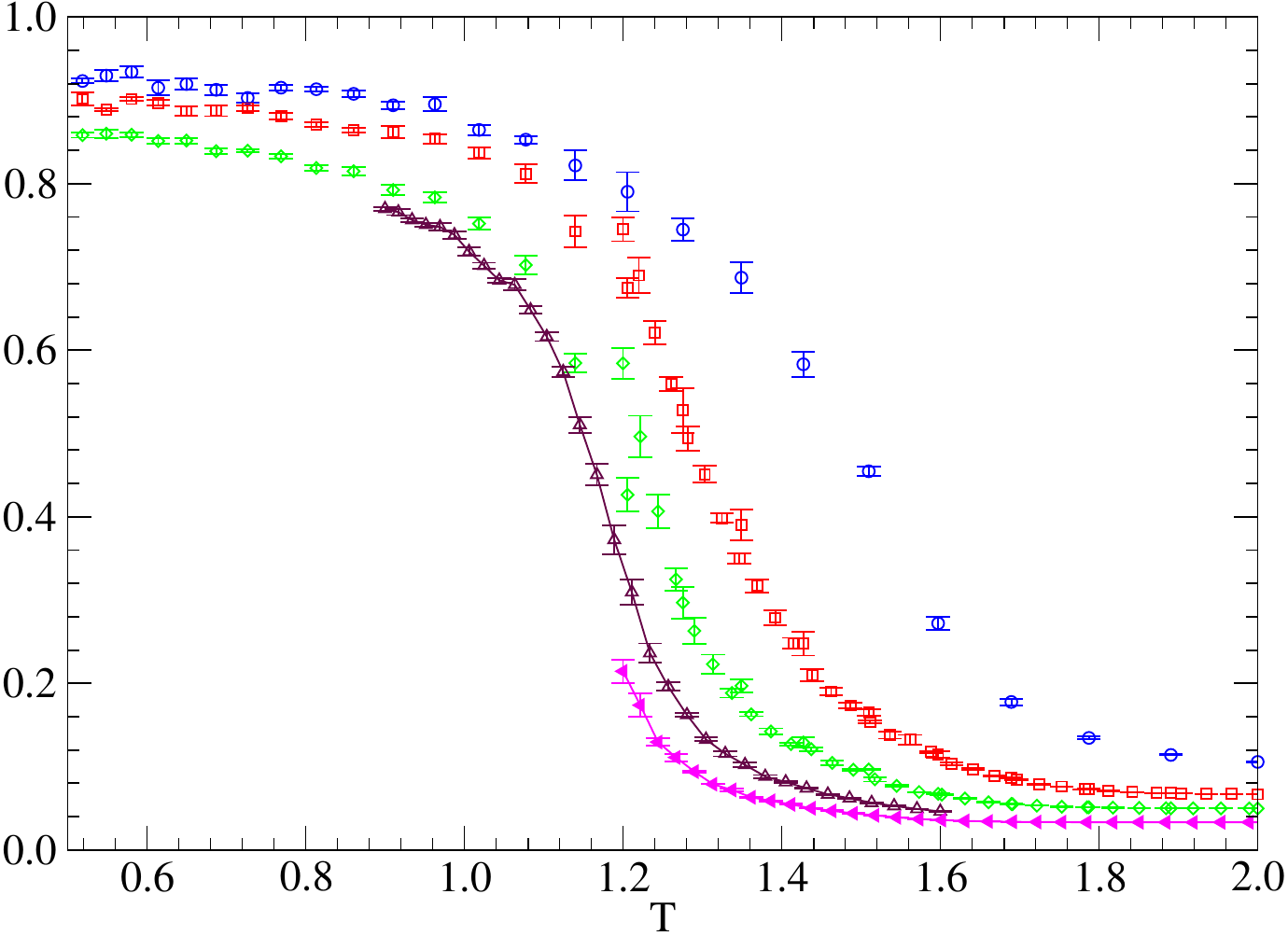}
    \includegraphics[width=0.8\columnwidth]{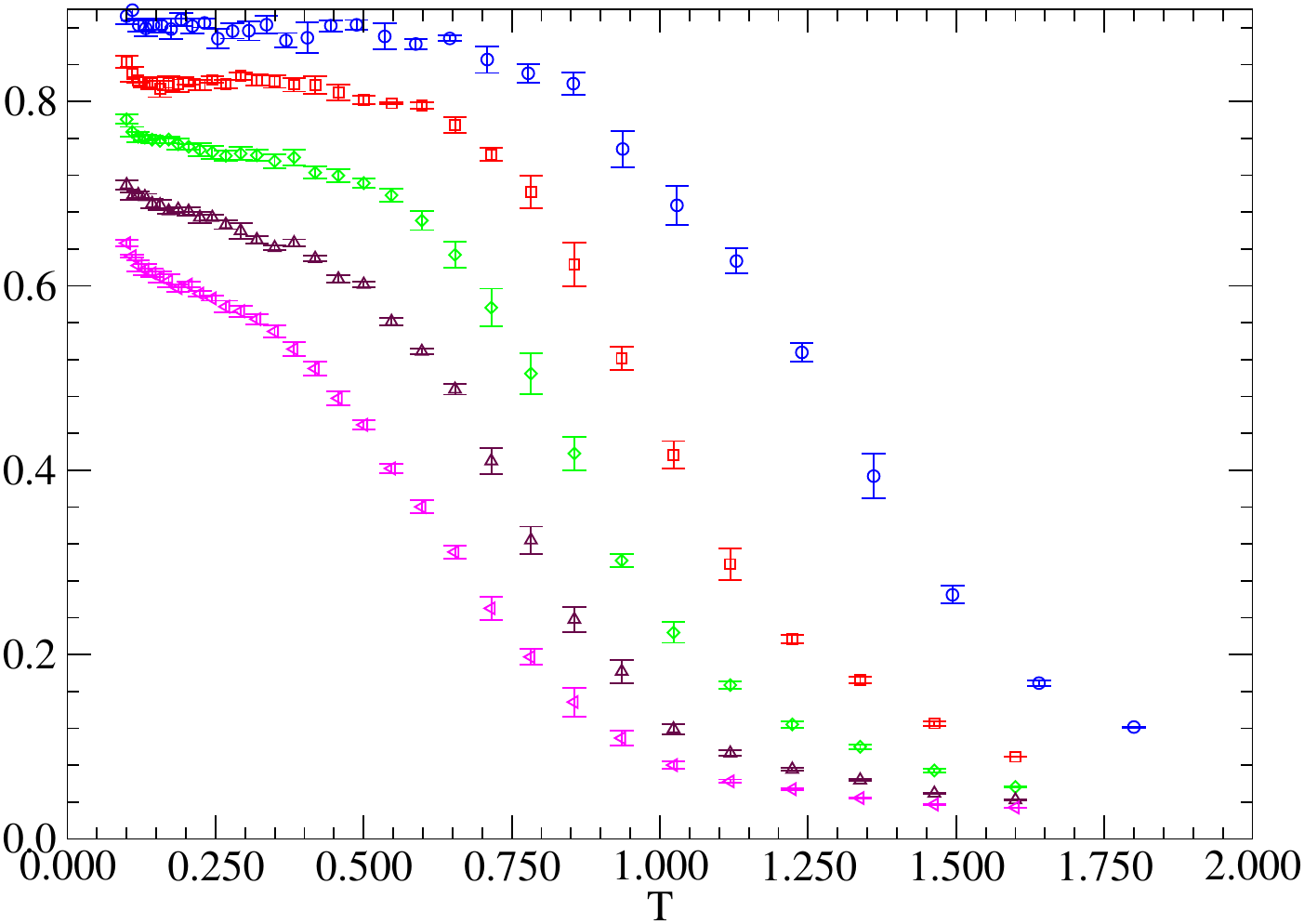}
    \includegraphics[width=0.8\columnwidth]{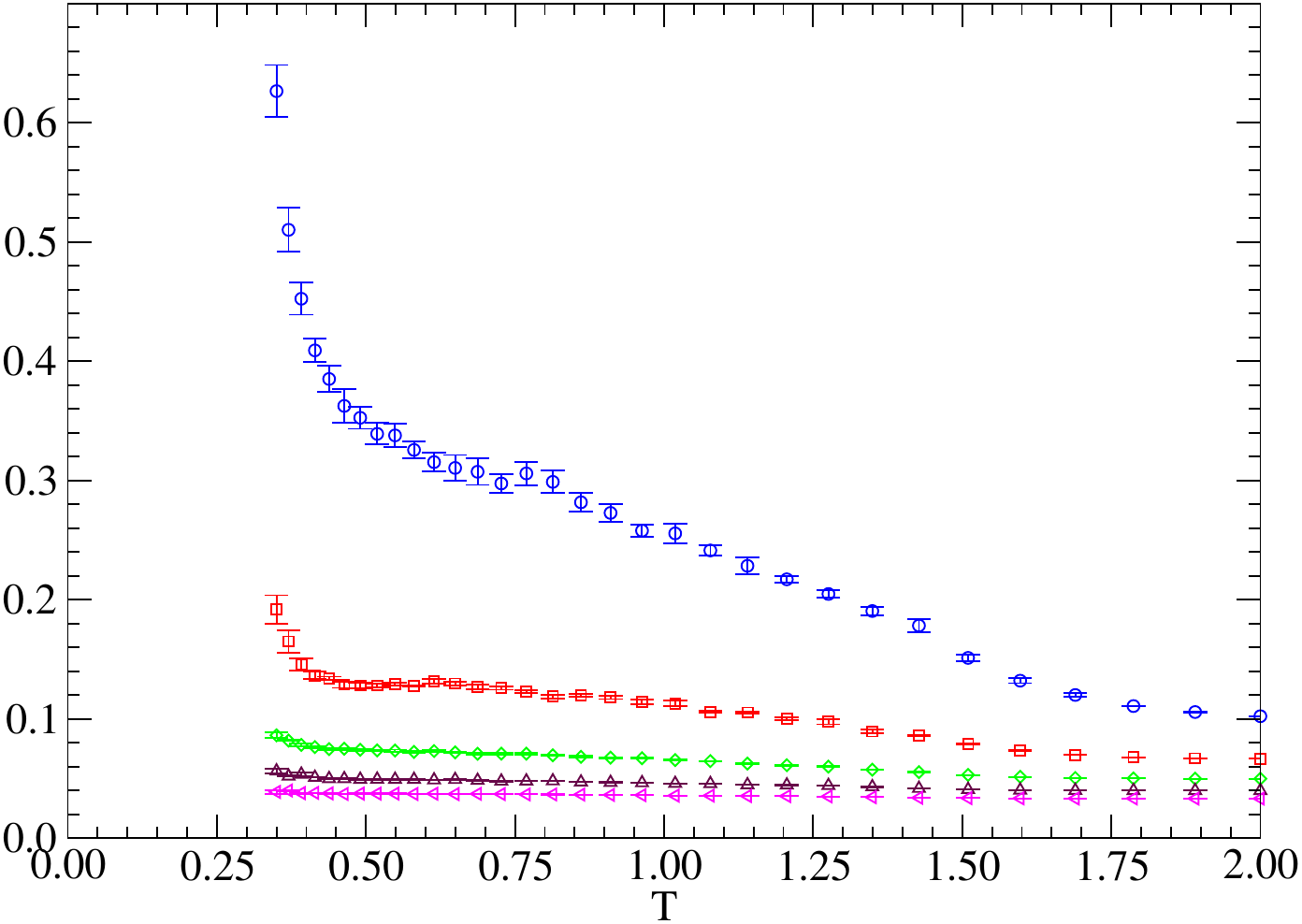}
     \caption{Temperature behavior of the average Polyakov line at $p = 0.0$ (top), $p = 6\%$ (second), $p = 7\%$ (third), and p$ = 9\%$ (fourth) on $8^3$ (blue circle), $12^3$ (red square), $16^3$ (green diamond), $20^3$ (maroon up-triangle), and $24^4$ (magenta left-triangle) for symmetric depolarizing noise random coupled-plaquette gauge model.}
    \label{fig:order-parameter}
\end{figure}

\begin{figure}
    \centering
    \includegraphics[width=0.8\columnwidth]{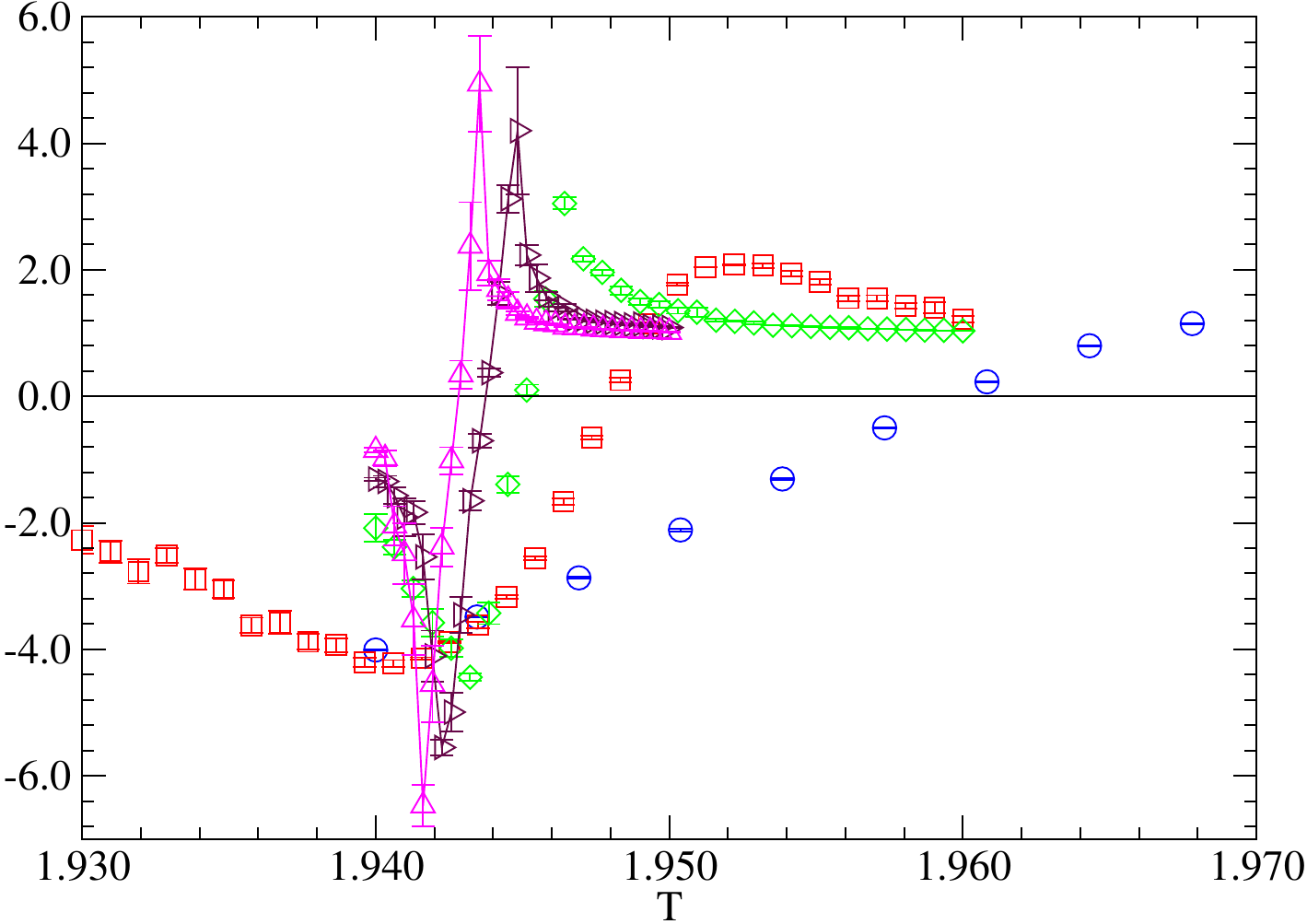}
    \includegraphics[width=0.8\columnwidth]{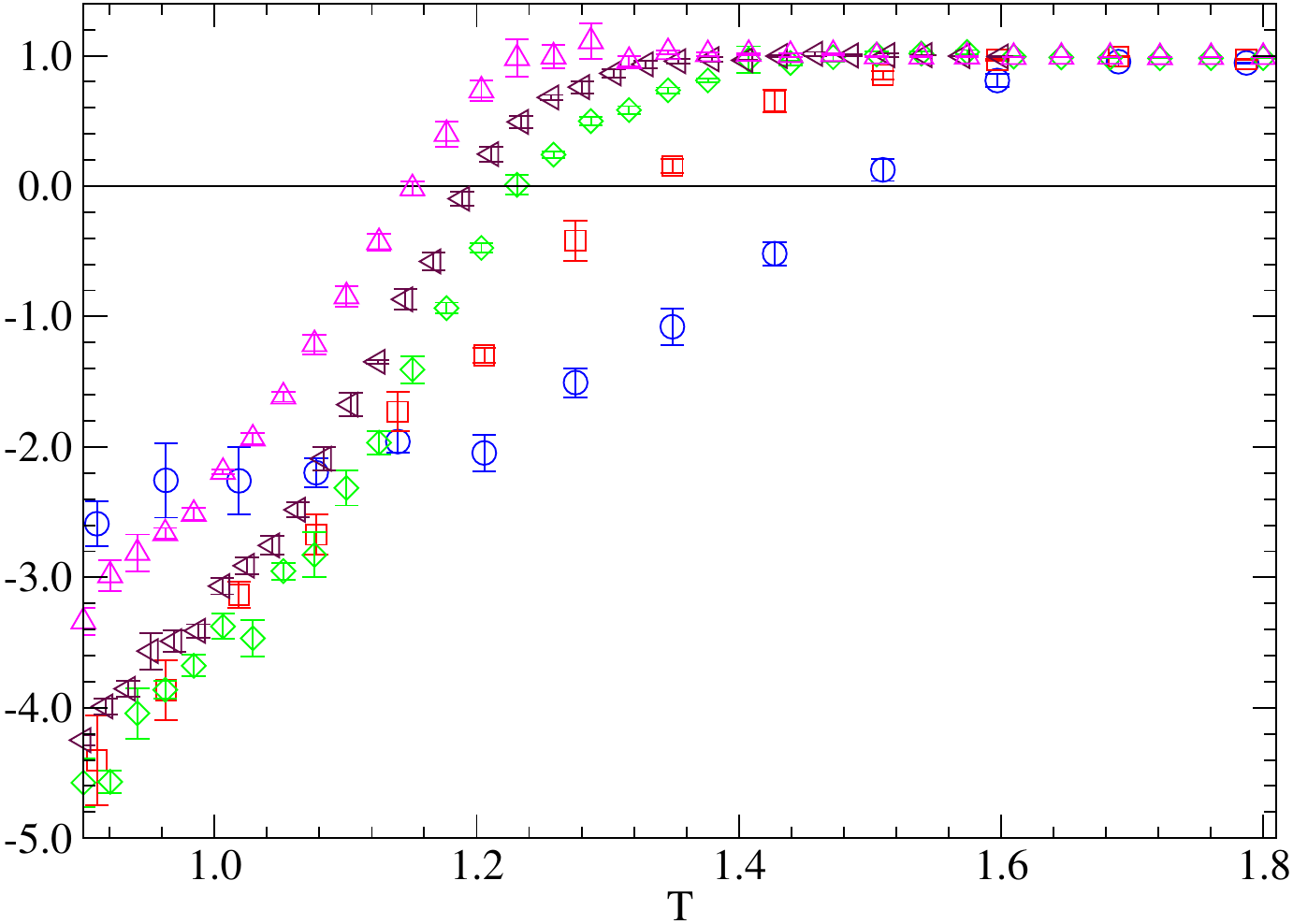}
    \includegraphics[width=0.8\columnwidth]{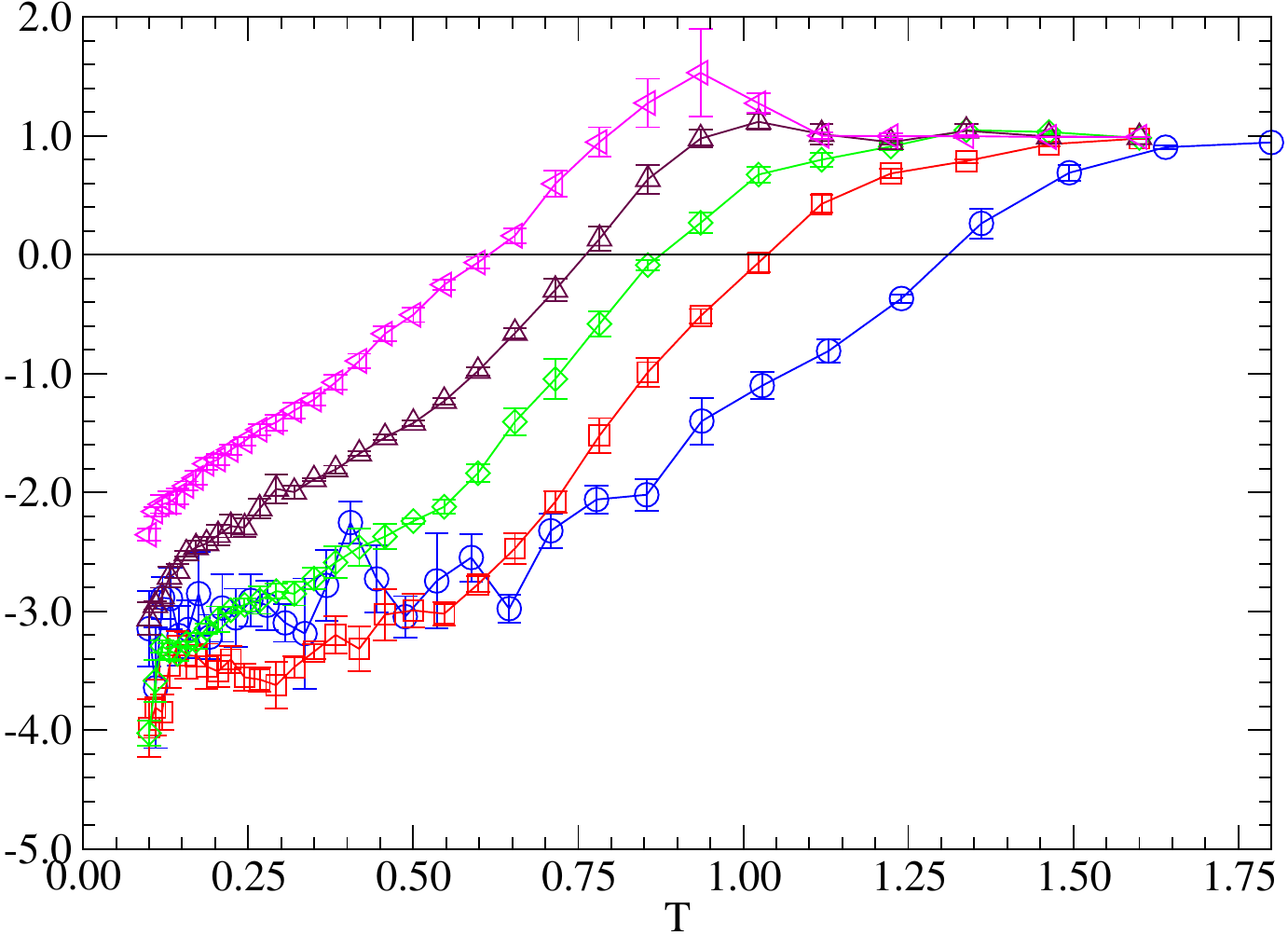}
    \includegraphics[width=0.8\columnwidth]{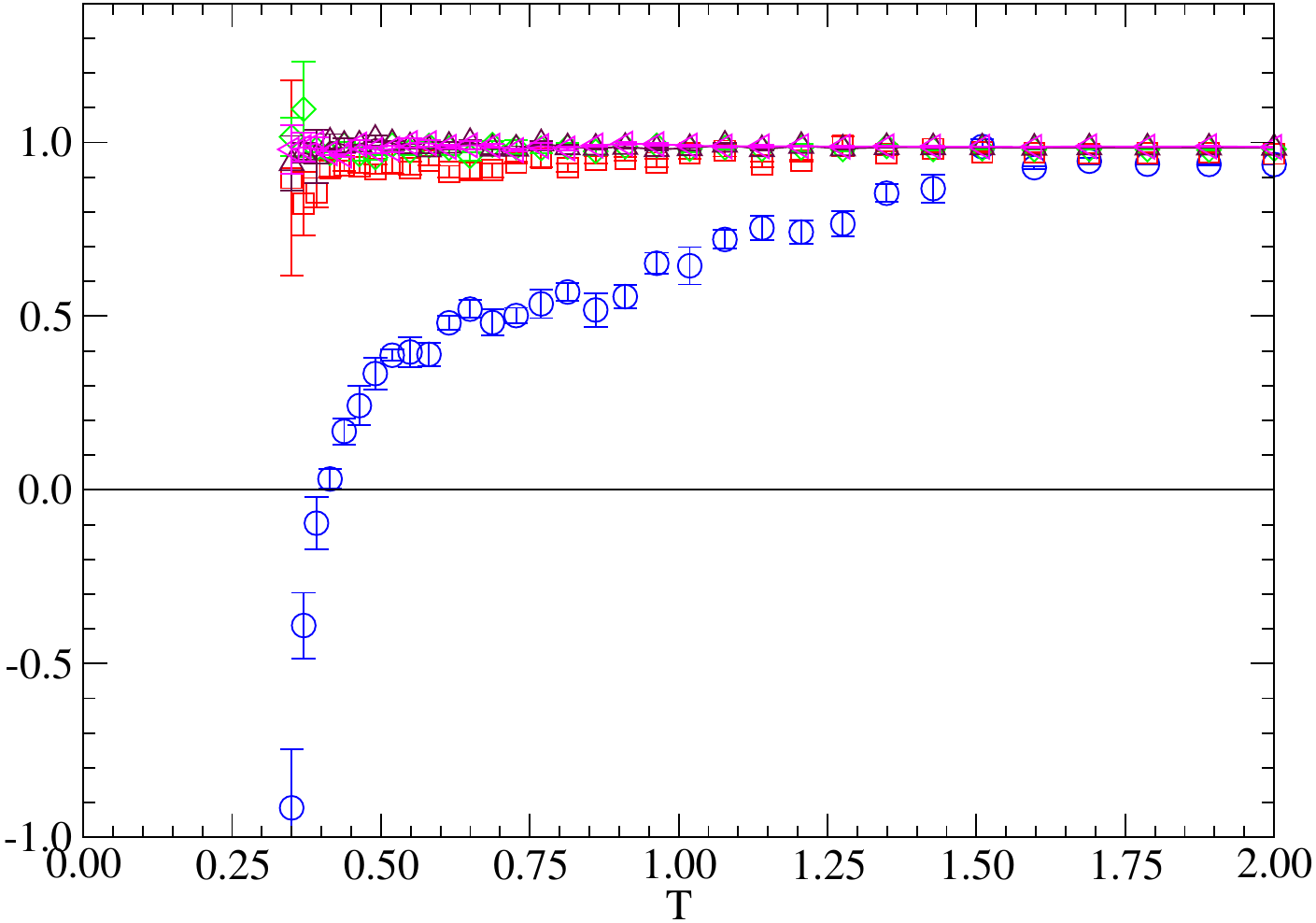} 
    \caption{Temperature behavior of the third order cumulant of the Polyakov line at $p = 0.0$ (first), $p = 6\%$ (second), $p = 7\%$ (third), and $p = 9\%$ (fourth). Symbols and colors are the same as in Fig.~\ref{fig:order-parameter}.}
    \label{fig:B3}
\end{figure}

\begin{figure}
    \centering
    \includegraphics[width=0.8\columnwidth]{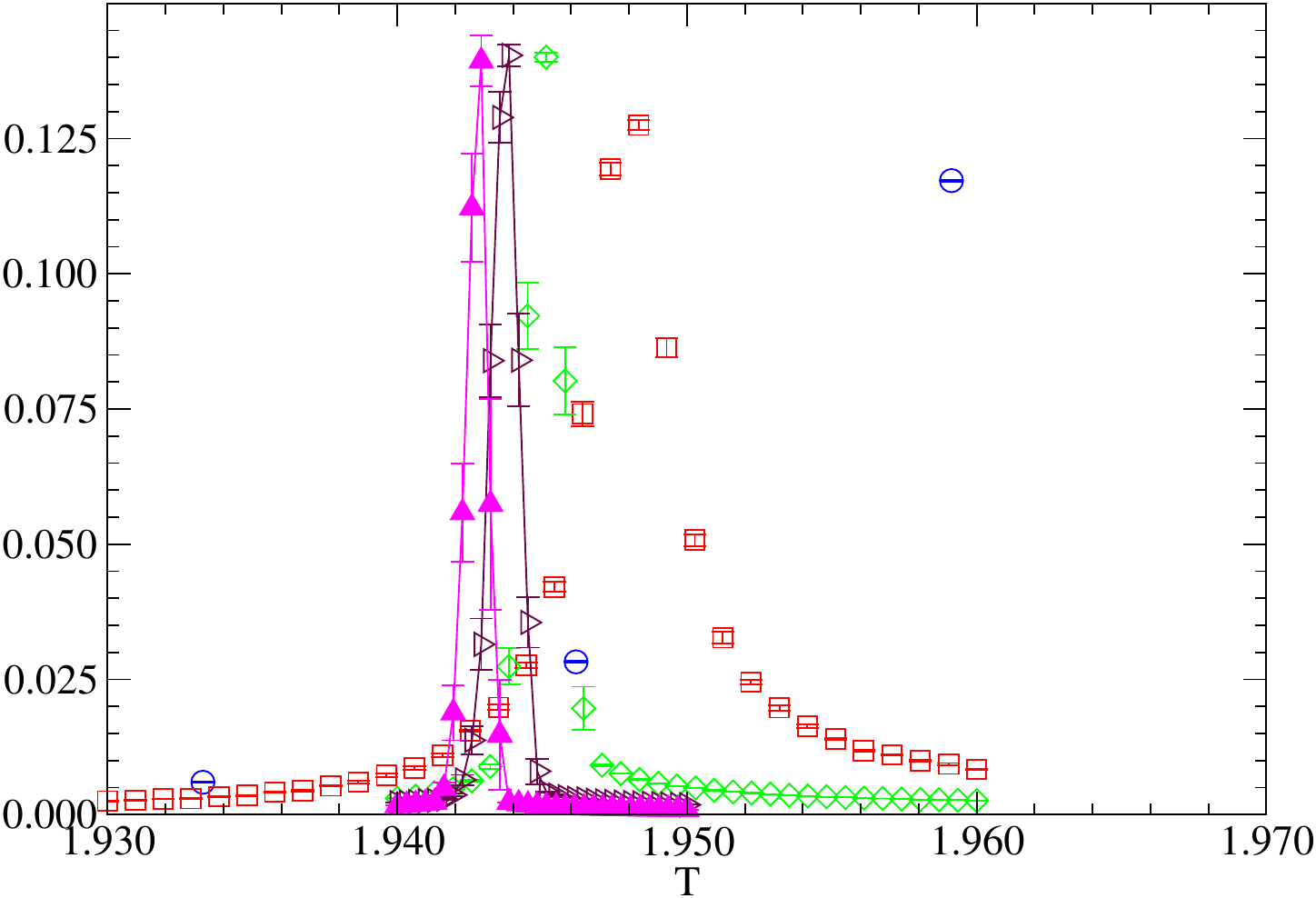}
    \includegraphics[width=0.8\columnwidth]{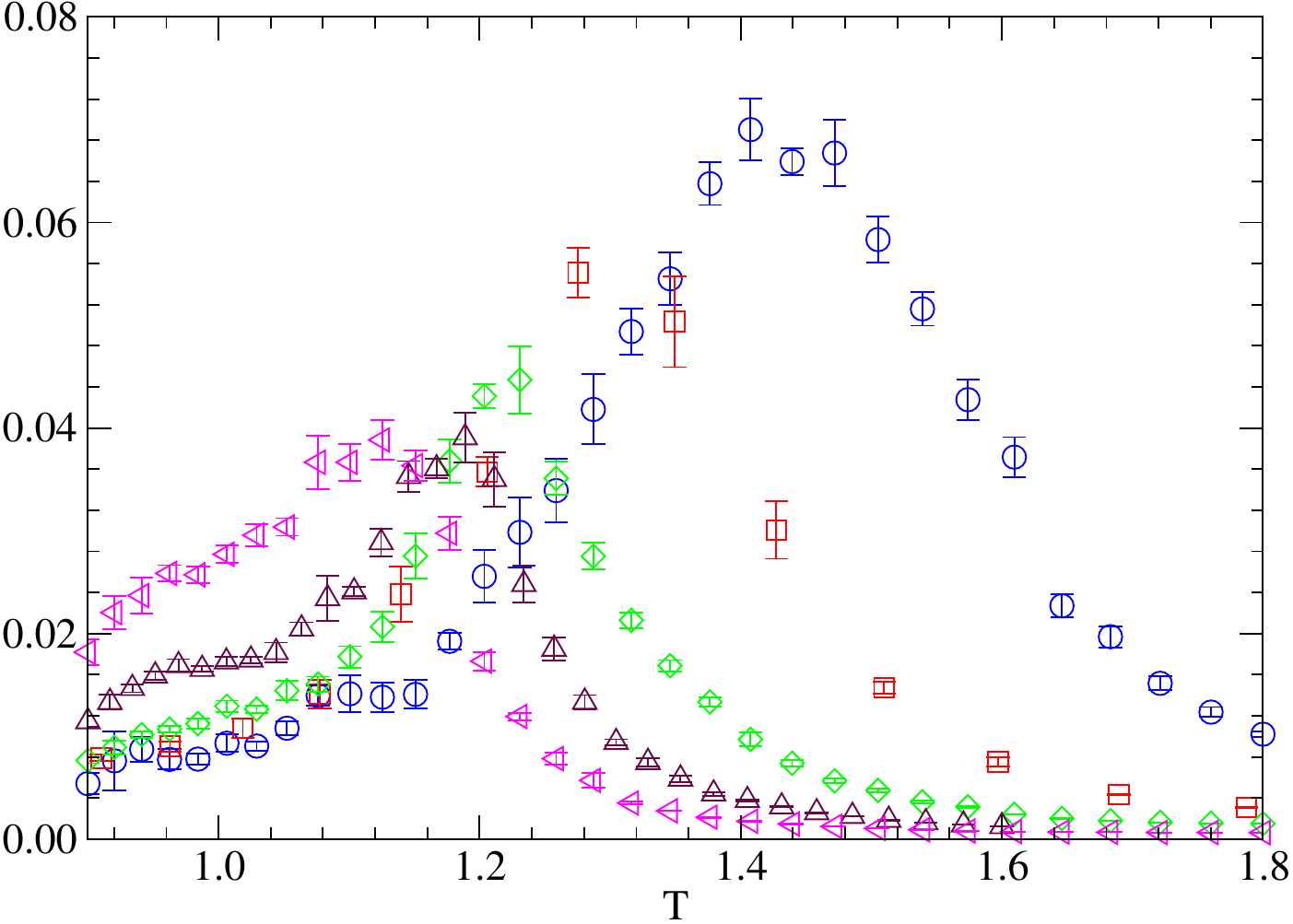}
    \includegraphics[width=0.8\columnwidth]{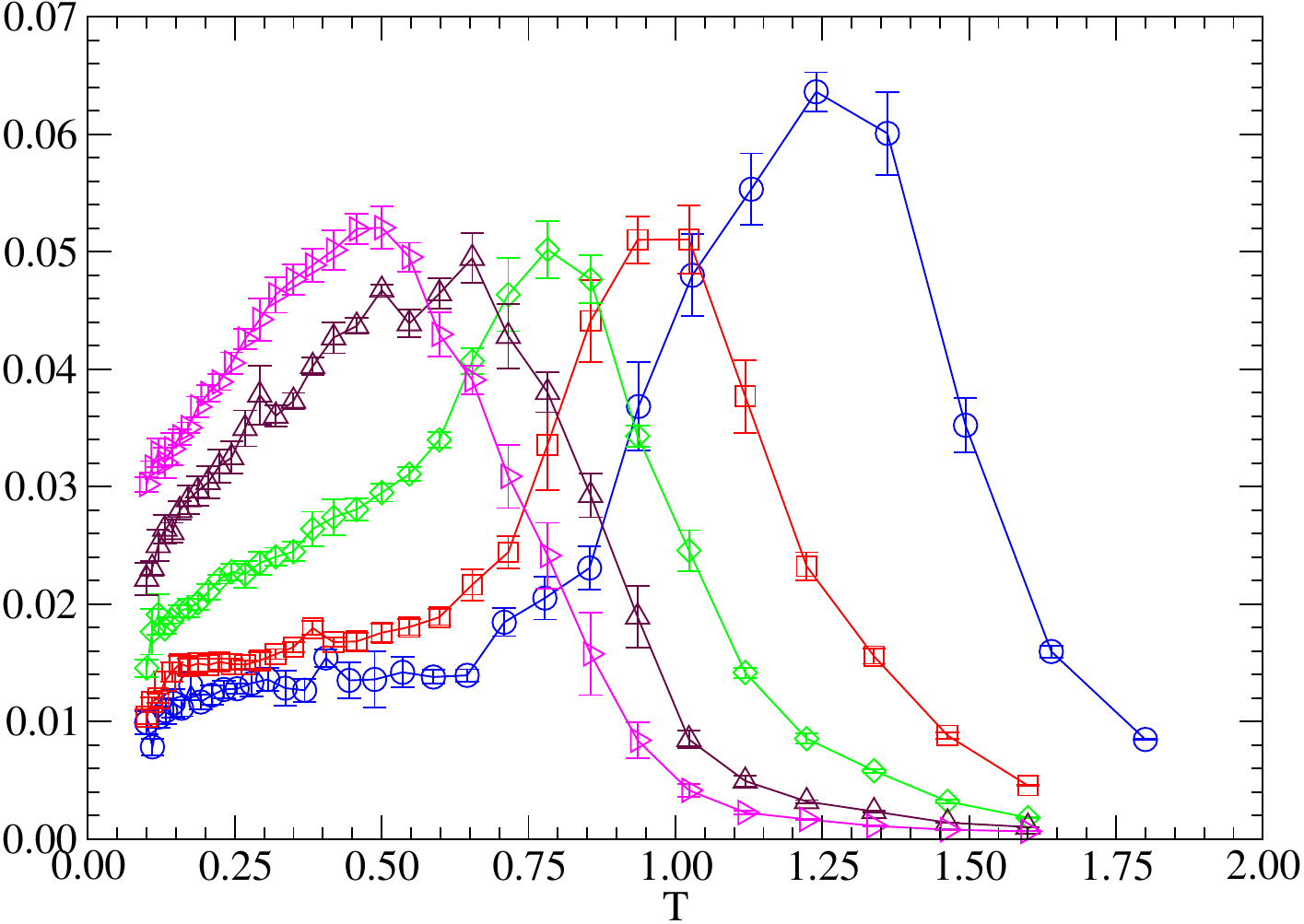}
    \includegraphics[width=0.8\columnwidth]{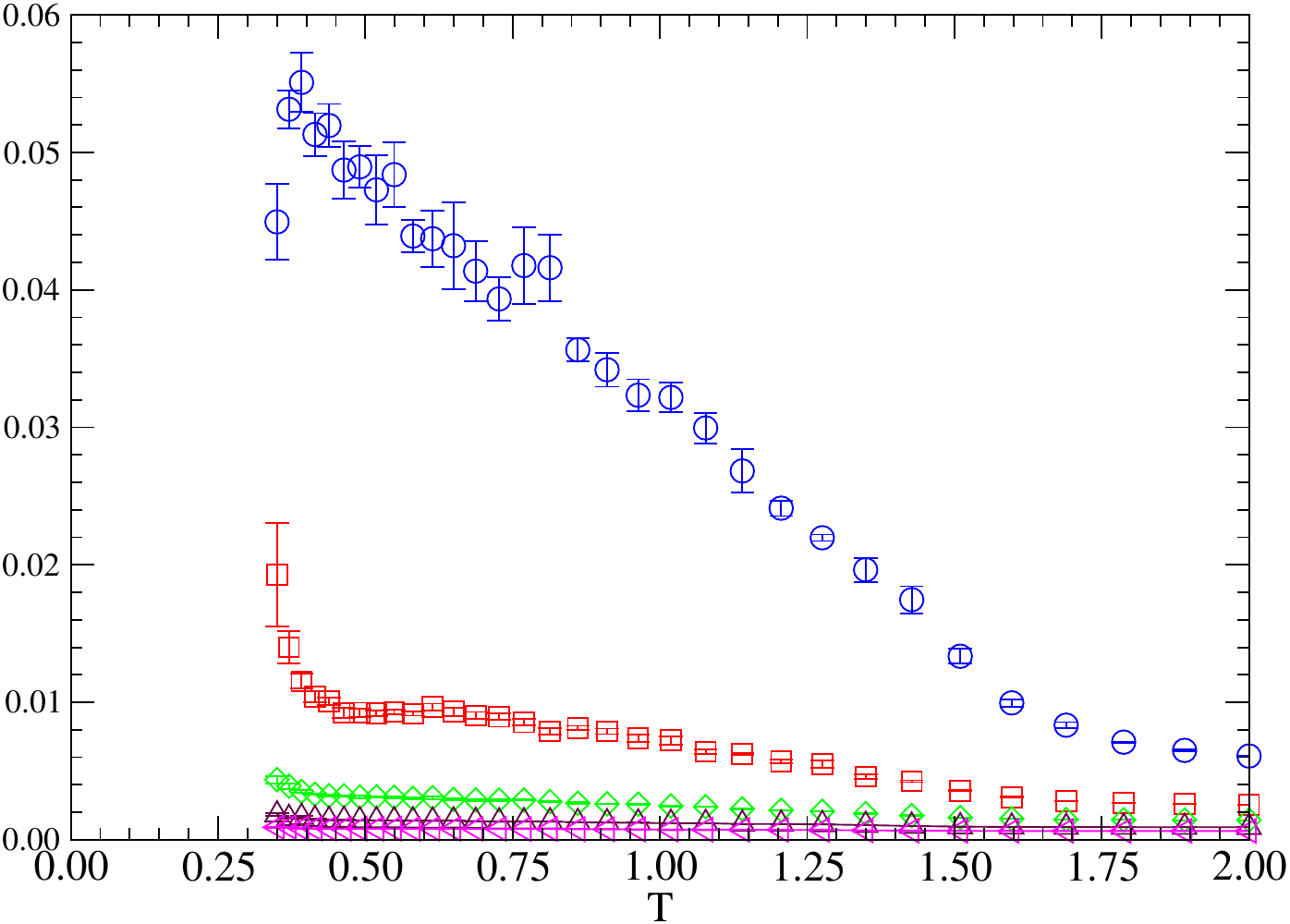}     
    \caption{Temperature behavior of the susceptibility of the Polyakov line at $p = 0.0$ (top), $p = 6\%$ (second), $p = 7\%$ (third), and $p = 9\%$ (fourth). Symbols and colors are the same as in Fig.~\ref{fig:order-parameter}.}
    \label{fig:susc}
\end{figure}

\begin{figure}
    \centering
    \includegraphics[width=0.9\columnwidth]{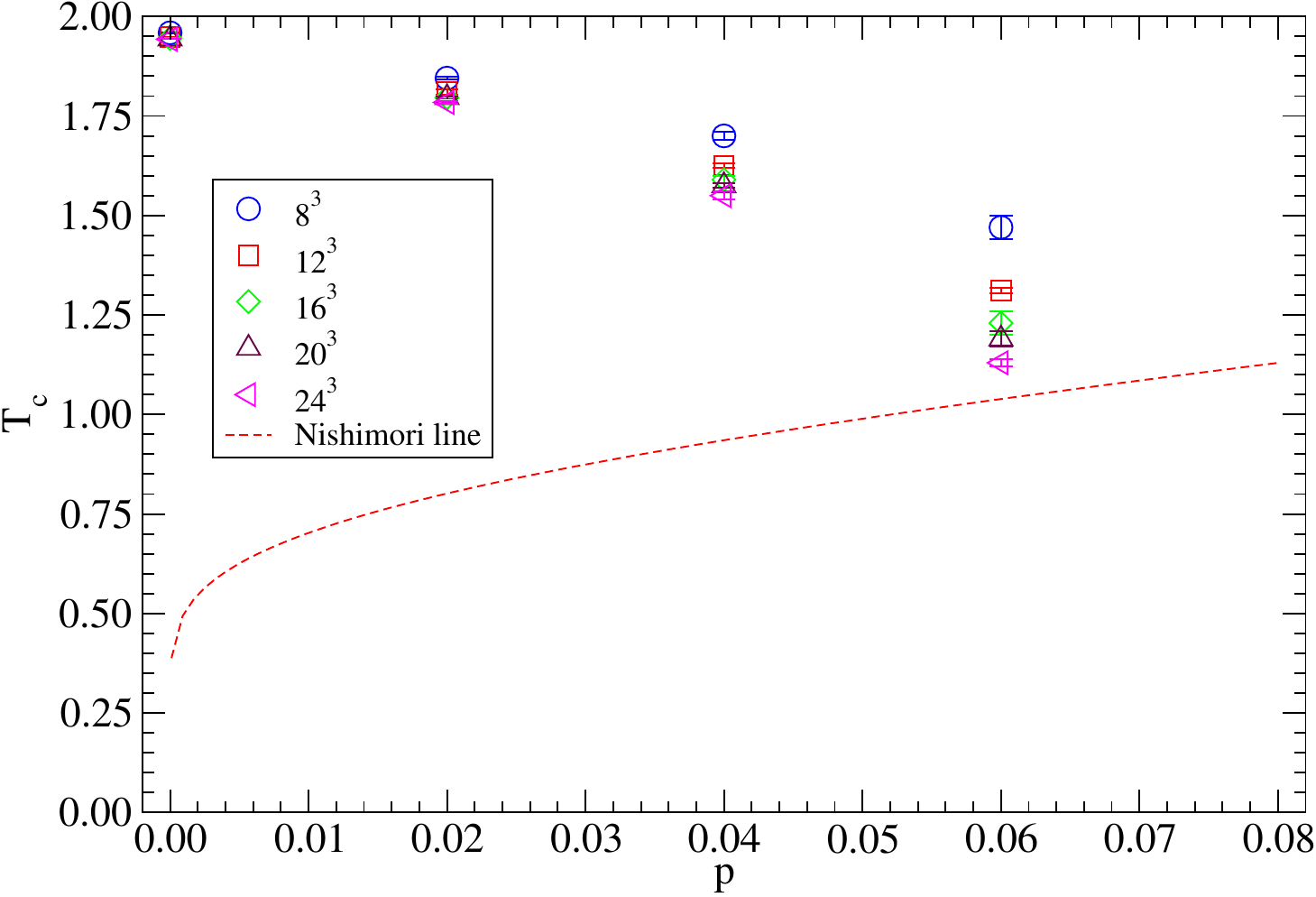}    
    \caption{The phase diagram for the case where the symmetric depolarizing noise level ($p$) is equal to the measurement error rate ($q$) from the Monte Carlo simulation of symmetric depolarizing noise Random Coupled-Plaquette Gauge model (RCPGM).}
    \label{fig:p-q_phase}
\end{figure}

\begin{figure}
    \centering
    \includegraphics[width=0.9\columnwidth]{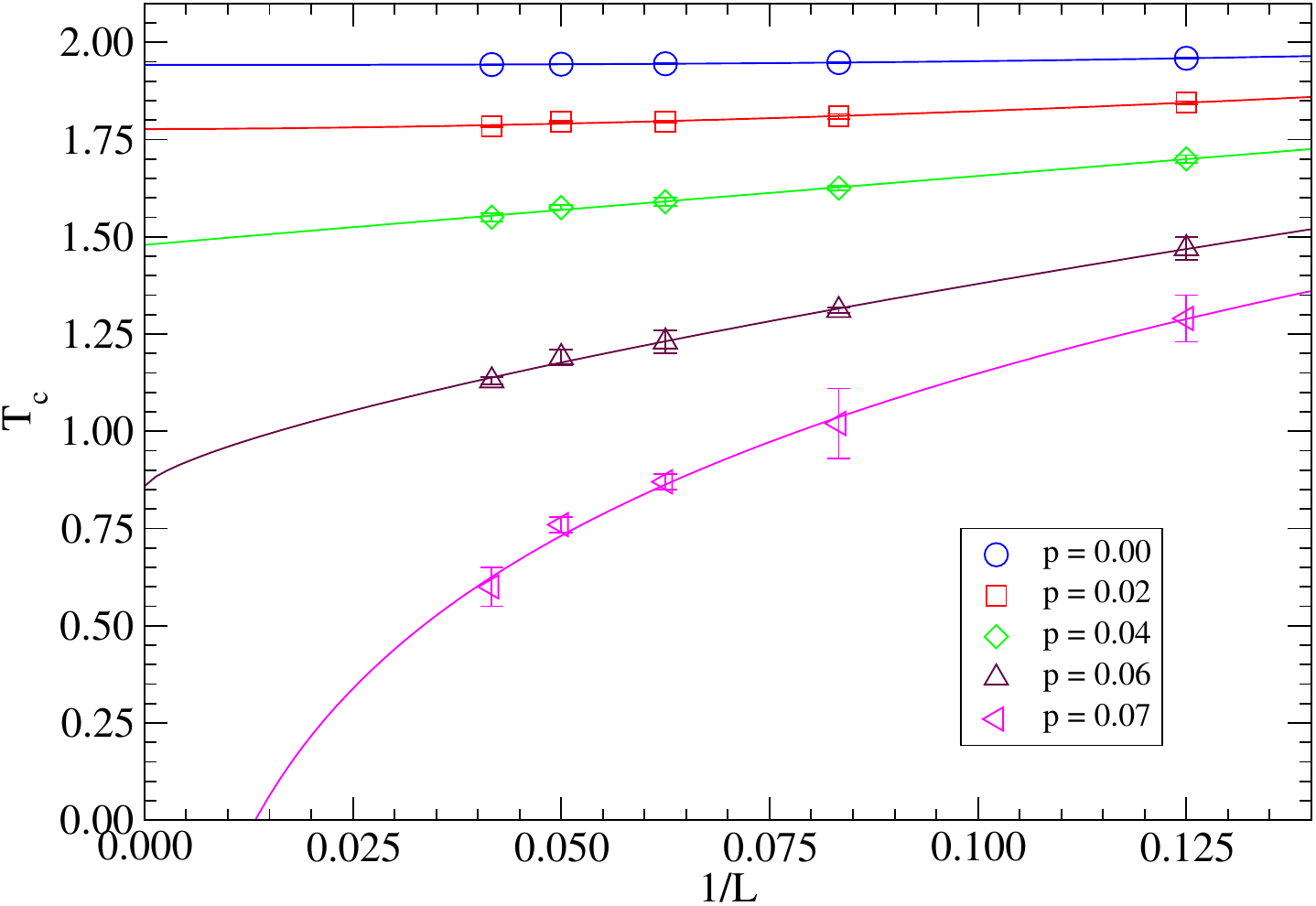}
    \caption{Critical temperature of the symmetric depolarizing RCPGM against inverse system size for varying noise strength $p$. Lines are interpolations by fitting to a power law function with offset. A crossing point manifestly larger than zero of a curve with the $y$-axis indicates a finite critical temperature in the large system limit, i.e.~the corresponding noise strength being below threshold where error correction is beneficial.}
    \label{fig:depolarizing_RCPGM_fit_vs_V}
\end{figure}

Figure~\ref{fig:order-parameter} shows the average Polyakov line behavior, and Fig.~\ref{fig:B3} shows the third order cumulant ($B_3$), and Fig.~\ref{fig:susc} shows the susceptibility ($\chi$) at various noise levels on different lattice volumes. Then the transition temperature is mapped out and is summarized in Fig.~\ref{fig:p-q_phase} and Fig.~\ref{fig:depolarizing_RCPGM_fit_vs_V} depicts the transition temperature in ~the large volume limit. These figures show that the threshold probability ($p_c$) is bigger than $p = 6\%$ and is smaller than $p = 7\%$: the average Polyakov line, $B_3$ and $\chi$ from $p = 9\%$ (top right figure in Fig. \ref{fig:order-parameter}, \ref{fig:B3}, and \ref{fig:susc}) do not show a transition. On the other hand, the data for $p = 7\%$ show a transition (bottom right figure in Fig. \ref{fig:order-parameter}, \ref{fig:B3}, and \ref{fig:susc}) but the transition temperature at the large finite volume limit (Fig. \ref{fig:depolarizing_RCPGM_fit_vs_V}) does not exist. 

We thus report a threshold of $p_c=6\%$, which is to be understood as accurate to the first significant digit. The important comparison is to find out the threshold value for the identical noise model when applying the well-known uncoupled RPGM. For this, we take the marginal $X$-error rate (i.e. $X$ or $Y$) $\pr(X)=2p/3$ and the syndrome error rate $q=p$. This leads to an anisotropic RPGM, which is less well studied compared to the standard result for $q=p$. Harrington conjectured a relationship of the form $p^2q=p_c^3$, which is supported by numerical results~\cite{Harrington2004}. Plugging in our noise rates into this relationship results in a threshold of $p_c=4.3\%$. We thus conclude that accounting for $Y$-error correlations by promoting the gauge theory from the RPGM to the RCPGM leads to a substantial improvement of the threshold value from $4.3\%$ to $6\%$, i.e. a relative improvement of $40\%$. When comparing to the perfect syndrome readout case, where promoting the RBIM to the R8VM led to a relative increase of $15\%$ in the threshold value, our result suggests that accounting for $Y$-correlations is even more relevant in the more realistic scenario of noisy syndrome readout. 

\section{Circuit-level noise in the toric code}
\label{sec:circuit-level-noise-toric-code}
After the noise model overview in Sec.~\ref{sec:noise-models}, circuit-level noise requires further detailed analysis, which we will present in the following section in order to derive an effective noise model.
\subsection{Toric code circuit schedule and unit cell}
\begin{figure}
    \centering
    \includegraphics[width=\columnwidth]{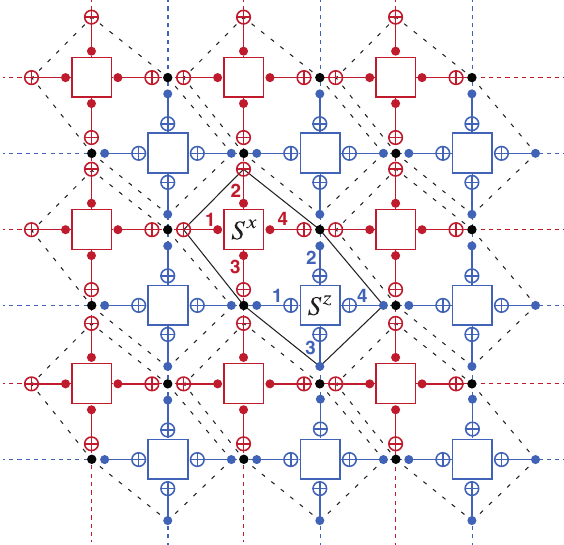}
\caption{Toric code lattice with CNOT gates indicated, including the sequence in which the CNOT gates are applied (1,2,3,4). The unit cell encompasses one horizontal and one vertical data qubit and one $X$-syndrome and one $Z$-syndrome ancilla. We assume translation invariance, on a finite system the top and bottom as well as the left and right boundary would be identified to implement the torus. Since this error model evidently leads to unreliable syndrome information we have to repeat this schedule in time, leading to a three-dimensional syndrome volume. The two preceding steps (preparation and Hadamard/idling as well as the final two steps (Hadamard/idling and measurement) are omitted here for readability. In total, this implements one round of syndrome measurements in eight time-steps.}
    \label{fig:toric_code_unit_cell}
\end{figure}

We will use a schedule which accomplishes one round of parity check measurements on the entire toric code (Fig.~\ref{fig:toric_code_unit_cell}) in 8 steps, using the check circuit illustrated in Fig.~\ref{fig:stabilizer_measurement_circuit_circuit_level}. These are comprised of two single qubit operations, four CNOT gates and two single qubit operations. Note that idling, i.e. waiting for a time-step is also a noisy operation.  Under the assumption that operations on disjoint subsets of qubits can be parallelized, a possible schedule is to apply the four CNOTs of both $X$- and $Z$-syndrome in the order "west-north-south-east" across the entire square lattice. As a side-note we remark that for practical applications, the surface code can be slightly modified, which is known as the rotated surface code~\cite{Horsman2012}. In that case one would have to modify the schedule in order to preserve the code distance~\cite{Tomita2014}, however for the "unrotated" version we use here this is not the case~\cite{Manes2023}.

\subsection{Noise model reduction}
By injecting single errors~\footnote{Note that a single CNOT error, i.e. an event with probability $p$ can have support on two qubits.} at all locations in the unit cell, we can infer their effect on the syndrome and the data qubits at the end of the cycle by means of error propagation. We compiled an exhaustive list that shows all circuit errors and their effect in Appendix~\ref{app:location_list}. In principle, the resulting plethora of distinct syndrome patterns would lead us to perform the combinatorics to find the corresponding space time equivalences which would become the variables of a statistical mechanical model. While we do not see fundamental obstacles in this attempt, we realized that the resultant statistical model that fully accounts for all syndrome patterns will stray very far from known statistical models as it will contain a large number of (on the order of 40) interactions on a non-standard lattice or interaction graph. We therefore leave it as an open problem to find this full circuit noise statistical model and instead opt to focus on building an (approximate) relationship to the random coupled-plaquette gauge model we have established above. We thus opt to simplify the noise model by performing a reduction. 

Towards this, we define a reduced set of error mechanisms and corresponding syndrome patterns, which our statistical model will be able to genuinely account for. For the syndrome patterns that fall outside of this subset, we then attribute the error probability of that event to all overlapping error mechanisms that are contained in the reduced set. For example, let us imagine we were to define data bit-flip and syndrome errors as the reduced set, i.e. one that would map to the RPGM on a cubic lattice. Here, an exemplary event that e.g. triggers one syndrome bit at a given time-step and the diagonally adjacent syndrome bit at the next time-step with some given probability $p$, 
would contribute to three event probabilities under the reduction, namely one in each principal lattice direction, separately increasing the weight of horizontal as well as diagonal as well as temporal edges by $p$. 

This achieves our primary goal to reduce the error mechanism set. While admittedly this reduction is not well controlled, i.e. it is not strictly over- or underestimating error rates and in the end error correction threshold results, we would like to discuss the example of the well understood repetition code circuit noise case. Here, the effect of circuit noise can be fully compressed into introducing one additional error mechanism, namely the circuit error sitting between the two required CNOT gates~\cite{Vodola2022}. This leads to a diagonal error edge, promoting the underlying syndrome lattice from a square to a triangular lattice. For the extremal case, where this is the only error process with some probability $r$, fully accounting for this leads to a threshold of $r_{\mathrm{th.}}=50\%$. When defining pure data and syndrome errors as the reduced error mechanism set, the previously described error would be "broken up" into a data error and a syndrome error both with the given probability $r$. This model then immediately can be understood as the well-known case of the uniform random bond Ising model on a square lattice which leads to a threshold of $10.9\%$. This provides some intuition that breaking up error processes potentially overestimates them and then correspondingly depresses thresholds on a reduced noise model compared to a full noise model.

\begin{figure}
    \centering
\includegraphics[width=\columnwidth]{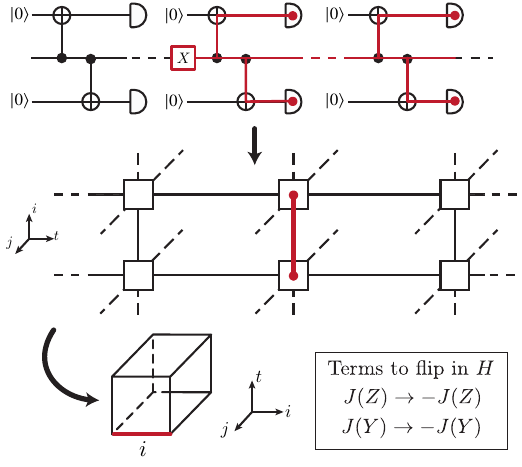}    
    \caption{ Translating circuit errors to edges: an $X$-error on a data qubit flips the two adjacent syndrome bits, which corresponds to a "space-like" edge on the syndrome difference graph in the $i-$direction. Note that in quantum circuit convention time flows to the right while in the cubic lattice it flow upwards (such that the spatial lattice in the xy-plane corresponds to one time-step of the toric code syndrome lattice). Gates where the error does not propagate are omitted for simplicity. The picture is completely analogous for data qubits in the spatial $j-$direction. Note that the error graph shown here is not identical with the Hamiltonian lattice. To arrive at the latter, the shown edge would correspond to flipping the sign of the conjugate bonds in the Hamiltonian, i.e. $J(Y)$ and $J(Z)$ according to the prescription in Sec.~\ref{sec:RCPGM}.}
    \label{fig:i_edge_circuit_unitcell}
\end{figure}

\begin{figure}
    \centering
\includegraphics[width=\columnwidth]{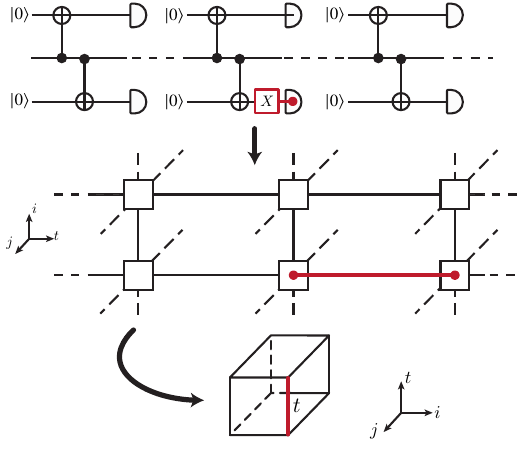}
    \caption{
    Translating circuit errors to edges: a measurement error on the ancilla qubit flips the measurement outcome at one time-step, which corresponds to two flips of the syndrome difference graph, i.e. a "time-like" edge on the syndrome (difference) graph in the $t-$direction. Note that in quantum circuit convention time flows to the right while in the cubic lattice it flow upwards (such that the spatial lattice in the xy-plane corresponds to one time-step of the toric code syndrome lattice). Gates where the error does not propagate are omitted for simplicity.}
    \label{fig:t_edge_circuit_unitcell}
\end{figure}

\subsection{Target noise model: Independent $XZ$ noise plus syndrome noise}
\label{indepXZ}
Let us start by defining as target model the case of independent $X$ and $Z$ noise as well as syndrome noise (syndrome bits flip independently with some error rate). When starting from circuit-level noise and performing the above attribution for all circuit-errors to leading order in circuit-noise strength $p$, we find 
\begin{align}
\pr(X_h) = \frac{8p}{3}  + \frac{48p}{15}\\
\pr(X_v) = \frac{8p}{3}  + \frac{32p}{15}\\
\pr(q) = \frac{8p}{3}  + \frac{48p}{15}
\label{prob_RPM}
\end{align}
where $h$ and $v$ label refer to horizontal and vertical qubits in the 2d-sublattice and $q$ indicates a syndrome error.
Note that we write contributions from single and two-qubit errors separately to guide the reader. This indicates e.g. that there are four single-qubit locations, on each one there are two out of the three Paulis contributing, leading to an overall  $8p/3$ (there is always one Pauli that commutes with the measurement and is hence "invisible"). In addition, there are e.g. 48 locations for the effective measurement error $pr(q)$. Due to non-trivial error propagations, this evades an immediate intuition, one can obtain these by adding all locations in the corresponding table (see App.~\ref{app:location_list}), e.g. the $t_Z$ column for $pr(q)$.

\begin{figure}
    \centering
    \includegraphics[width=0.8\columnwidth]{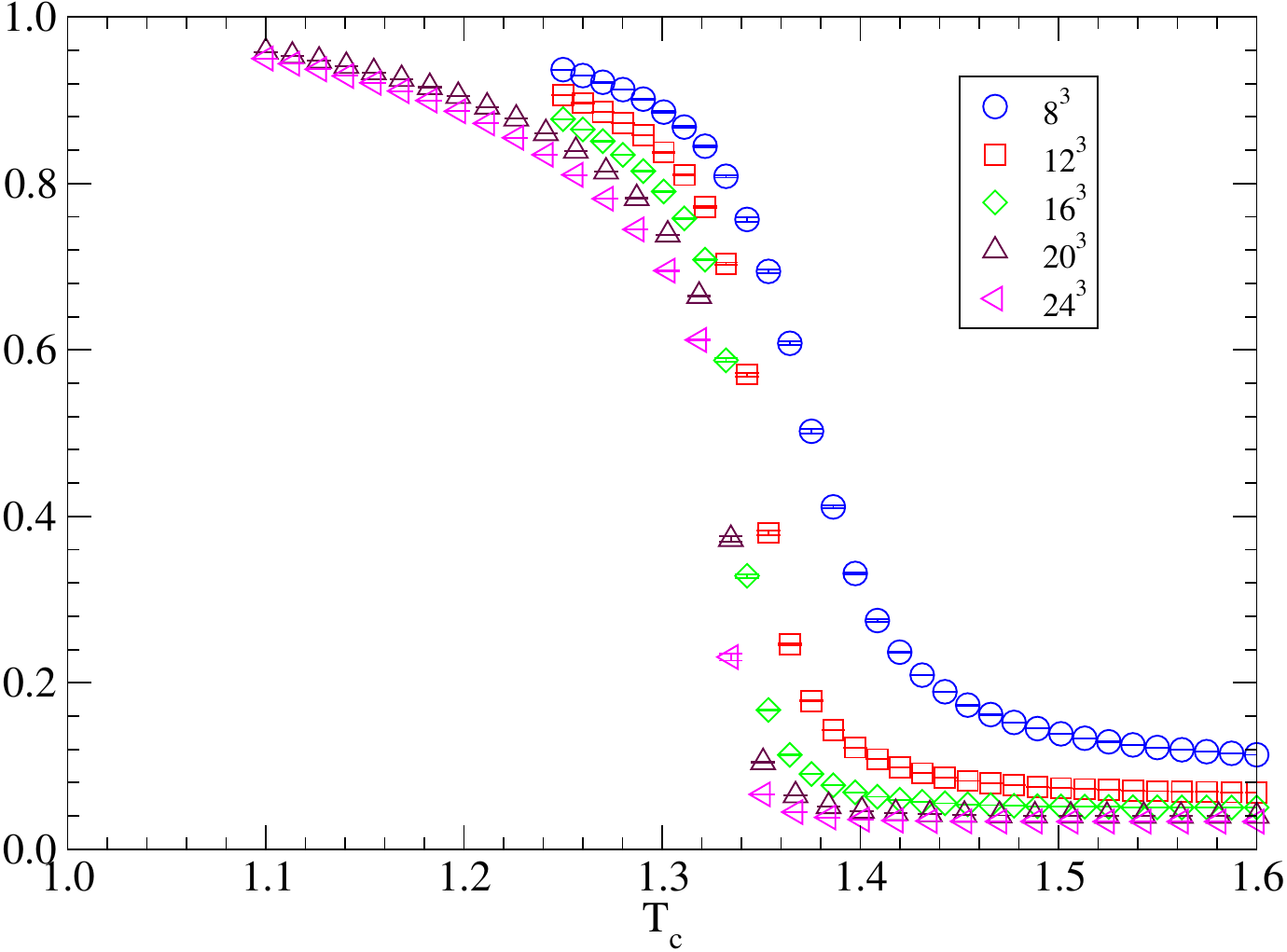}
    \includegraphics[width=0.8\columnwidth]{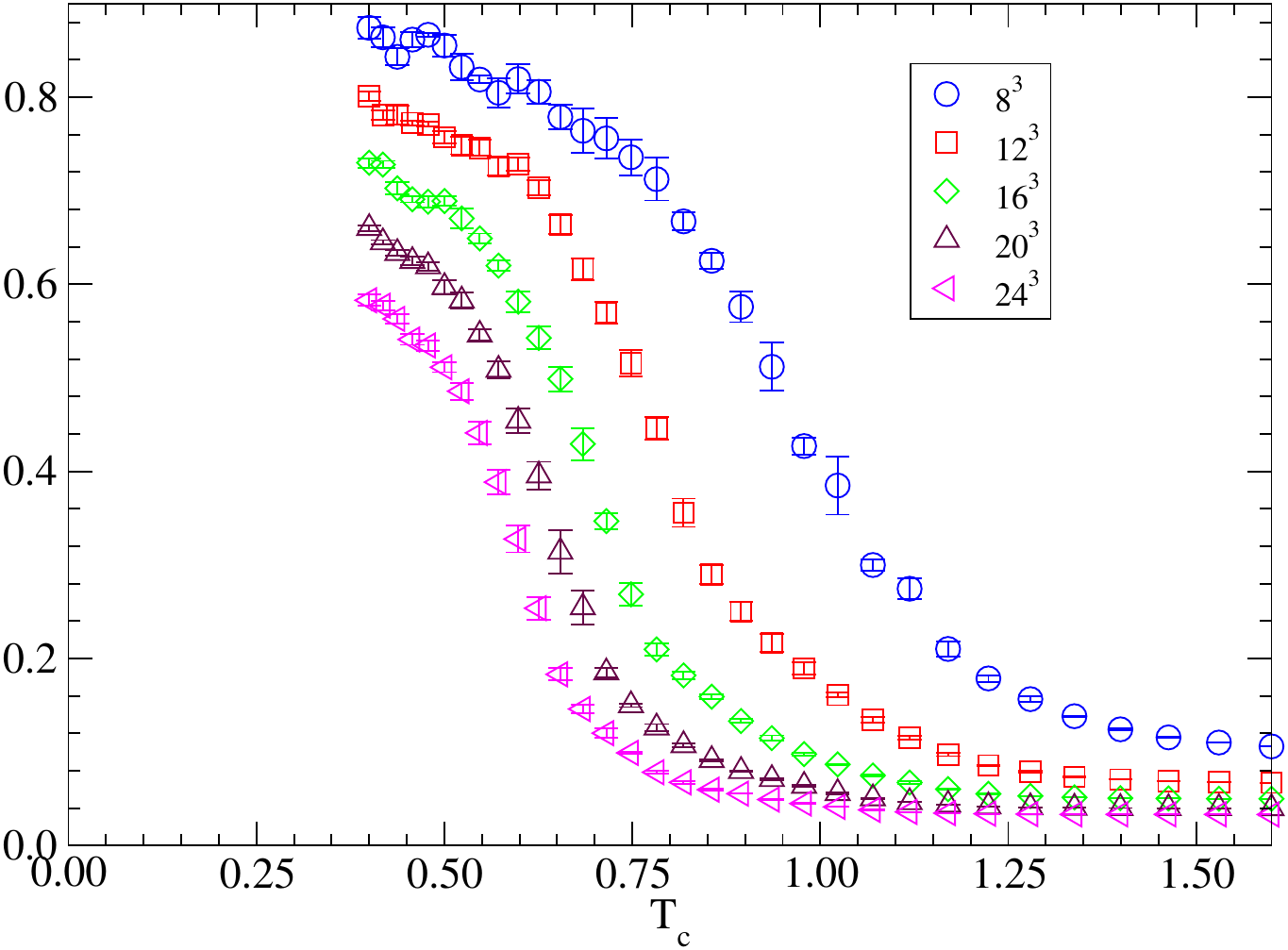}
    \includegraphics[width=0.8\columnwidth]{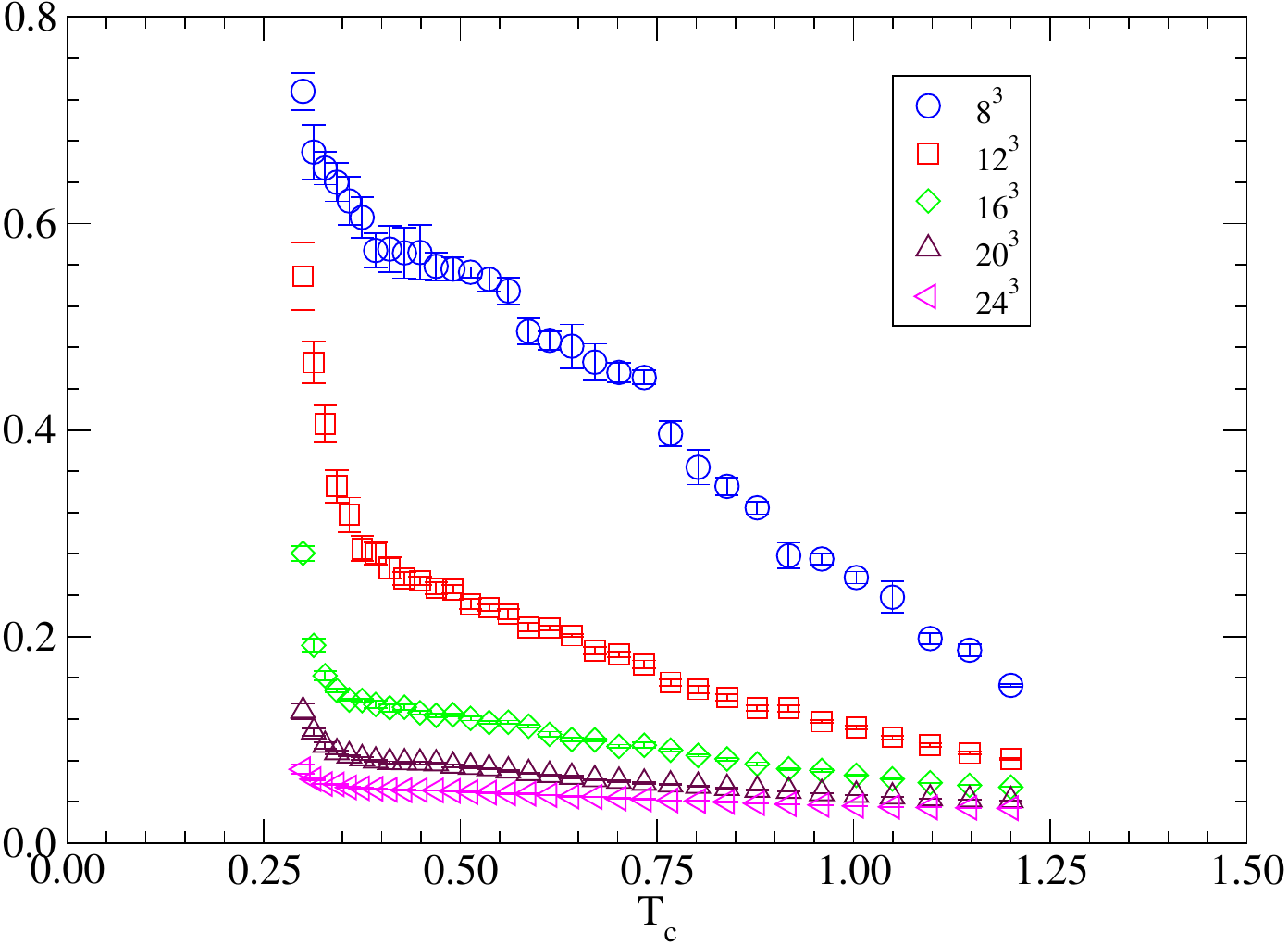}
    \caption{Temperature behavior of the average Polyakov line with $p = 1.70 \times 10^{-5}$ (top), $p = 0.682\%$ (middle), and $p = 0.852\%$ (bottom) for anisotropic random plaquette gauge model on $8^3$ (blue circle), $12^3$ (red square), $16^3$ (green diamond), $20^3$ (maroon up-triangle), and $24^4$ (magenta left-triangle). From Eq. \ref{prob_RPM}, $\pr (X_h) = \frac{88p}{15} = 0.01\%$ (top), $4\%$ (middle), and $5\%$ (bottom) respectively.}
    \label{fig:order-parameter_RPM}
\end{figure}

\begin{figure}
    \centering
    \includegraphics[width=0.8\columnwidth]{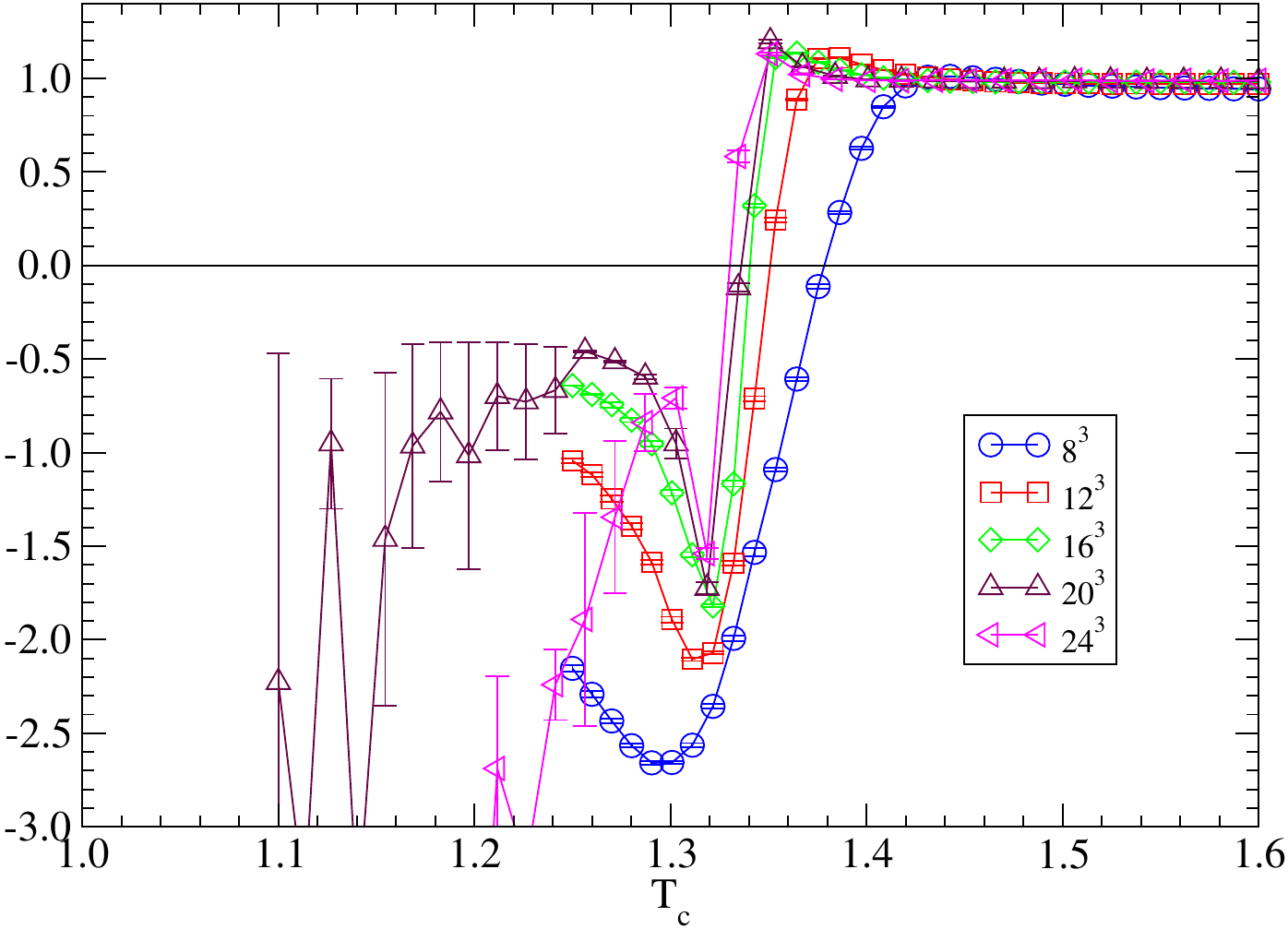}
    \includegraphics[width=0.8\columnwidth]{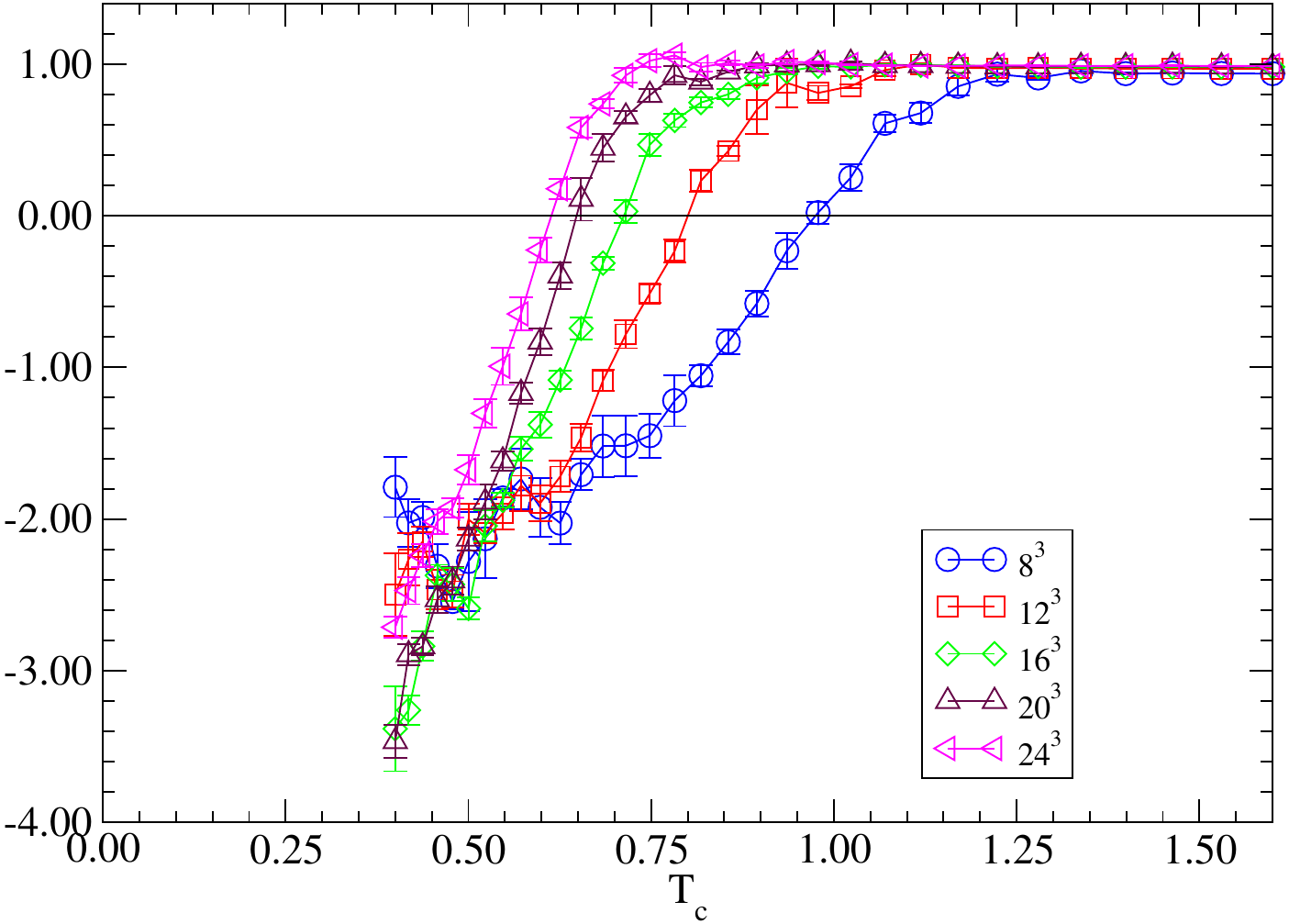}
    \includegraphics[width=0.8\columnwidth]{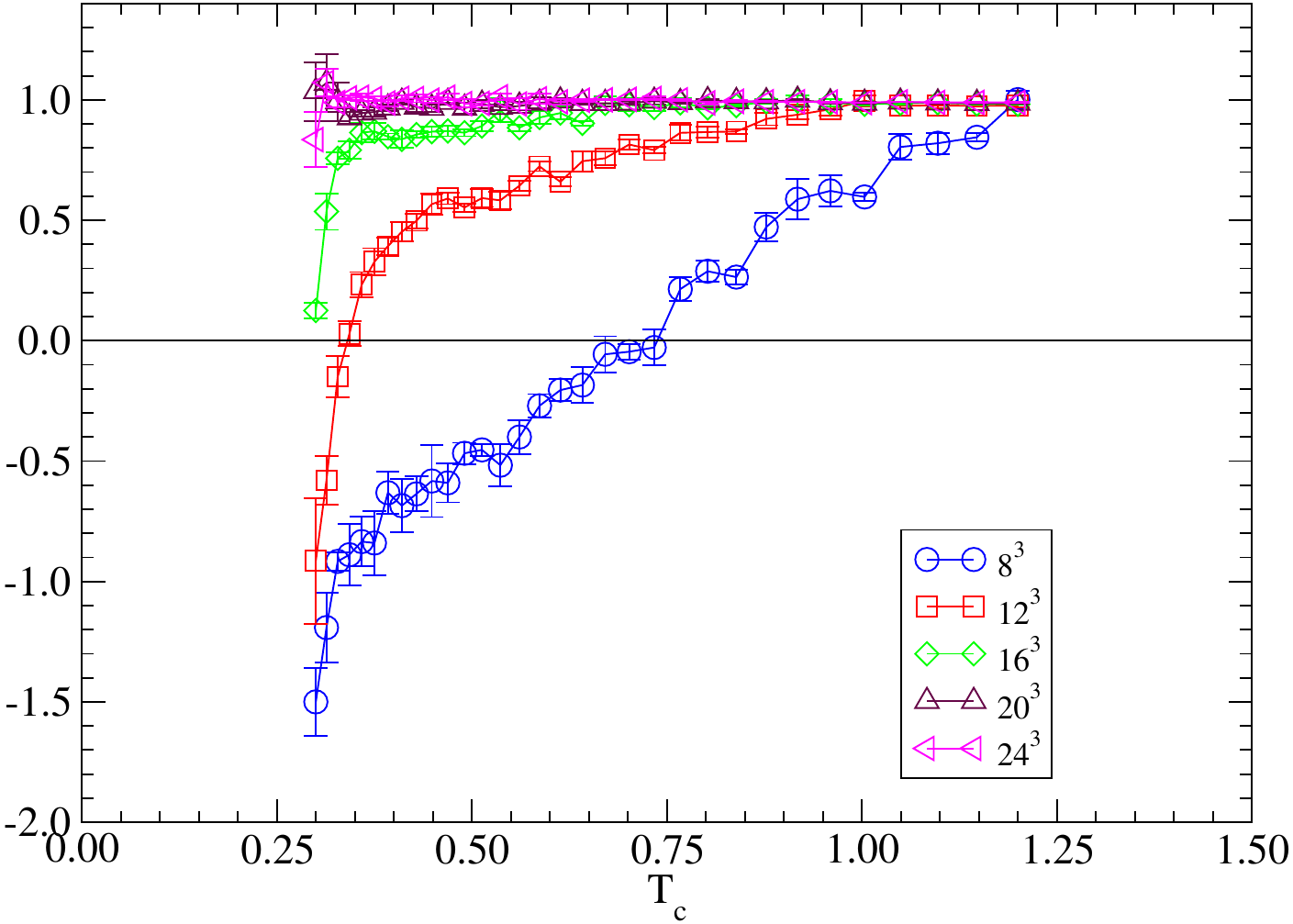} 
    \caption{Temperature behavior of the third order cumulant of Polyakov line with $p = 1.70 \times 10^{-5}$ (top), $p = 0.682\%$ (middle), and $p = 0.852\%$ (bottom) for the anisotropic random plaquette gauge model. Note that $\pr (X_h) = \frac{88p}{15} = \pr (q)$. Symbols and colors are the same as in Fig.~\ref{fig:order-parameter_RPM}.}
    \label{fig:B3_RPM}
\end{figure}

\begin{figure}
    \centering
    \includegraphics[width=0.8\columnwidth]{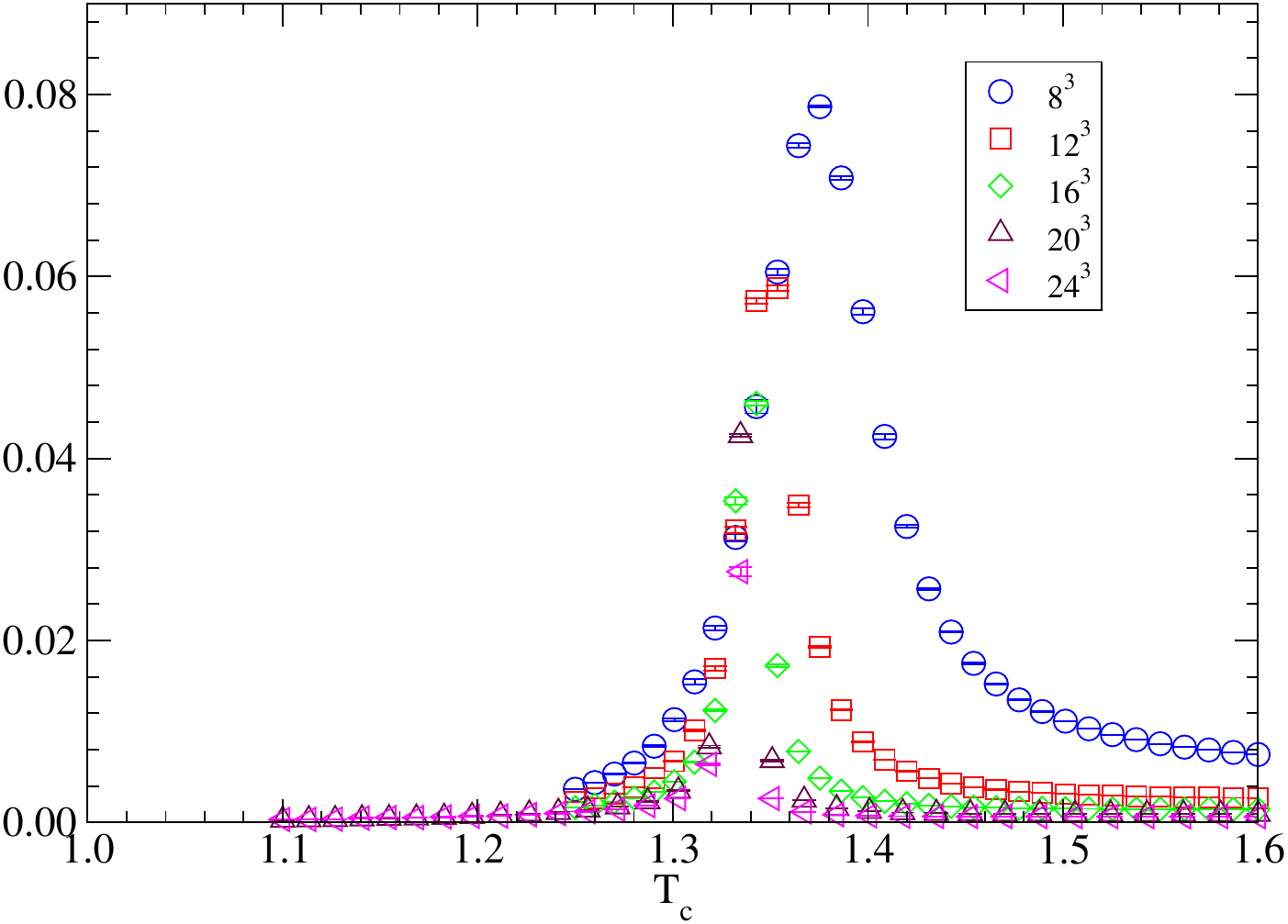}
    \includegraphics[width=0.8\columnwidth]{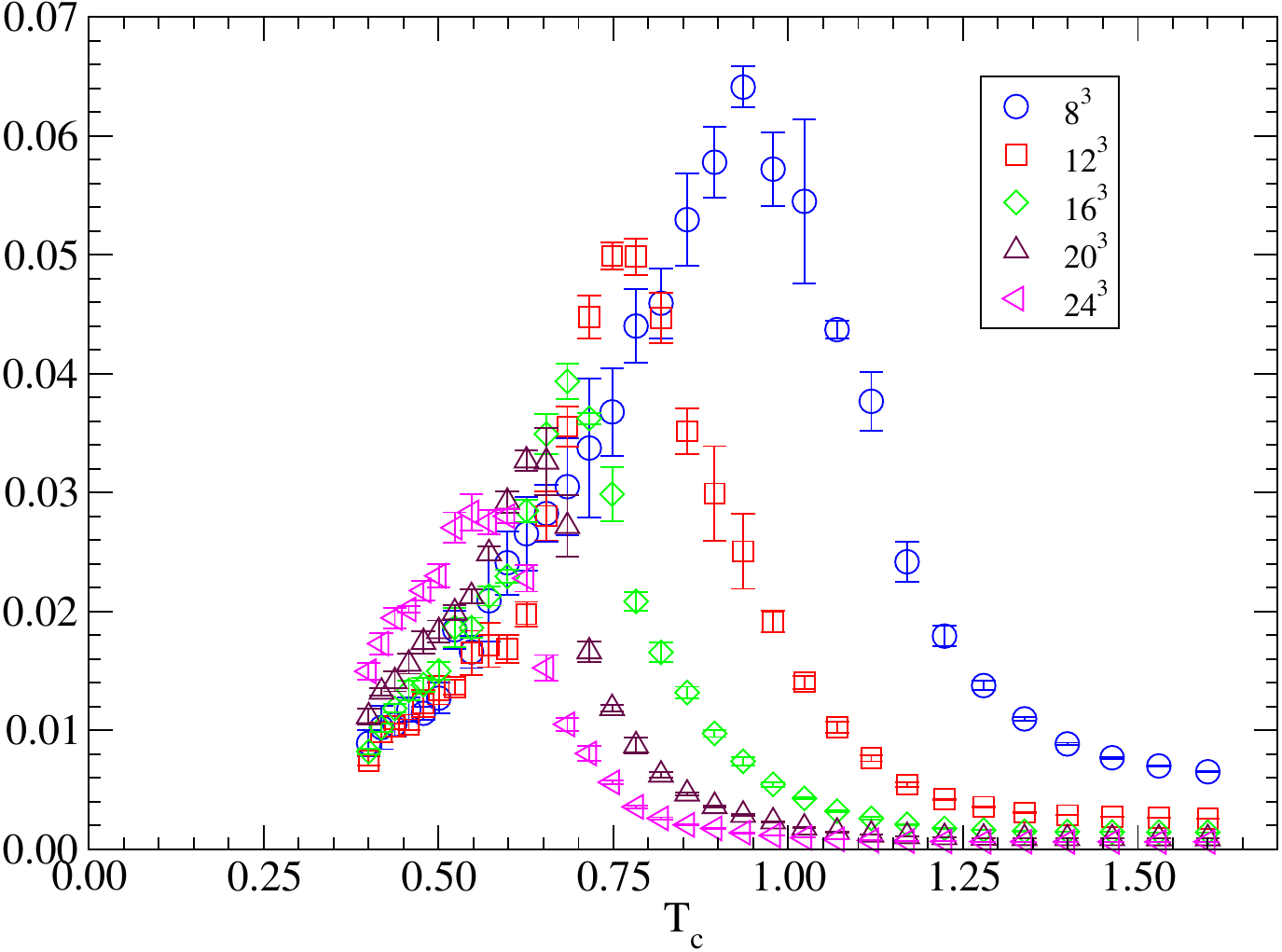}
    \includegraphics[width=0.8\columnwidth]{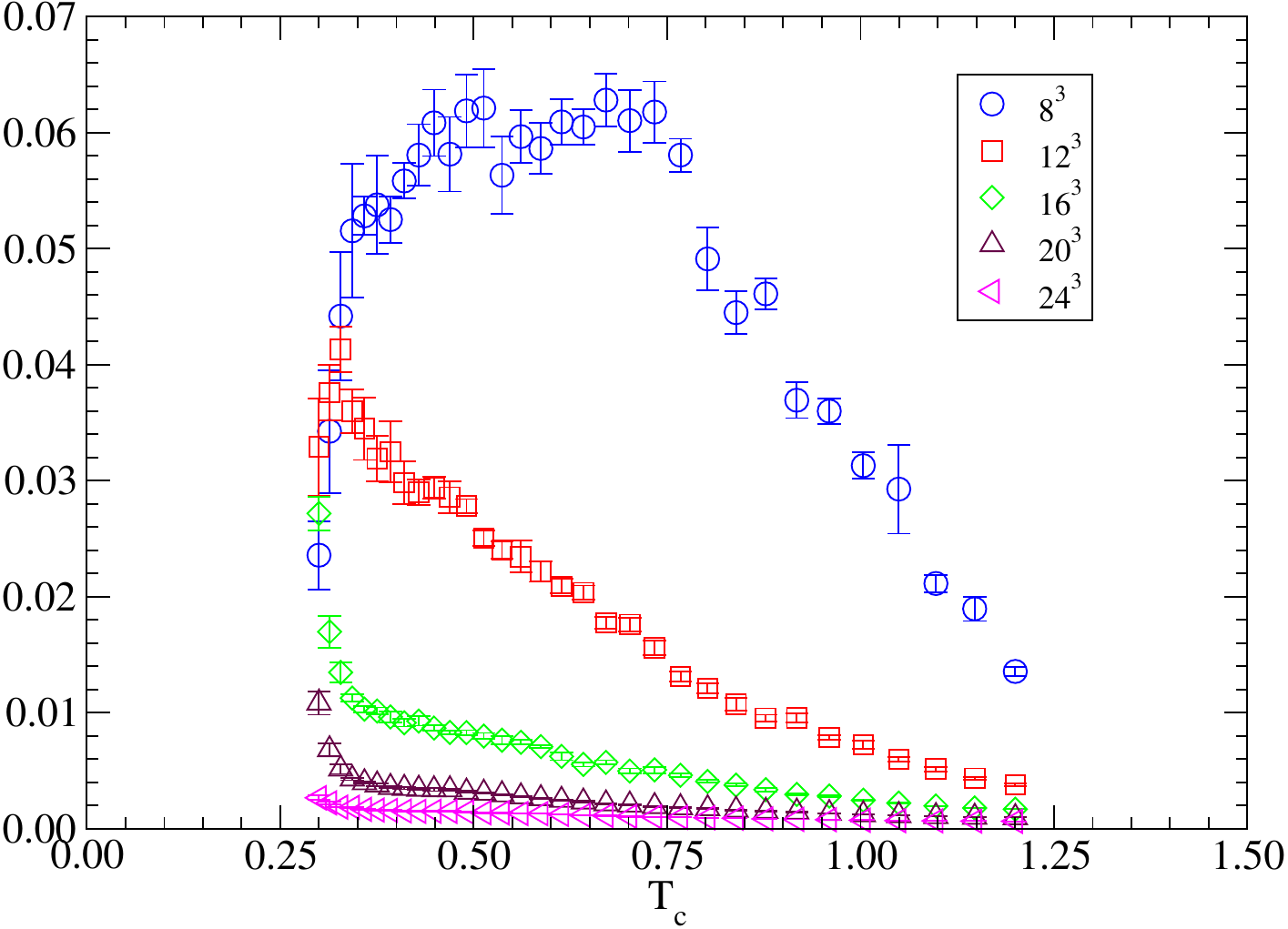}     
    \caption{Temperature behavior of the susceptibility of the Polyakov line with $p = 1.70 \times 10^{-5}$ (top), $p = 0.682\%$ (middle), and $p = 0.852\%$ (bottom) for the anisotropic random plaquette gauge model. Note that $\pr (X_h) = \frac{88p}{15} = \pr (q)$. Symbols and colors are the same as in Fig.~\ref{fig:order-parameter_RPM}.}
    \label{fig:susc_RPM}
\end{figure}

\begin{figure}
    \centering
    \includegraphics[width=0.9\columnwidth]{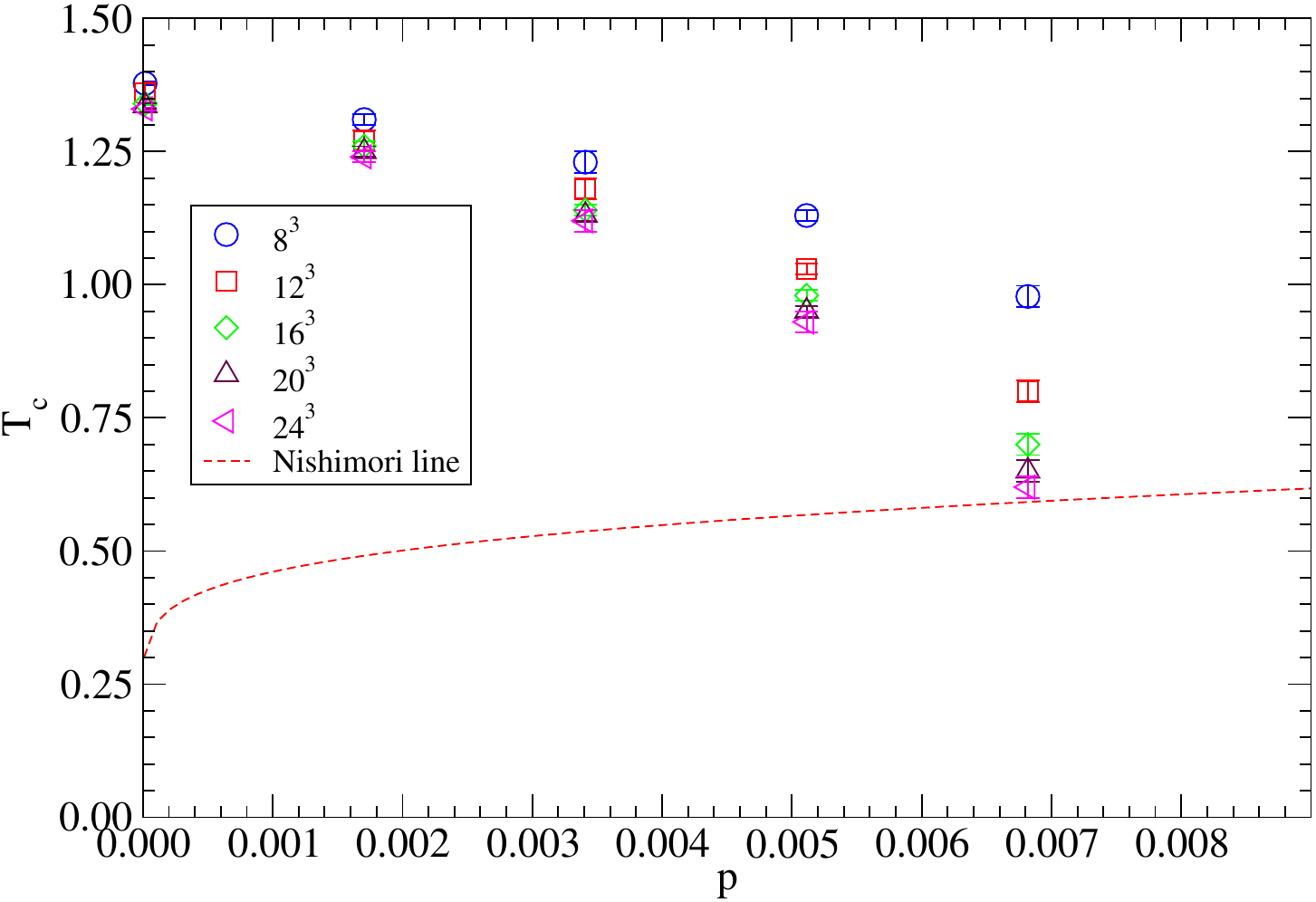}    
    \caption{The phase diagram for the case where the depolarizing noise level ($\pr (X_h) = \frac{88p}{15}$) is equal to the measurement error rate ($ \pr (q) = \frac{88p}{15}$) from the Monte Carlo simulation of the anisotropic Random Plaquette Gauge Model (RPGM).}
    \label{fig:p-q_phase_RPM}
\end{figure}

\begin{figure}
    \centering
    \includegraphics[width=0.9\columnwidth]{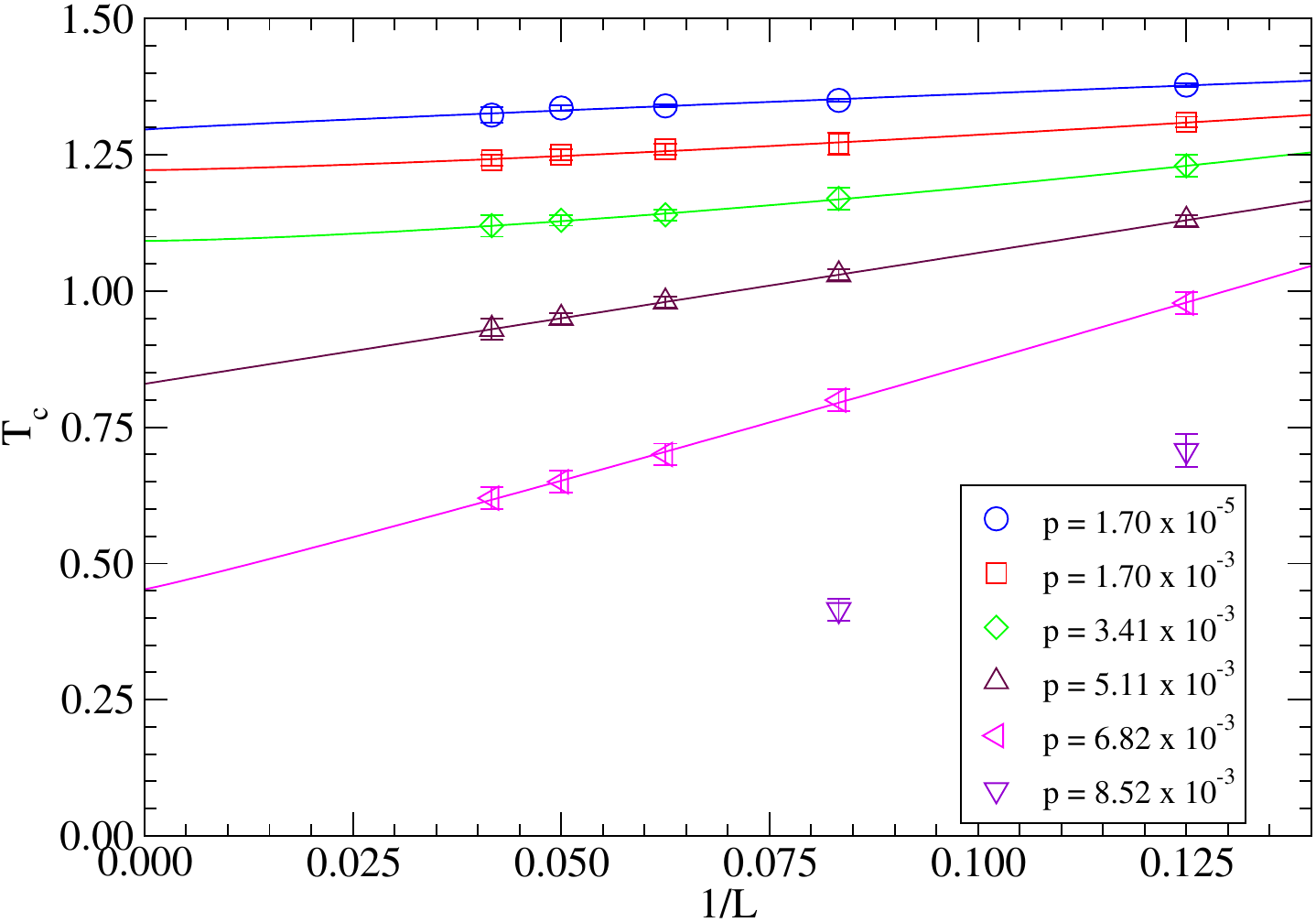}
    \caption{Critical temperature of the anisotropic RPGM against inverse system size for varying noise strength $p$. Lines are interpolations by fitting to a power law function with offset. A crossing point manifestly larger than zero of a curve with the $y$-axis indicates a finite critical temperature in the large system limit, i.e. the corresponding noise strength being below threshold where error correction is beneficial.}
    \label{fig:aniso_RPM_fit_vs_V}
\end{figure}

Fig. \ref{fig:order-parameter_RPM} shows the average value of Polyakov line from Monte Carlo simulation of the mapped anisotropic RPGM. Fig. \ref{fig:B3_RPM} and \ref{fig:susc_RPM} are the third order cumulant and susceptibility of the Polyakov line, respectively. The average Polyakov line, the third order cumulant and the Polyakov line susceptibility for the MC simulation with $p = 0.852\%$ do not show a first order phase transition on lattice volumes larger than $16^3$. On the other hand, MC simulation with $p = 0.682\%$ shows a finite temperature first order transition in the large lattice volume limit. Fig. \ref{fig:p-q_phase_RPM} is the summary phase diagram for the independent $XZ$ noise plus syndrome noise case (i.e., anisotropic RPGM) from Monte Carlo data. Fig. \ref{fig:aniso_RPM_fit_vs_V} displays the transition temperatures from finite volume simulation as a function of noise probability $p$ together with fits to
\begin{equation}
    T_c(L)= aL^{-b}+T_c,
\end{equation}
a power-law with an offset~\cite{Kubica2018}, where $L$ is the 1-dimensional size of the lattice volume ($L^3$). We conclude that $p = 0.852\%$ lies above the error threshold probability and $p = 0.682\%$ lies below the threshold. That is, MC simulation suggests $p_\mathrm{th} \sim 0.682\%$.

\subsection{Target noise model: Anisotropic asymmetric depolarizing noise plus syndrome noise}
\label{Depolar}

\begin{figure}
    \centering
\includegraphics[width=\columnwidth]{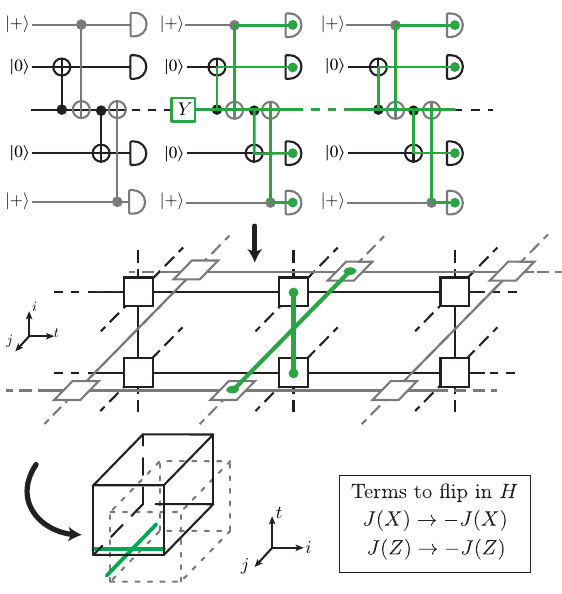}    
    \caption{Translating circuit errors to edges: a $Y-$error on the data qubit flips the adjacent syndrome bits on both $X$ and $Z$ syndrome lattices, which corresponds to four flips on the syndrome difference graph, i.e. a "space-like" hyper-edge on the syndrome (difference) graph at a crossing of edges of the respective sublattices. Note that in quantum circuit convention time flows to the right while in the cubic lattice it flow upwards (such that the spatial lattice in the ij-plane corresponds to one time-step of the toric code syndrome lattice). Gates where the error does not propagate are omitted for simplicity. Note again that the error graph is distinct from the Hamiltonian lattice, to arrive at the latter one has to flip the bonds conjugate to the error (cf. Sec~\ref{sec:RCPGM}), i.e. $J(X)$ and $J(Z)$ in this case. }
    \label{fig:t_edge_circuit_unitcell}
\end{figure}

Given the apparent success of Monte Carlo investigation of the target noise model ansatz, we now define the main target model in this study: (asymmetric) depolarizing noise on the data qubits with noise rates $\pr (X), \pr (Y)$ and $ \pr (Z)$ and syndrome noise with probability $q$. The reduction technique is the same as above, except that now $Y$ errors are not broken up further into a separate $X$ and $Z$ errors but enter the target model. Again starting from circuit-level noise in the given circuit schedule above, we find the following effective noise rates:
\begin{align}
\pr(X_h) = \frac{4p}{3}  + \frac{32p}{15}\\
\pr(X_v) = \frac{4p}{3}  + \frac{16p}{15}\\
\pr(Y_h) = \pr(Y_v) = \frac{4p}{3}  + \frac{16p}{15}\\
\pr(Z_h) = \frac{4p}{3}  + \frac{16p}{15}\\
\pr(Z_v) = \frac{4p}{3}  + \frac{32p}{15}\\
\pr(q) = \frac{8p}{3}  + \frac{48p}{15}
\label{prob_aniso}
\end{align}
Note that the $X$ error rate is consistent with the previous subsection when marginalizing (where the $Y$ adds to $X$ rate). 

\begin{figure}
    \centering
    \includegraphics[width=0.8\columnwidth]{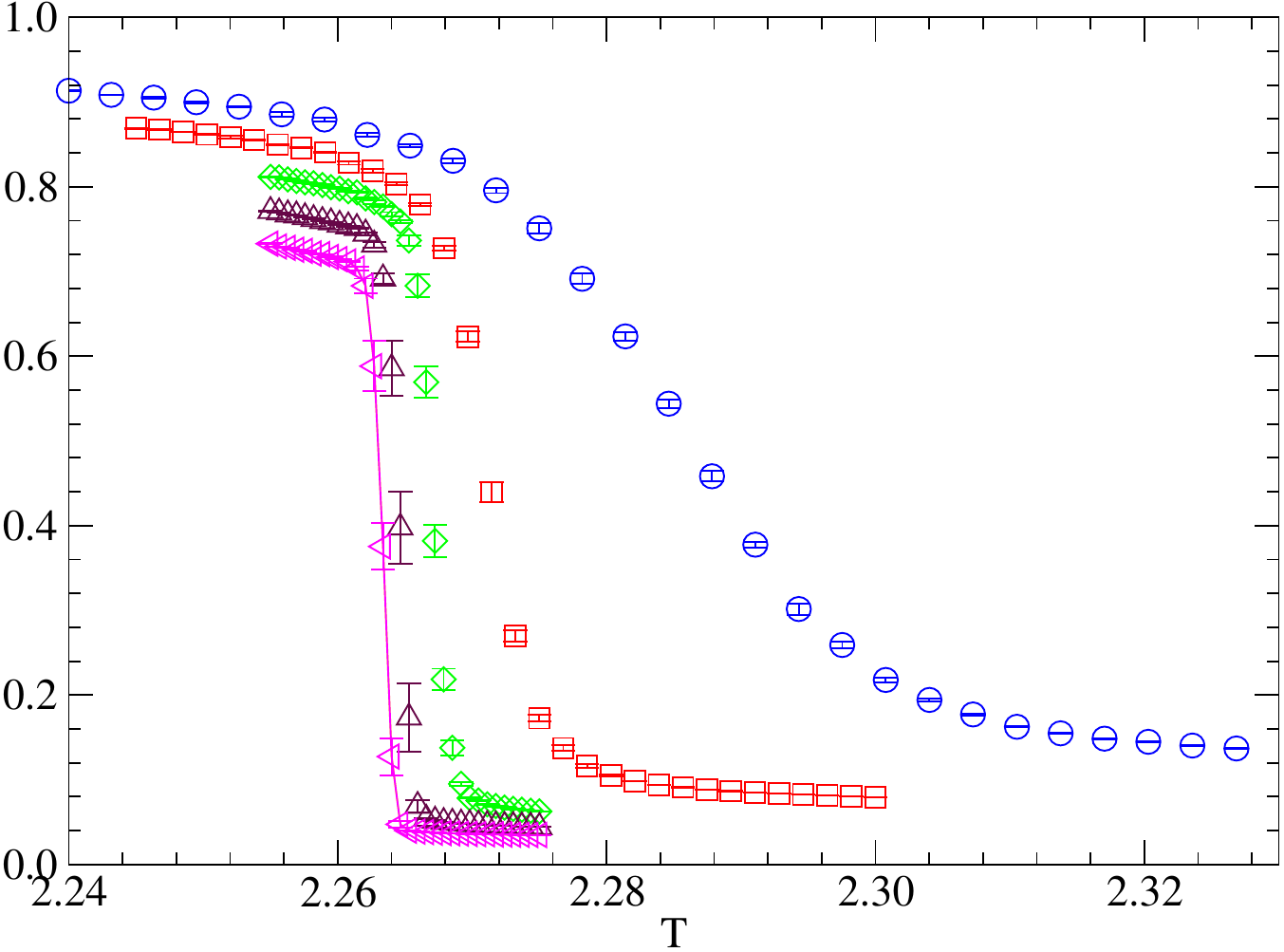}
 \includegraphics[width=0.8\columnwidth]{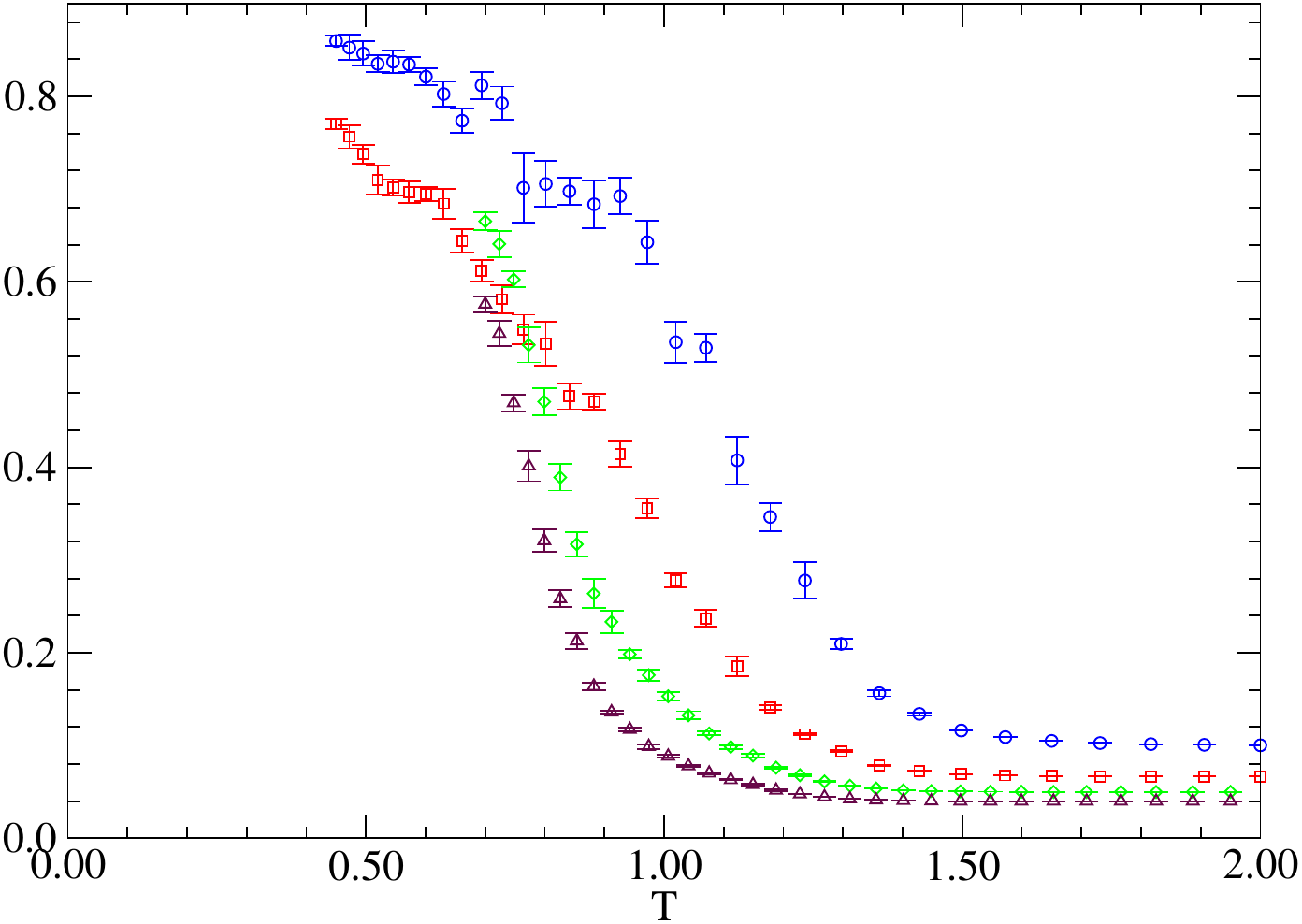} \includegraphics[width=0.8\columnwidth]{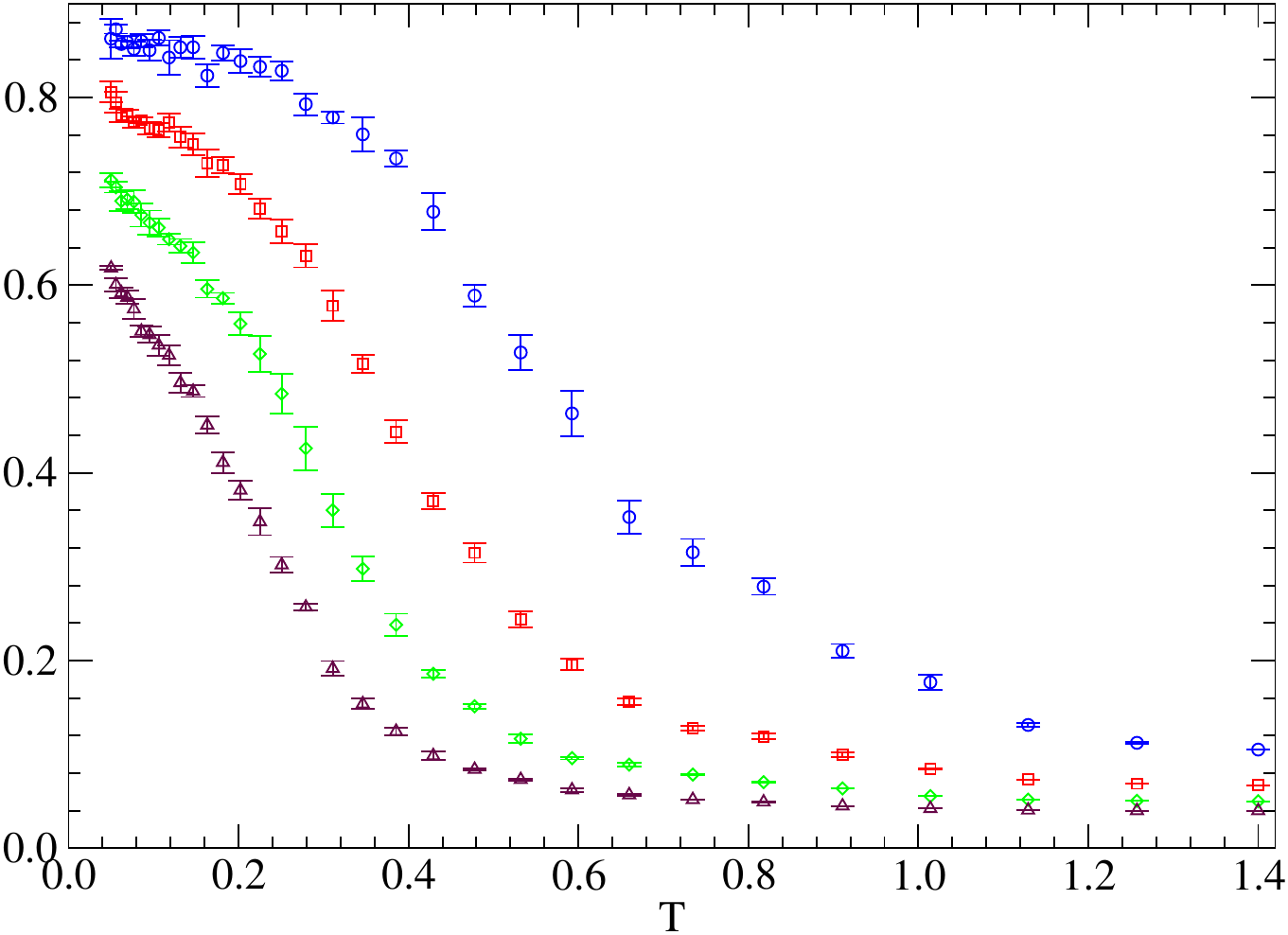}
 \includegraphics[width=0.8\columnwidth]{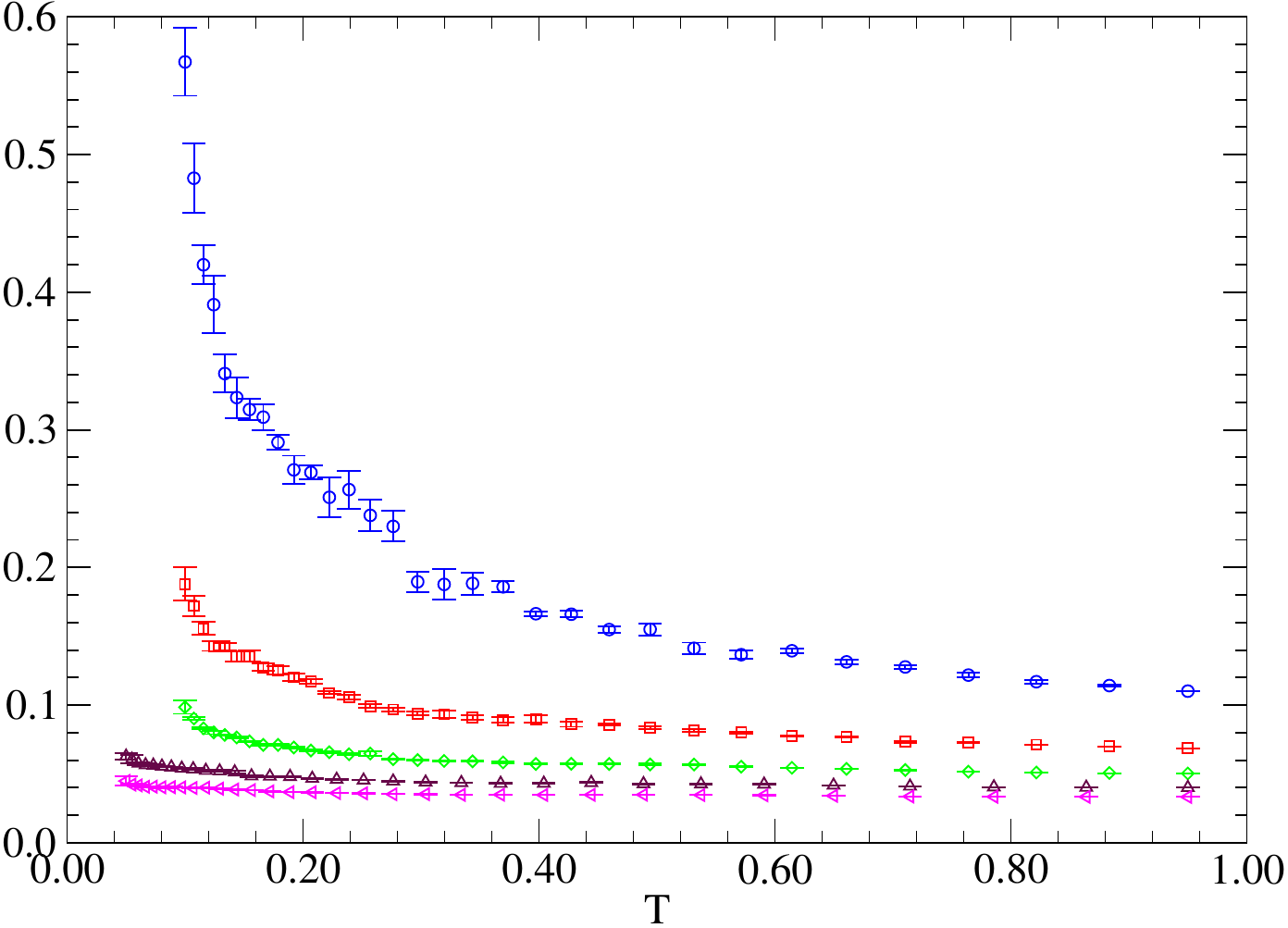}
    \caption{Temperature behavior of average Polyakov line at $p = 2.88\times 10^{-5}$ (first), $p = 1.44\%$ (second) and $p = 1.87\%$ (third) and $p = 2.31\%$ (fourth), for the anisotropic circuit level noise RCPGM on $8^3$ (blue circle), $12^3$ (red square), $16^3$ (green diamond), $20^3$ (maroon up-triangle), and $24^3$ (magenta left-triangle). From Eq. \ref{prob_aniso}, $\pr (X_h) = \frac{52p}{15} = 0.01\%$ (top left), $8\%$ (top right), $5\%$ (bottom left), and $6.5\%$ (bottom right) respectively.}
    \label{fig:circuit-order-parameter}
\end{figure}

\begin{figure}
    \centering
    \includegraphics[width=0.8\columnwidth]{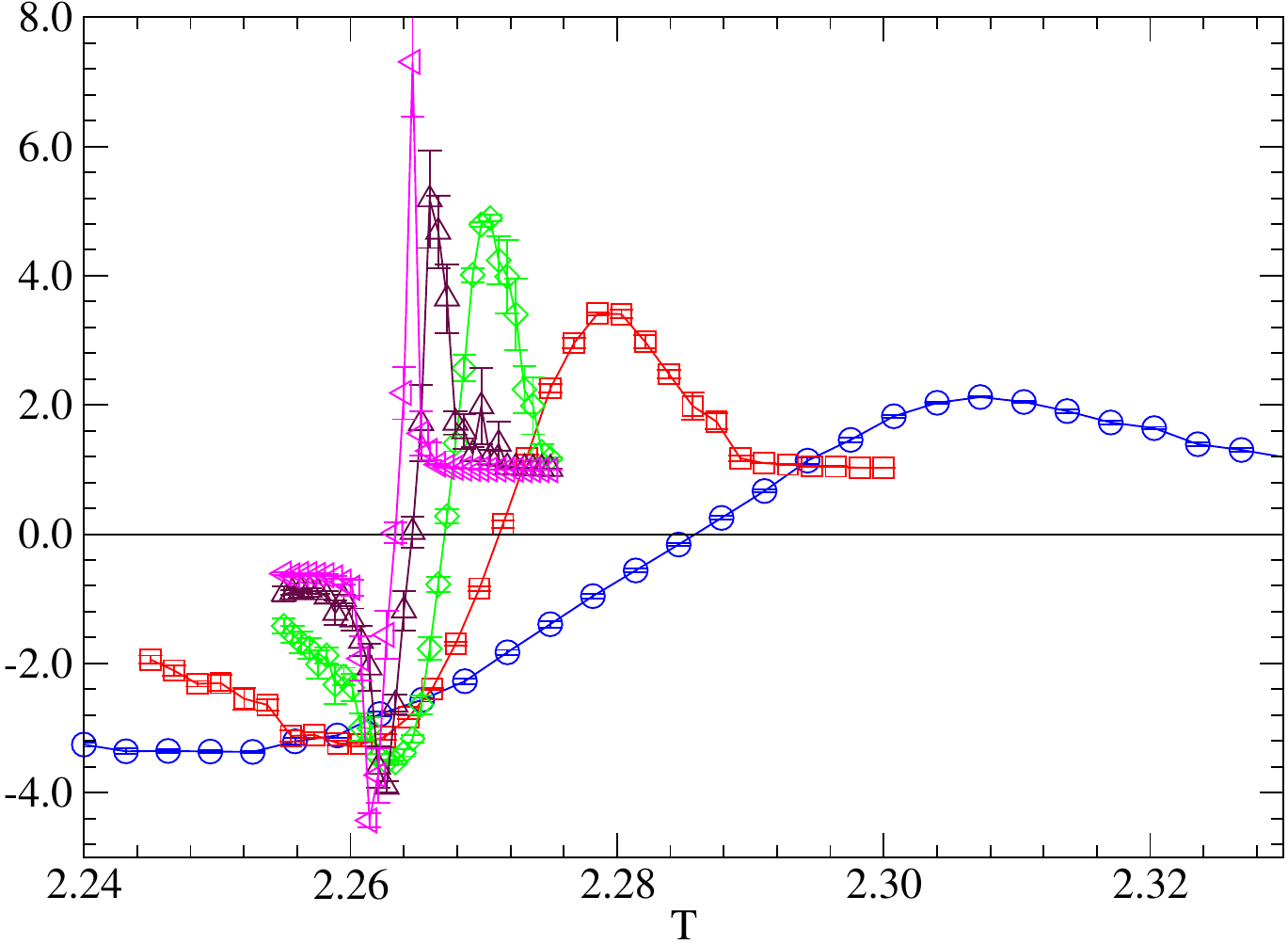}    
    \includegraphics[width=0.8\columnwidth]{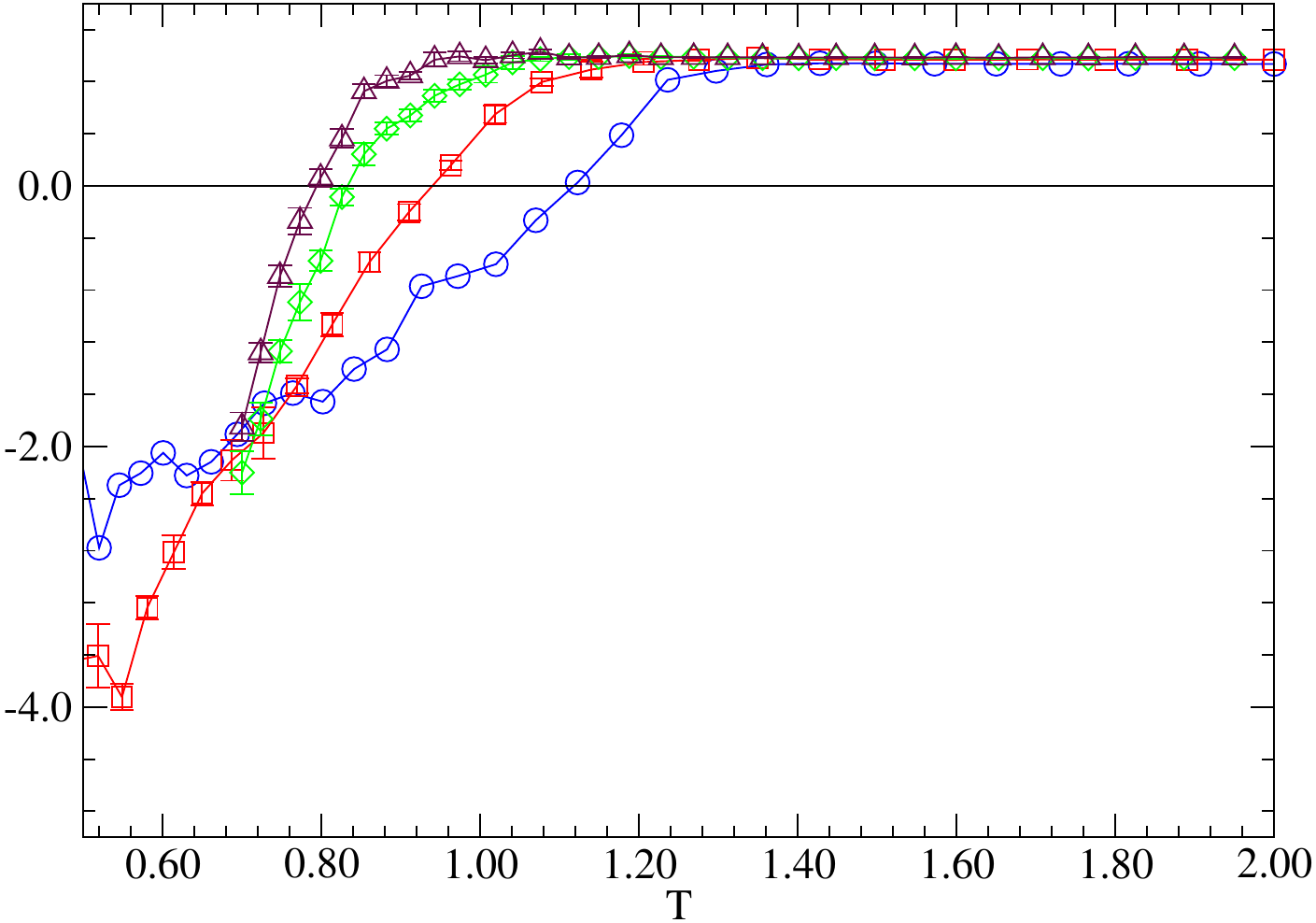} 
    \includegraphics[width=0.8\columnwidth]{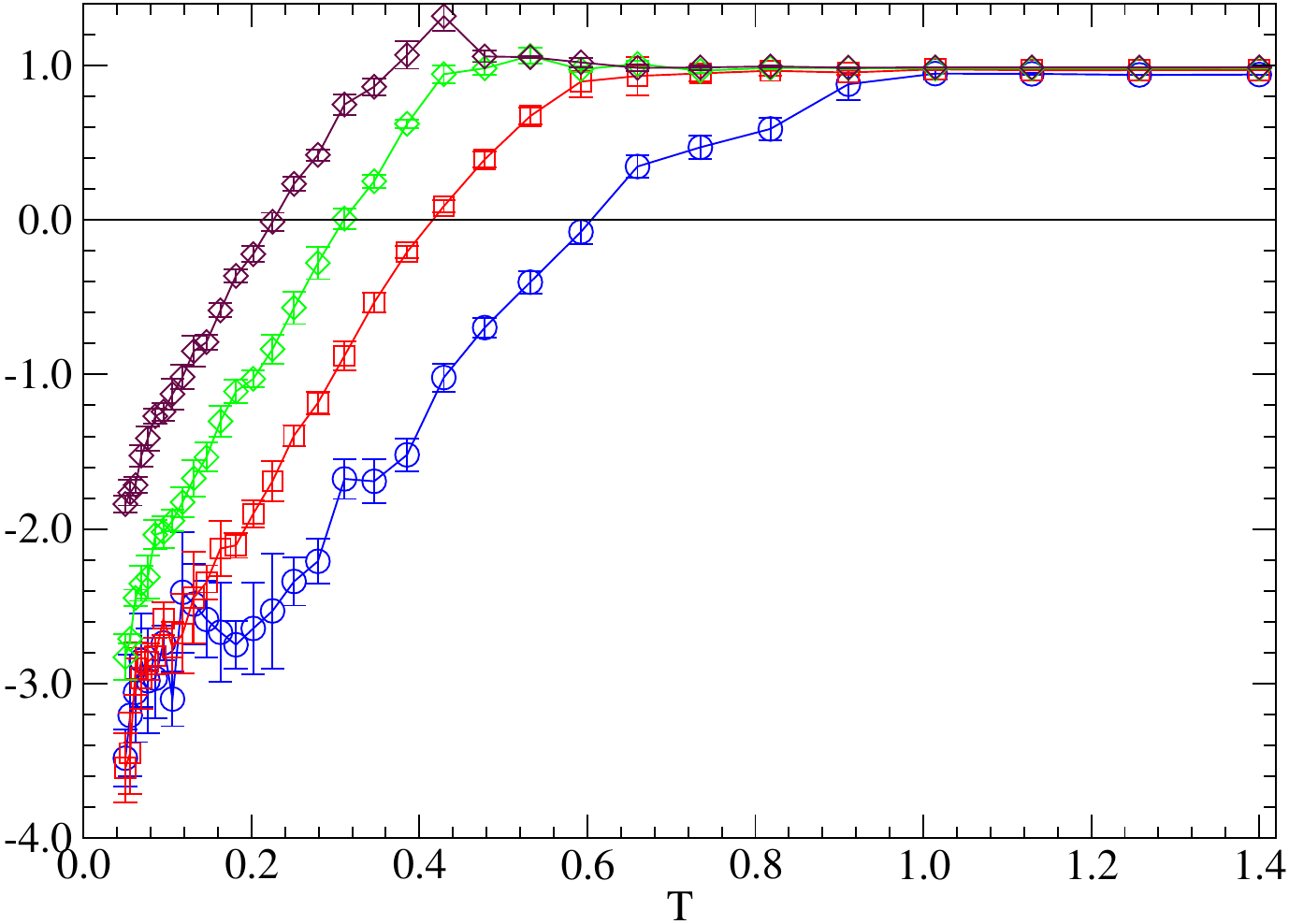}
    \includegraphics[width=0.8\columnwidth]{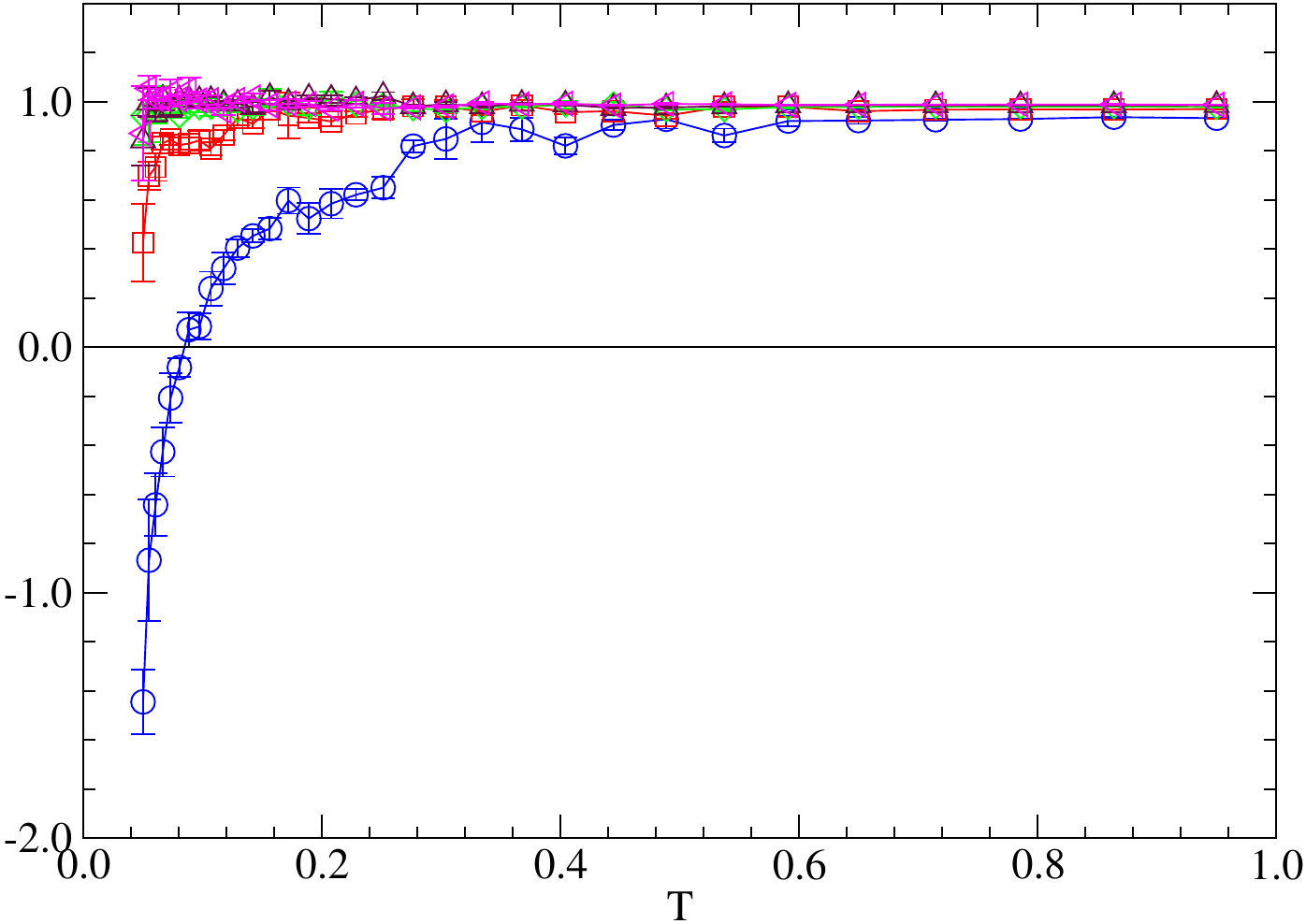} 
    \caption{Temperature behavior of the third order cumulant of the Polyakov line at $p = 2.88\times 10^{-5}$ (first), $p = 1.44\%$ (second), $p = 1.87\%$ (third) and $p = 2.31\%$ (fourth) for the circuit level noise RCPGM. Symbols and colors are the same as in Fig.~\ref{fig:circuit-order-parameter}}
    \label{fig:circuit-B3}
\end{figure}

\begin{figure}
    \centering
    \includegraphics[width=0.8\columnwidth]{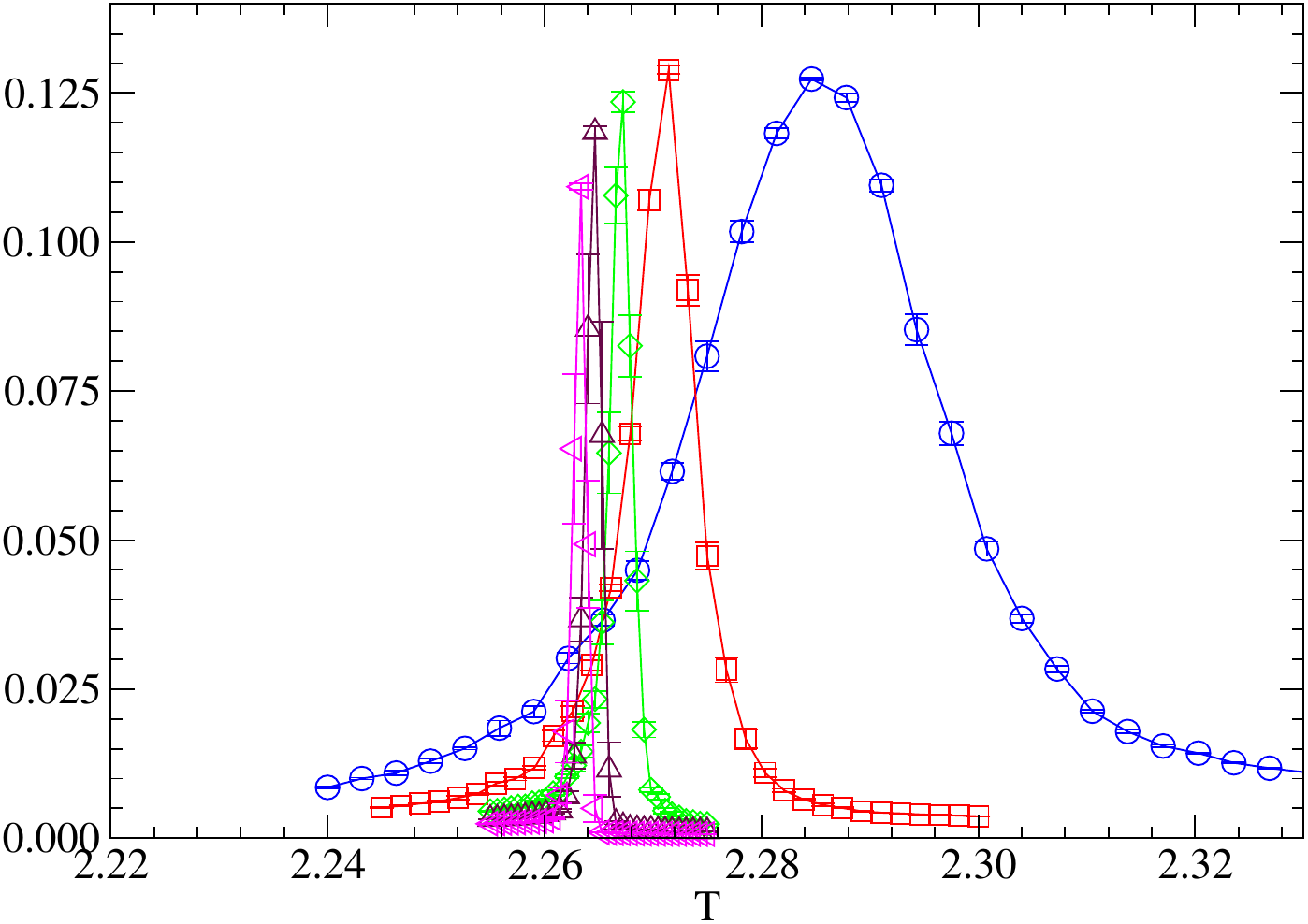}
    \includegraphics[width=0.8\columnwidth]{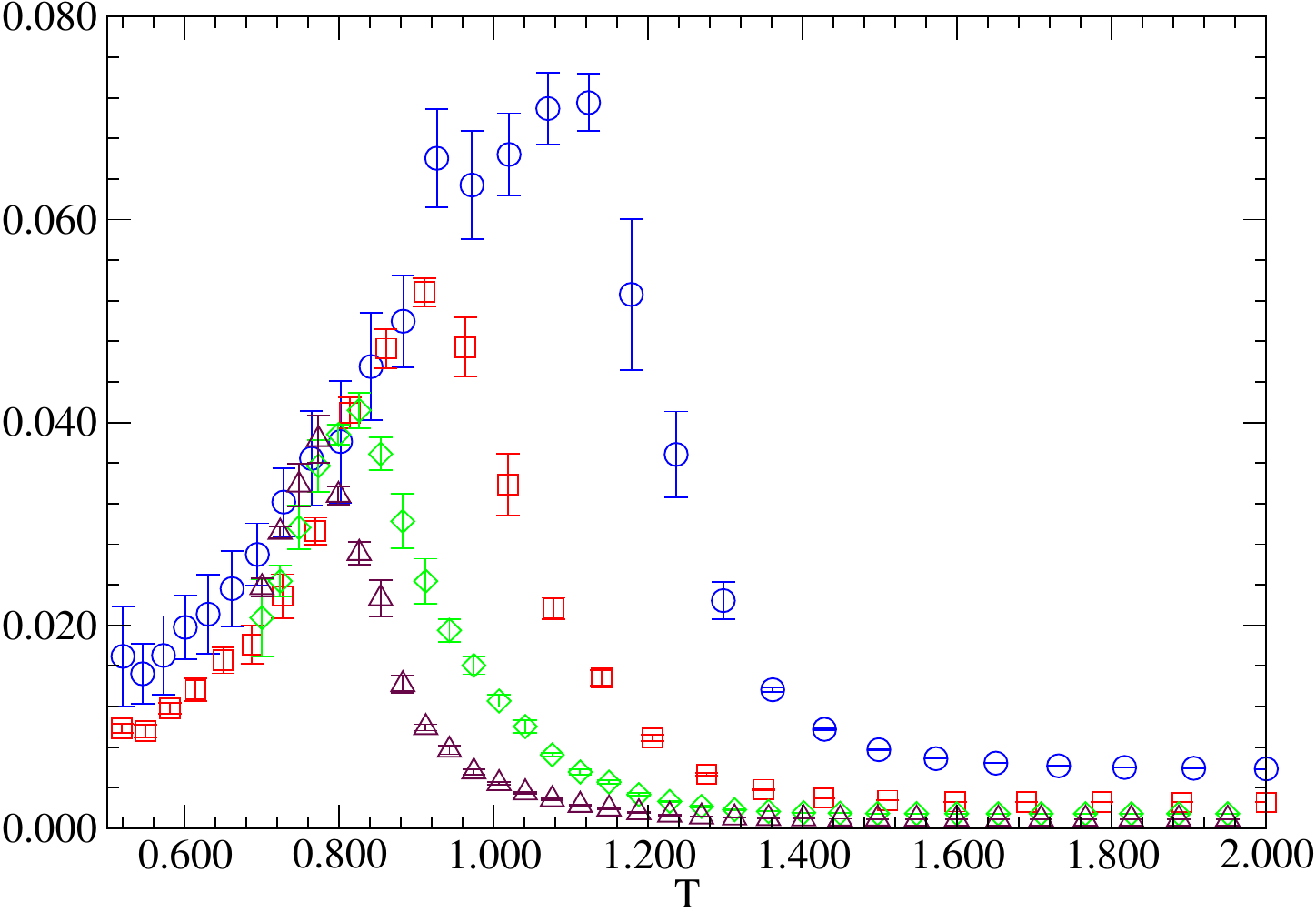} \includegraphics[width=0.8\columnwidth]{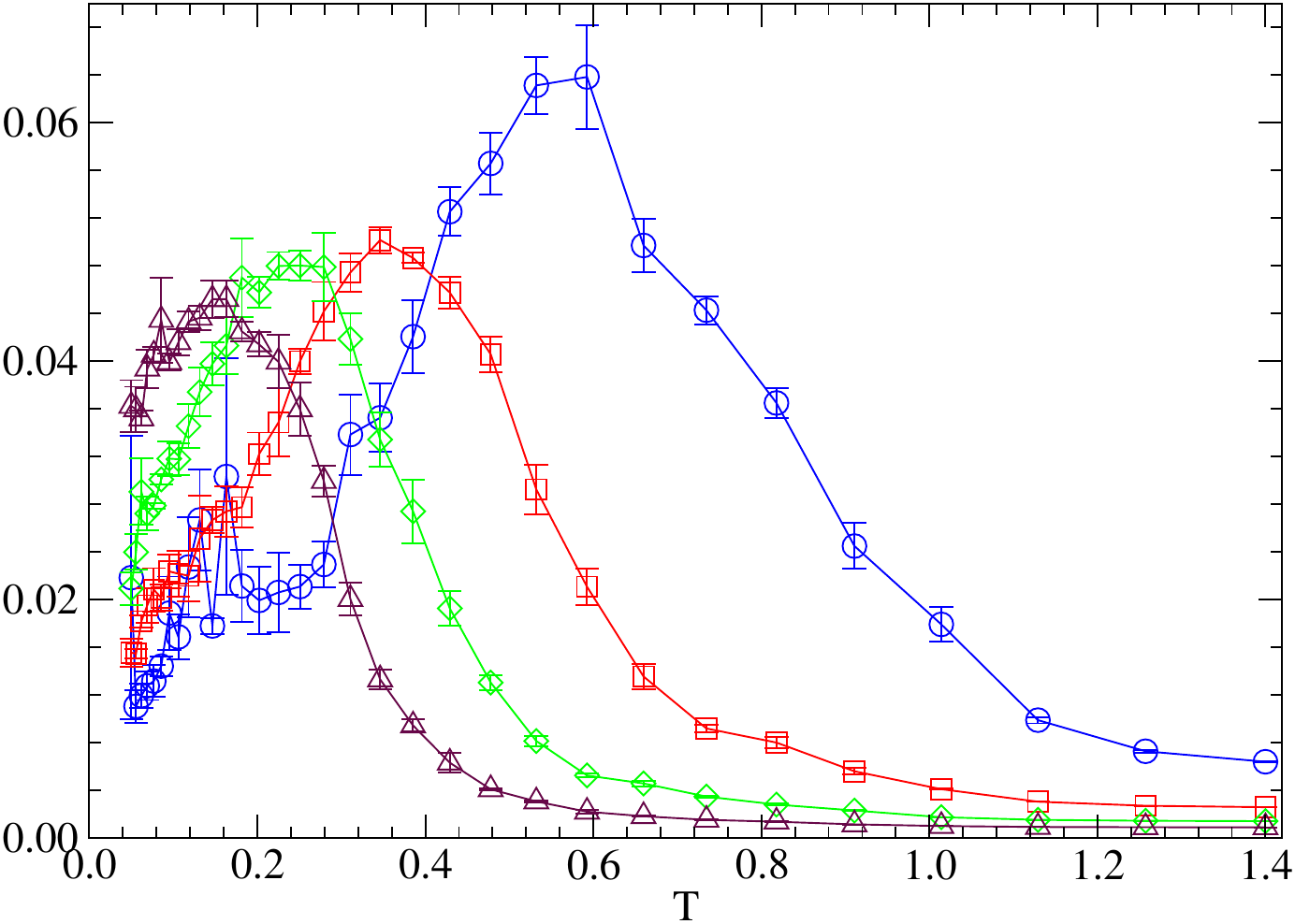}
    \includegraphics[width=0.8\columnwidth]{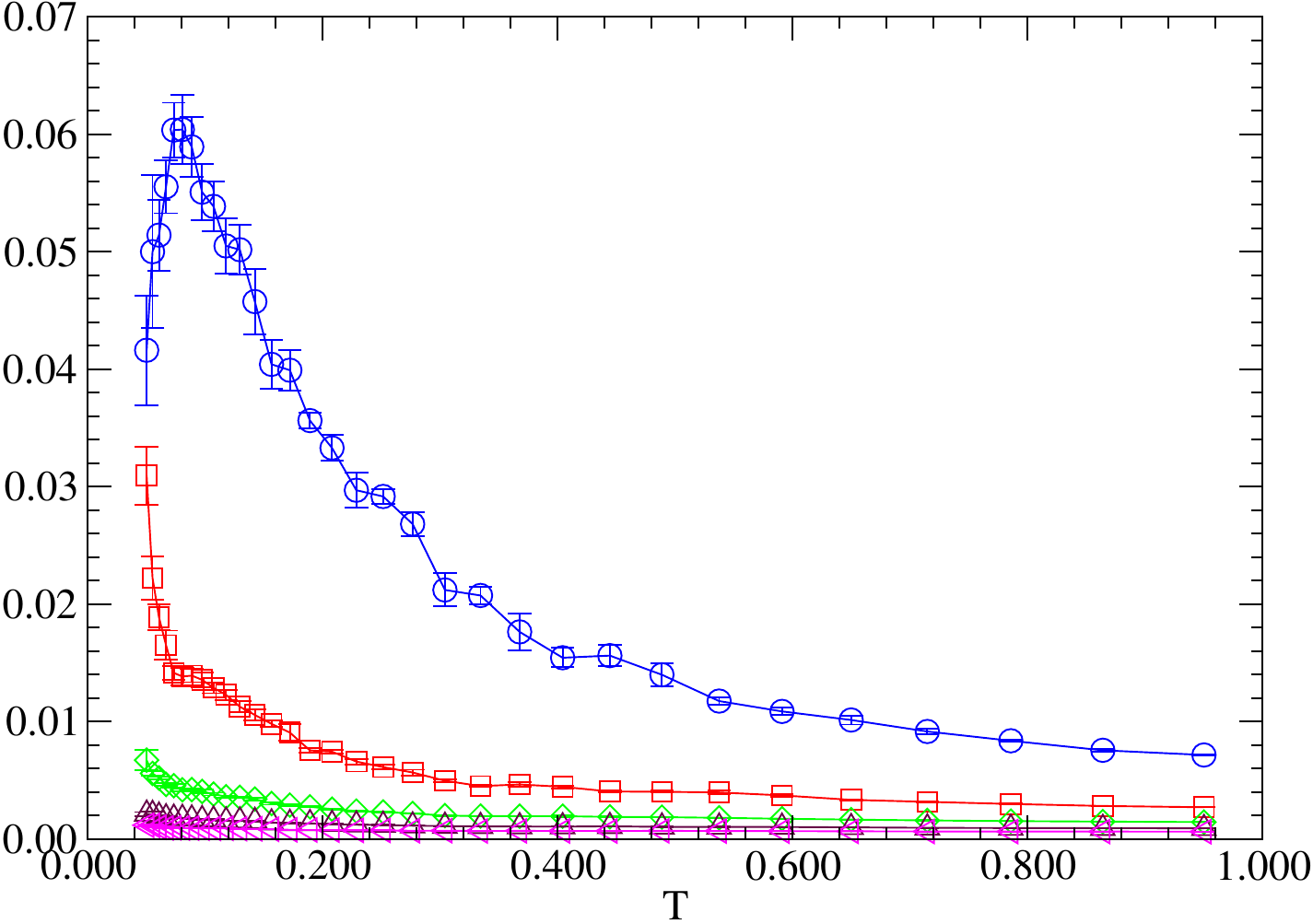}    
    \caption{Temperature behavior of the susceptibility of the Polyakov line at $p = 2.88\times 10^{-5}$ (first), $p = 1.44\%$ (second), $p = 1.87\%$ (third) $p = 2.31\%$ (fourth) for the circuit level noise RCPGM. Symbols and colors are the same as in Fig.~\ref{fig:circuit-order-parameter}}
    \label{fig:circuit-susc}
\end{figure}

The results from Monte Carlo simulation of this statistical physics model, RCPGM with anisotropic couplings (which corresponds to the realistic circuit-level noise case), are shown in Fig. \ref{fig:circuit-order-parameter} (the average Polyakov line), Fig. \ref{fig:circuit-B3} ($B_3)$, and Fig. \ref{fig:circuit-susc} ($\chi$). The average Polyakov line behaves similarly to the symmetric depolarizing noise case. There is a well-defined transition temperature in the infinite volume limit at $p = 2.88\times 10^{-5}$ (top left in Fig. \ref{fig:circuit-order-parameter}). Well above the threshold error probability (e.g., at $p = 2.31\%$ (top right in Fig. \ref{fig:circuit-order-parameter})), the average Polyakov line does not show a transition as the simulation volume increases. Below and near the threshold probability (e.g., $p = 1.44\%$ (bottom in Fig. \ref{fig:circuit-order-parameter})), the order parameter still retains a transition.

\begin{figure}
    \centering
    \includegraphics[width=0.9\columnwidth]{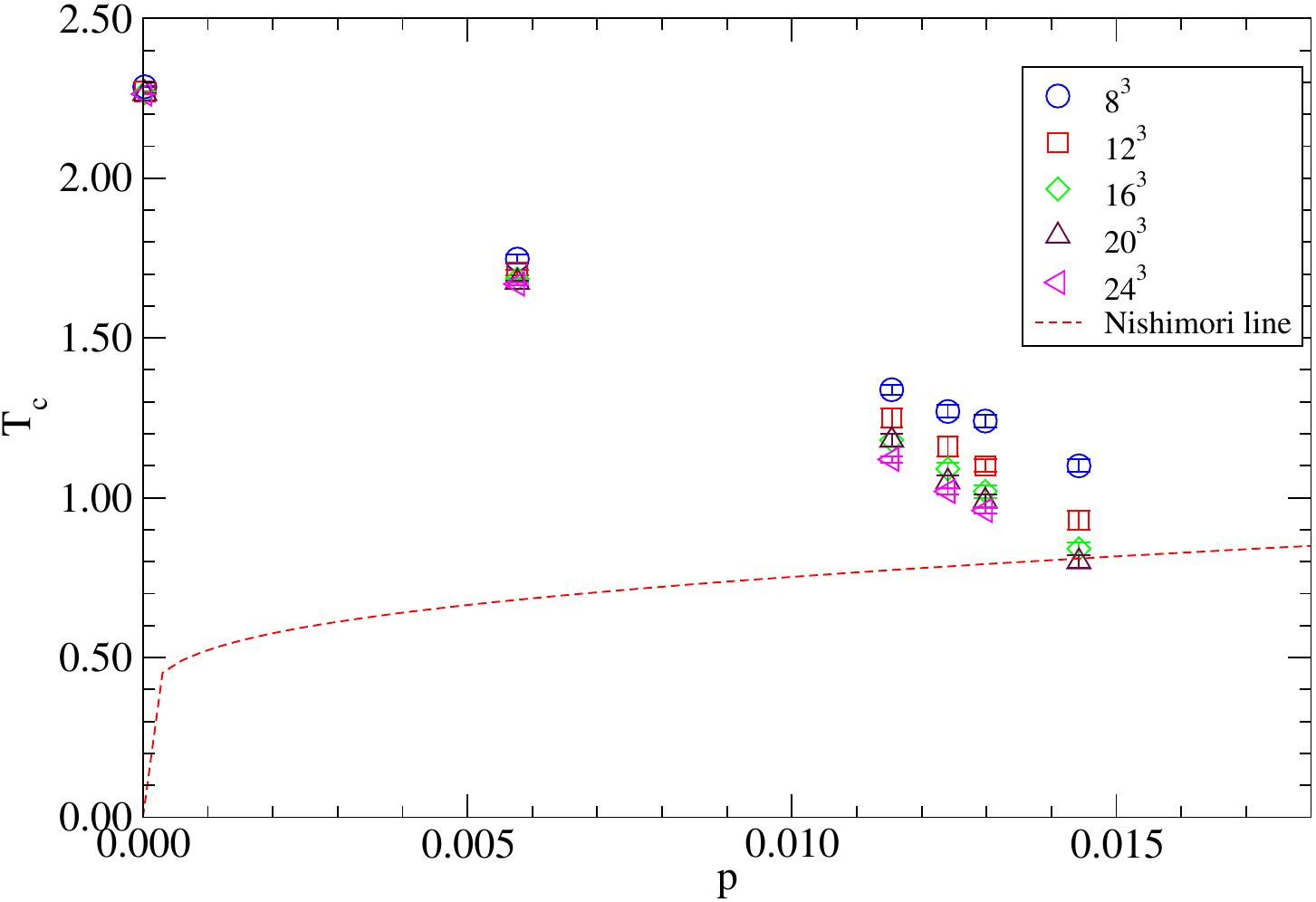}    
    \caption{MC phase diagram for the circuit-level noise RCPGM}
    \label{fig:aniso_phase}
\end{figure}

\begin{figure}
    \centering
    \includegraphics[width=0.9\columnwidth]{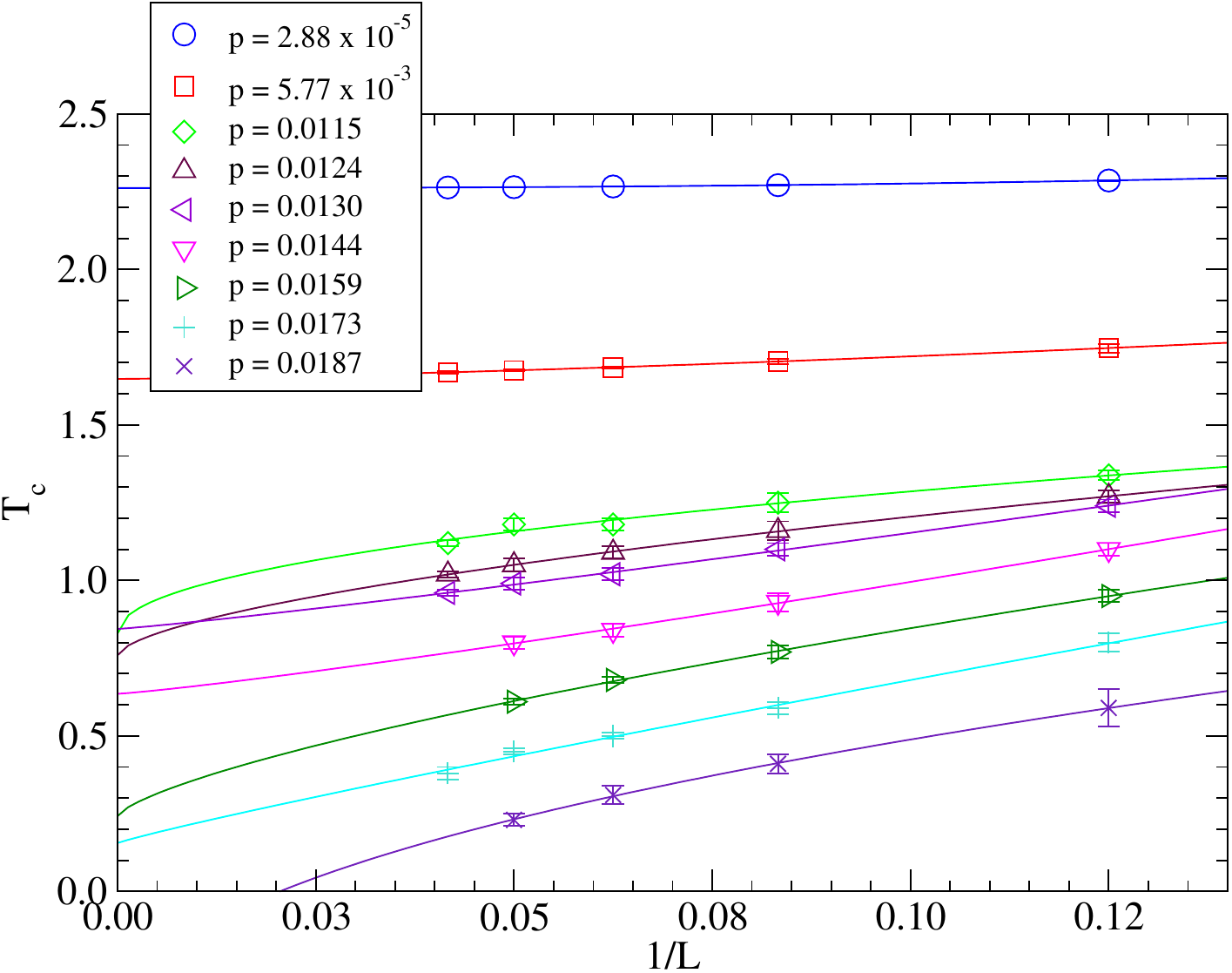}
    \caption{Critical temperature of the anisotropic RCPGM against inverse system size for varying noise strength $p$. Lines are interpolations by fitting to a power law function with offset. A crossing point manifestly larger than zero of a curve with the $y$-axis indicates a finite critical temperature in the large system limit, i.e. the corresponding noise strength being below threshold where error correction is beneficial.}
    \label{fig:anisotropic_RCPGM_fit_vs_V}
\end{figure}
Via the behavior of the order parameter, the third order cumulant ($B_3$) and the susceptibility ($\chi$) at various noise levels on different lattice volumes, the transition temperature is mapped out and is summarized in Fig. \ref{fig:aniso_phase}. In Fig.~\ref{fig:anisotropic_RCPGM_fit_vs_V}, we plot the critical temperature as a function of the inverse of 1-dimensional system size together with fits to a power-law with an offset, $T_c (L) = a L^{-b} + T_c (\infty)$ ~\cite{Kubica2018}, in which one can estimate the large volume limit of the transition temperatures. 

Figures \ref{fig:circuit-order-parameter}, \ref{fig:circuit-B3} and \ref{fig:circuit-susc} show that clearly there is no transition at $p = 2.31\%$ (top right in each figure). At $p = 1.59\%, 1.73\%$ and $1.87\%$, the average Polyakov line, $B_3$, and $\chi$ show a transition. However, large volume limit of the transition temperature does not exist unlike at $p = 1.44\%$: as the MC simulation lattice volume increases, the transition temperature keep decreasing to ever lower temperature without reaching a non-zero temperature. Thus, Monte Carlo simulation suggests that the threshold error probability for the realistic circuit noise together with syndrome noise is $p \simeq 1.44\%$.

\section{Comparison to other decoding methods}
\begin{table}[h]
\begin{tabular}{c|c|c|c|c|c}
    \#steps & init./meas. & idling?& method  & threshold &  Ref.\\
    \hline\\
     8 & BF     &yes& MWPM& $0.502\%$ & \cite{Stephens2014}\\
     5(6)\footnote{5 steps on data, 6 steps on ancilla, since data only gets one joint idling  for init and measurement locations} & DEP & yes&MWPM & $0.82\%$  & \cite{Higgott2023Fragile}\\
    5(6) & DEP &  yes&belief-MWPM & $0.94\%$ & \cite{Higgott2023Fragile}\\
     6 & BF & yes& MWPM & $0.9\%$    & \cite{Wang2010}\\
     6 & DEP      & yes & MWPM & $0.75\%$ & \cite{Raussendorf2007}\\
     ?\footnote{The given paper does not precisely define the error model. However, Ref.~\cite{Fowler2009} has $p_i,p_m,p_r,p_2$, explicitly excluding single-qubit gate errors, which are assumed to be compiled away into the entangling gates.} & ?      & yes& MWPM & $0.5\%$& \cite{Fowler2013}\\ 
     ? & ?      & yes & correlated MWPM & $\approx 0.65\%$ & \cite{Fowler2013}\\ 
      6& BF& no& MWPM & $0.78\%$ & \cite{Piveteau2023}\\
      6 &  BF & no &tensor network & $0.8\%$ & \cite{Piveteau2023}\\

\end{tabular}
\caption{Overview over literature results regarding circuit-level noise models thresholds and decoders used. Note that the number of time-steps depends on whether authors assume initialization and measurement in the X-basis or not (the latter requiring two additional time-steps for Hadamards). Furthermore the initialization and measurement errors are either modeled as bit-flip (BF) with probability $p$ or as depolarizing (DEP), the latter implies a correspondingly lower error rate of a flip rate $2p/3$. For some works we failed to extract the exact noise model from the manuscript.  that~\cite{Wang2010} reports time-to-failure, which has fallen out of fashion and the same authors transitioned to reporting logical failure rates per syndrome volume, which seems to represent thresholds more accurately.}
\label{tab:cln-decoder-thresholds}
\end{table}

Notable strides towards decoding the surface code under circuit-level noise 
beyond weight matching decoding were taken before. A recent in-depth study was performed in~\cite{Higgott2023Fragile}. The authors define the circuit-level noise model with all errors as depolarizing, however they assume initialization and measurement are available in the $X$ basis as well (corresponding to a 6-step, with the modification that initialization and measurement count as only half a location). 
They report that their belief-matching decoder achieves $p_{th}=0.94\%$ compared to MWPM at $p_{th}=0.82\%$. Wang et al.~\cite{Wang2010} reported a threshold of $0.9\%$ under MWPM, where they have a slightly different circuit-noise model which has initialization and measurement flip with probability $p$ (opposed to $2p/3$ above) alongside the depolarizing idle, single and two-qubit gates. In Ref.~\cite{Heim2016}, the authors take this as a reference point to present a maximum-likelihood decoder. In an approach somewhat complementary to our work, they marginalize the effects of circuit-level noise to X- and Z-syndrome separately. They perform time-to-failure simulation on a circuit-noise model ("6-step SN-circuit" inspired by~\cite{Wang2010}) and report a threshold of around $1.5\%$. Barring differences in details of simulation method and  noise model, this indicates substantial room for improvement on the best known threshold with practical decoders.\section{Discussion and outlook}
In this work, we have established the random coupled-plaquette gauge model (RCPGM) as the relevant statistical model to describe the decoding problem of the surface code under depolarizing data and bit-flip measurement noise. This new statistical model is capable of capturing Y-error correlations between both syndrome lattices in a setting that describes the realistic situation of unreliable syndrome information. We have mapped out the phase diagram of this quenched disordered model with Parallel Tempering Monte Carlo simulations and reported a threshold of $6\%$ under uniform noise strength. Given that the previously known uncoupled random plaquette gauge model yields a threshold of $4.3\%$ in the identical noise model, this demonstrates the importance of treating $Y$-errors appropriately, leading to a relative improvement of $40\%$ in threshold value.

For the case of circuit-level noise, we have employed a reduction technique which has allowed us to use the RCPGM as a proxy for this experimentally relevant noise setting. We have performed Parallel Tempering Monte Carlo simulations for the resulting anisotropic RCPGM, which show a threshold value of up to $1.4\%$. We have benchmarked against the corresponding uncoupled anisotropic RPGM, which shows a threshold of $0.7\%$ under the otherwise identical noise scenario. This reinforces and even surpasses the findings for phenomenological case by showing a relative improvement of doubling the threshold value.\\

Our result establishes that there is substantial room for improvement beyond state of the art efficient decoders for the surface code under realistic noise. This is encouraging for research towards efficient decoders that aim at incorporating syndrome lattice correlations induced by Y-errors and/or take code degeneracy into account, most notably those improving the leading decoding algorithm of minimum weight perfect matching. As a very recent example in this direction, substantial improvements were achieved by belief-matching, where enhancing MWPM with belief propagation leads to significant performance improvements compared to standard MWPM~\cite{Higgott2023Fragile}. For an overview over scalable decoder literature results and reported thresholds see Tab.~\ref{tab:cln-decoder-thresholds}.\\

Note that we have found these drastic improvements already under a reduction of circuit-level noise, where we have approximately captured the effects of circuit-level noise with a heuristic simplification in order to compress the plethora of different noise processes. Incorporating further effects of circuit noise into an even more complex statistical mechanical model beyond the RCPGM could push these limits even further. We leave it for future work to generalize the model and simulation techniques in this direction.

For phenomenological noise, the statistical mechanics mapping carries over to topological color codes rather straightforwardly. For coherent noise on the surface code, recent progress has shown that this can be incorporated for code capacity~\cite{Venn2023} and phenomenological noise~\cite{Marton2023}. For circuit-level noise, the picture is much less clear.  E.g.~when considering more general codes such as the topological color code, the syndrome extraction circuit itself can induce high weight errors by dangerous error propagation, such that one has to resort e.g.~to flag qubits~\cite{Chamberland2020_low_degree,Chamberland2020_trivalent}, which would have to be incorporated into a statistical mechanics picture. It would be highly desirable to improve our understanding of the threshold of color codes under circuit-level noise, which are less well understood. In particular, a related question of the influence of the circuit design and schedule on the code threshold under circuit noise would be an interesting avenue to explore starting from our work. Lastly, the picture for finite rate qLDPC codes offering a number of logical qubits growing with system size is only recently being touched on even for phenomenological syndrome noise~\cite{Placke2023}. Tt would be extremely valuable to explore the situation for circuit noise on finite rate codes, which will be vital to drive down the overhead required for fault-tolerant quantum computation~\cite{Gottesman2013}.

\begin{acknowledgements}
We thank Eliana Fiorelli for discussions at the inital phase of this project. SK is supported by the National Research Foundation of Korea under grant NRF-2021R1A2C1092701 funded by the Korean government (MEST) 
and by the Institute of Information \& Communication Technology Planning \& Evaluation grant funded by the Korean government (Ministry of Science and ICT) (IITP-2024-RS-2024-00437191). SK would like to thank IQI (RWTH Aachen University) and the Forschungszentrum J\"ulich for the support for the visit to the IQI in the initial phase of this work. M.R.~and M.M. acknowledge support for this research, which is also part of the Munich Quantum Valley (K-8), which is supported by the Bavarian state government with funds from the Hightech Agenda Bayern Plus, as well as support by the  Deutsche Forschungsgemeinschaft (DFG, German Research Foundation) under Germany’s Excellence Strategy Cluster of Excellence Matter and Light for Quantum Computing (ML4Q) EXC 2004/1 390534769. M.M.~furthermore acknowledges funding by the U.S. ARO Grant No. W911NF-21-1-0007, funding from the European Union’s Horizon Europe research and innovation programme under grant agreement No 101114305 (“MILLENION-SGA1” EU Project), and the ERC Starting Grant QNets through Grant No.~804247.
\end{acknowledgements}

\appendix

\section{Exhaustive error location tables}
\label{app:location_list}
\noindent

 Here we list the effect of every possible depolarizing error following any of the eight CNOTs in the schedule. The superscript indicates which stabilizer the CNOT implements, e.g. $\mathrm{CNOT}^X_1$ is the first CNOT of the X-stabilizer circuit. The errors are given in the first column in the order data-ancilla, the subsequent columns refer to the edges in the 3D unit cell of the X-Z-syndrome volume, where e.g. $i_Z$ and $j_Z$ are the two spatial edges in the Z-syndrome lattice and $t_Z$ the time-like edge, which can either be triggered (1) or not (0). For example, the first column then reads, that placing an X-error on the X-syndrome ancilla qubit after the first CNOT has the effect of triggering the $i_Z$ edge, i.e. it is equivalent to a data qubit error.  As can be seen by the CNOT propagation rules, this error is equivalent to the error $\mathrm{XX}$ before the first CNOT, i.e. a regular data X-error and an X-flip on the ancilla, where the latter is inconsequential since it is acting on its eigenstate $\ket{+}=H\ket{0}$. As another example, we can retrieve the case presented in Fig.~\ref{fig:t_edge_circuit_unitcell} as the error $\mathrm{IX}$ on $\mathrm{CNOT}_4^Z$.\\

\begin{tabular}{c|c|c|c|c|c|c}
 $\mathrm{CNOT}_1^X$  & $i_Z$ & $j_Z$ & $t_Z$ & $i_X$ & $j_X$ & $t_X$\\
 \hline
IX & 1 & 0 & 0 & 0 & 0 & 0\\
IY & 1 & 0 & 0 & 0 & 0 & 1\\
IZ & 0 & 0 & 0 & 0 & 0 & 1\\
XI & 1 & 0 & 0 & 0 & 0 & 0\\
XX & 0 & 0 & 0 & 0 & 0 & 0\\
XY & 0 & 0 & 0 & 0 & 0 & 1\\
XZ & 1 & 0 & 0 & 0 & 0 & 1\\
YI & 1 & 0 & 0 & 0 & 1 & 1\\
YX & 0 & 0 & 0 & 0 & 1 & 1\\
YY & 0 & 0 & 0 & 0 & 1 & 0\\
YZ & 1 & 0 & 0 & 0 & 1 & 0\\
ZI & 0 & 0 & 0 & 0 & 1 & 1\\
ZX & 1 & 0 & 0 & 0 & 1 & 1\\
ZY & 1 & 0 & 0 & 0 & 1 & 0\\
ZZ & 0 & 0 & 0 & 0 & 1 & 0\\
\end{tabular}

\begin{tabular}{c|c|c|c|c|c|c}
 $\mathrm{CNOT}_2^X$  & $i_Z$ & $j_Z$ & $t_Z$ & $i_X$ & $j_X$ & $t_X$\\
 \hline
IX & 1 & 1 & 1 & 0 & 0 & 0\\
IY & 1 & 1 & 1 & 0 & 0 & 1\\
IZ & 0 & 0 & 0 & 0 & 0 & 1\\
XI & 0 & 1 & 1 & 0 & 0 & 0\\
XX & 1 & 0 & 0 & 0 & 0 & 0\\
XY & 1 & 0 & 0 & 0 & 0 & 1\\
XZ & 0 & 1 & 1 & 0 & 0 & 1\\
YI & 0 & 1 & 1 & 1 & 0 & 1\\
YX & 1 & 0 & 0 & 1 & 0 & 1\\
YY & 1 & 0 & 0 & 1 & 0 & 0\\
YZ & 0 & 1 & 1 & 1 & 0 & 0\\
ZI & 0 & 0 & 0 & 1 & 0 & 1\\
ZX & 1 & 1 & 1 & 1 & 0 & 1\\
ZY & 1 & 1 & 1 & 1 & 0 & 0\\
ZZ & 0 & 0 & 0 & 1 & 0 & 0\\
\end{tabular}

\begin{tabular}{c|c|c|c|c|c|c}
 $\mathrm{CNOT}_3^X$  & $i_Z$ & $j_Z$ & $t_Z$ & $i_X$ & $j_X$ & $t_X$\\
 \hline
IX & 1 & 0 & 0 & 0 & 0 & 0\\
IY & 1 & 0 & 0 & 0 & 0 & 1\\
IZ & 0 & 0 & 0 & 0 & 0 & 1\\
XI & 0 & 1 & 1 & 0 & 0 & 0\\
XX & 1 & 1 & 1 & 0 & 0 & 0\\
XY & 1 & 1 & 1 & 0 & 0 & 1\\
XZ & 0 & 1 & 1 & 0 & 0 & 1\\
YI & 0 & 1 & 1 & 1 & 0 & 0\\
YX & 1 & 1 & 1 & 1 & 0 & 0\\
YY & 1 & 1 & 1 & 1 & 0 & 1\\
YZ & 0 & 1 & 1 & 1 & 0 & 1\\
ZI & 0 & 0 & 0 & 1 & 0 & 0\\
ZX & 1 & 0 & 0 & 1 & 0 & 0\\
ZY & 1 & 0 & 0 & 1 & 0 & 1\\
ZZ & 0 & 0 & 0 & 1 & 0 & 1\\
\end{tabular}

\begin{tabular}{c|c|c|c|c|c|c}
$\mathrm{CNOT}_4^X$  & $i_Z$ & $j_Z$ & $t_Z$ & $i_X$ & $j_X$ & $t_X$\\
\hline
IX & 0 & 0 & 0 & 0 & 0 & 0\\
IY & 0 & 0 & 0 & 0 & 0 & 1\\
IZ & 0 & 0 & 0 & 0 & 0 & 1\\
XI & 1 & 0 & 0 & 0 & 0 & 0\\
XX & 1 & 0 & 0 & 0 & 0 & 0\\
XY & 1 & 0 & 0 & 0 & 0 & 1\\
XZ & 1 & 0 & 0 & 0 & 0 & 1\\
YI & 1 & 0 & 0 & 0 & 1 & 0\\
YX & 1 & 0 & 0 & 0 & 1 & 0\\
YY & 1 & 0 & 0 & 0 & 1 & 1\\
YZ & 1 & 0 & 0 & 0 & 1 & 1\\
ZI & 0 & 0 & 0 & 0 & 1 & 0\\
ZX & 0 & 0 & 0 & 0 & 1 & 0\\
ZY & 0 & 0 & 0 & 0 & 1 & 1\\
ZZ & 0 & 0 & 0 & 0 & 1 & 1\\
\hline
\end{tabular}

\begin{tabular}{c|c|c|c|c|c|c}
 $\mathrm{CNOT}_1^Z$  & $i_Z$ & $j_Z$ & $t_Z$ & $i_X$ & $j_X$ & $t_X$\\
 \hline
IX & 0 & 0 & 1 & 0 & 0 & 0\\
IY & 0 & 0 & 1 & 1 & 0 & 0\\
IZ & 0 & 0 & 0 & 1 & 0 & 0\\
XI & 0 & 1 & 1 & 0 & 0 & 0\\
XX & 0 & 1 & 0 & 0 & 0 & 0\\
XY & 0 & 1 & 0 & 1 & 0 & 0\\
XZ & 0 & 1 & 1 & 1 & 0 & 0\\
YI & 0 & 1 & 1 & 1 & 0 & 0\\
YX & 0 & 1 & 0 & 1 & 0 & 0\\
YY & 0 & 1 & 0 & 0 & 0 & 0\\
YZ & 0 & 1 & 1 & 0 & 0 & 0\\
ZI & 0 & 0 & 0 & 1 & 0 & 0\\
ZX & 0 & 0 & 1 & 1 & 0 & 0\\
ZY & 0 & 0 & 1 & 0 & 0 & 0\\
ZZ & 0 & 0 & 0 & 0 & 0 & 0\\
\end{tabular}

\begin{tabular}{c|c|c|c|c|c|c}
 $\mathrm{CNOT}_2^Z$  & $i_Z$ & $j_Z$ & $t_Z$ & $i_X$ & $j_X$ & $t_X$\\
 \hline
IX & 0 & 0 & 1 & 0 & 0 & 0\\
IY & 0 & 0 & 1 & 1 & 1 & 1\\
IZ & 0 & 0 & 0 & 1 & 1 & 1\\
XI & 1 & 0 & 1 & 0 & 0 & 0\\
XX & 1 & 0 & 0 & 0 & 0 & 0\\
XY & 1 & 0 & 0 & 1 & 1 & 1\\
XZ & 1 & 0 & 1 & 1 & 1 & 1\\
YI & 1 & 0 & 1 & 0 & 1 & 1\\
YX & 1 & 0 & 0 & 0 & 1 & 1\\
YY & 1 & 0 & 0 & 1 & 0 & 0\\
YZ & 1 & 0 & 1 & 1 & 0 & 0\\
ZI & 0 & 0 & 0 & 0 & 1 & 1\\
ZX & 0 & 0 & 1 & 0 & 1 & 1\\
ZY & 0 & 0 & 1 & 1 & 0 & 0\\
ZZ & 0 & 0 & 0 & 1 & 0 & 0\\
\end{tabular}

\begin{tabular}{c|c|c|c|c|c|c}
 $\mathrm{CNOT}_3^Z$  & $i_Z$ & $j_Z$ & $t_Z$ & $i_X$ & $j_X$ & $t_X$\\
 \hline
IX & 0 & 0 & 1 & 0 & 0 & 0\\
IY & 0 & 0 & 1 & 1 & 0 & 0\\
IZ & 0 & 0 & 0 & 1 & 0 & 0\\
XI & 1 & 0 & 0 & 0 & 0 & 0\\
XX & 1 & 0 & 1 & 0 & 0 & 0\\
XY & 1 & 0 & 1 & 1 & 0 & 0\\
XZ & 1 & 0 & 0 & 1 & 0 & 0\\
YI & 1 & 0 & 0 & 0 & 1 & 1\\
YX & 1 & 0 & 1 & 0 & 1 & 1\\
YY & 1 & 0 & 1 & 1 & 1 & 1\\
YZ & 1 & 0 & 0 & 1 & 1 & 1\\
ZI & 0 & 0 & 0 & 0 & 1 & 1\\
ZX & 0 & 0 & 1 & 0 & 1 & 1\\
ZY & 0 & 0 & 1 & 1 & 1 & 1\\
ZZ & 0 & 0 & 0 & 1 & 1 & 1\\
\end{tabular}

\begin{tabular}{c|c|c|c|c|c|c}
 $\mathrm{CNOT}_4^Z$  & $i_Z$ & $j_Z$ & $t_Z$ & $i_X$ & $j_X$ & $t_X$\\
 \hline
IX & 0 & 0 & 1 & 0 & 0 & 0\\
IY & 0 & 0 & 1 & 0 & 0 & 0\\
IZ & 0 & 0 & 0 & 0 & 0 & 0\\
XI & 0 & 1 & 0 & 0 & 0 & 0\\
XX & 0 & 1 & 1 & 0 & 0 & 0\\
XY & 0 & 1 & 1 & 0 & 0 & 0\\
XZ & 0 & 1 & 0 & 0 & 0 & 0\\
YI & 0 & 1 & 0 & 1 & 0 & 0\\
YX & 0 & 1 & 1 & 1 & 0 & 0\\
YY & 0 & 1 & 1 & 1 & 0 & 0\\
YZ & 0 & 1 & 0 & 1 & 0 & 0\\
ZI & 0 & 0 & 0 & 1 & 0 & 0\\
ZX & 0 & 0 & 1 & 1 & 0 & 0\\
ZY & 0 & 0 & 1 & 1 & 0 & 0\\
ZZ & 0 & 0 & 0 & 1 & 0 & 0\\
\end{tabular}\\

\section{Details on the numerical analysis}

\subsection{Hamiltonian of statistical mechanics models}
\label{app:MonteCarlo}
Numerical study of various statistical physics models in this work is based on the variations of the following hamiltonian,
\begin{equation}
H = \sum_{i,j,k} \left[ H_X (i,j,k) + H_Y (i,j,k) + H_Z (i,j,k) \right]
\end{equation}
where
\begin{eqnarray}
& & H_X (i,j,k) = \nonumber \\
&-& {J_x(X)}(i,j,k) \sigma_y(i,j,k) \sigma_t(i,j+1,k)
\sigma_y(i,j,k+1) \sigma_t(i,j,k) \nonumber \\
    &-& {J_y(X)}(i,j,k) \sigma_t(i,j,k) \sigma_x(i,j,k+1) 
\sigma_t(i+1,j,k) \sigma_x(i,j,k) \nonumber \\
    &-& {J_q}^\sigma(i,j,k) \sigma_x(i,j,k) \sigma_y(i+1,j,k)
\sigma_x(i,j+1,k) \sigma_y(i,j,k) \nonumber \\
& & H_Z (i,j,k) = \nonumber \\
&-& {J_x(Z)}(i,j,k) \tau_y(i,j,k) \tau_t(i,j+1,k) \tau_y(i,j,k+1)
\tau_t(i,j,k) \nonumber \\
    &-& {J_y(Z)}(i,j,k) \tau_t(i,j,k) \tau_x(i,j,k+1) \tau_t(i+1,j,k)  
\tau_x(i,j,k) \nonumber \\
    &-& {J
_q}^\tau(i,j,k) \tau_x(i,j,k) \tau_y(i+1,j,k) \tau_x(i,j+1,k)
\tau_y(i,j,k) \nonumber \\
\end{eqnarray}
for the $X$-error and the $Z$-error and
\begin{eqnarray}
& & H_Y (i,j,k) = \nonumber \\
&-&{J_x(Y)}(i,j,k) \sigma_y (i,j,k) \sigma_t(i,j+1,k)
\sigma_y(i,j,k+1) \sigma_t(i,j,k) \nonumber \\
& & \tau_t(i+1,j,k) \tau_x(i+1,j,k+1)
\tau_t(i+2,j,k) \tau_x(i+1,j,k) \nonumber \\ 
&-& {J_y(Y)}(i,j,k) \sigma_t(i,j,k) \sigma_x(i,j,k+1)
\sigma_t(i+1,j,k) \sigma_x(i,j,k) \nonumber \\
& & \tau_y(i,j+1,k) \tau_t(i,j+2,k) 
\tau_y(i,j+1,k+1) \tau_t(i,j+1,k) \nonumber \\
\end{eqnarray}
for the $Y$ error, where $(i,j,k)$ denotes the location of lattice site in 3-dimension, and $x, y$ denotes the spatial direction and $t$ denotes the time direction. $\sigma_{x,y,t} (i,j,k)$ and $\tau_{x,y,z} (i,j,k)$ represent Ising spin degree of freedoms associated with the lattice site $(i,j,k)$ for the direction of $x,y,t$: $\sigma_x (i,j,k)$ lies between the site $(i,j,k)$ and the site $(i+1,j,k)$, and $J$'s denote the couplings. For non-coupled random plaquette model, one of $\sigma, \tau$ shall be dropped. For example, to study the independent $XZ$ noise plus syndrome noise model, we remove $\tau$ spin degrees of freedom from the Hamiltonian and choose $J_x (X) \neq J_y (X)$ as in Eq. \ref{prob_RPM}. For the artificial, symmetric depolarizing noise case of random coupled-plaquette model, we choose $J_x(X) = J_y(X) = J_x(Y) = J_y(Y) = J_x(Z) = J_y(Z) = J_q$. For the (asymmetric) depolarizing circuit-noise plus syndrome noise case of random coupled-plaquette model, $J_x(X) \neq J_y(X), J_x(Z) \neq J_y(Z)$ and $J_x(Y) = J_y (Y)$ are chosen as in Eq. \ref{prob_aniso}.

\subsection{Monte Carlo simulation parameters}

As mentioned in the main text, we perform initial thermalizing Monte Carlo simulation runs which consist of $10000 \times (N_{\rm met}$ Metropolis steps over the entire lattice volume$)$ at each temperature before starting Parallel Tempering MC simulations. 50 random sign configurations are drawn for all the Monte Carlo simulations listed in the following tables (Table \ref{tab:aniso_RPGM} -- \ref{tab:aniso_RCPGM2}). MC results are averaged over these random samples. $L$ denotes the 1-dimensional size of the system size ($L^3$). In each "sweep", $N_{\rm met}$ Metropolis step plus one parallel tempering step at the end of $N_{\rm met}$ Metropolis step is performed.  $N_{\rm sweep}$ is the total number of such sweeps. For example, MC simulation of $N_{\rm sweep} = 10000$, $N_{\rm met} = 100$ parameter results in $10^5$ Metropolis steps in total interleaved with 10000 parallel tempering steps. Temperature scan in parallel tempering follows
\begin{equation}
   r = \left(\frac{T_\mathrm{max}}{T_\mathrm{min}}\right)^\frac{1}{T_\mathrm{step} - 1}, \;\;\;\; \left(\frac{1}{T}\right)_{i} = r \; \left(\frac{1}{T}\right)_{i-1}, \label{tempering_ratio}
\end{equation}
starting from high temperature to low temperature with $T_0 = T_\mathrm{max}$ \cite{Earl2005}.

\begin{table}
\begin{tabular}{|c|c|c|c|c|c|c|}
\hline
 $p$  & $L$ & $N_{\rm sweep}$ & $N_{\rm met}$ & $T$ step & $T_{\rm min}$ & $T_{\rm max}$\\
 \hline\hline
$1.7 \times 10^{-5}$ & 8 & 10000 & 100 & 32 & 1.000 & 2.000\\ & 8 & 10000 & 100 & 32 & 1.100 & 1.600\\
          & 8 & 10000 & 100 & 32 & 1.250 & 1.600\\
          & 12 & 1000 & 100 & 32 & 1.000 & 2.000\\
          & 12 & 10000 & 100 & 32 & 1.100 & 1.600\\
          & 12 & 10000 & 100 & 32 & 1.250 & 1.600\\
          & 16 & 1000 & 100 & 32 & 1.000 & 2.000\\
          & 16 & 1000 & 100 & 32 & 1.100 & 1.600\\
          & 16 & 10000 & 100 & 32 & 1.250 & 1.600\\
          & 20 & 1000 & 100 & 32 & 1.000 & 2.000\\
          & 20 & 5000 & 100 & 32 & 1.000 & 2.000\\
          & 24 & 1000 & 100 & 32 & 1.000 & 2.000\\
          & 24 & 1000 & 100 & 32 & 1.250 & 1.600\\
\hline
$1.7 \times 10^{-3}$ & 8 & 10000 & 100 & 32 & 1.000 & 2.000\\ & 8 & 10000 & 100 & 32 & 1.100 & 1.600\\
          & 12 & 1000 & 100 & 32 & 1.000 & 2.000\\
          & 12 & 10000 & 100 & 32 & 1.100 & 1.600\\
          & 16 & 1000 & 100 & 32 & 1.000 & 2.000\\
          & 16 & 10000 & 100 & 32 & 1.100 & 1.600\\
          & 20 & 1000 & 100 & 32 & 1.000 & 2.000\\
          & 20 & 5000 & 100 & 32 & 1.100 & 1.600\\
          & 24 & 1000 & 100 & 32 & 1.000 & 2.000\\
          & 24 & 1000 & 100 & 32 & 1.180 & 1.240\\
\hline
$3.41 \times 10^{-3} $ & 8 & 1000 & 100 & 32 & 1.000 & 2.000\\ & 8 & 10000 & 100 & 32 & 1.050 & 1.500\\
                       & 12 & 1000 & 100 & 32 & 1.000 & 2.000\\
                       & 12 & 10000 & 100 & 32 & 1.050 & 1.500\\
                       & 16 & 1000 & 100 & 32 & 1.000 & 2.000\\
                       & 16 & 10000 & 100 & 32 & 1.050 & 1.500\\
                       & 20 & 1000 & 100 & 32 & 1.000 & 2.000\\
                       & 20 & 5000 & 100 & 32 & 1.050 & 1.500\\
                       & 24 & 1000 & 100 & 32 & 1.000 & 2.000\\
\hline                       
\end{tabular}\\
\caption{List of run parameters for anisotropic, independent $XZ$ noise RPGM Monte Carlo Simulations. MC results are discussed in \ref{indepXZ}.}
\label{tab:aniso_RPGM}
\end{table}

\begin{figure}
    \centering
    \includegraphics[width=0.9\columnwidth]{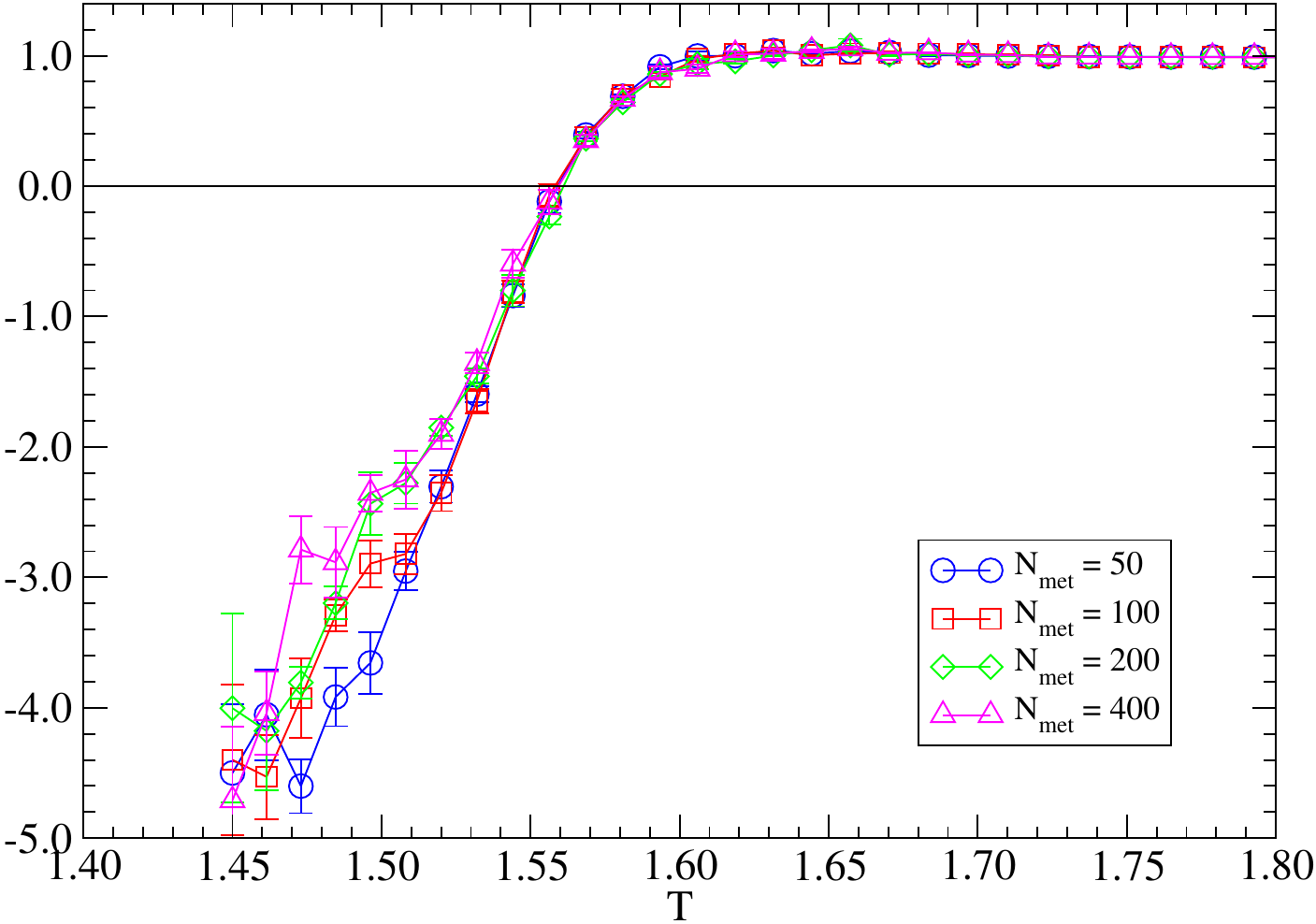}\\
    \includegraphics[width=0.9\columnwidth]{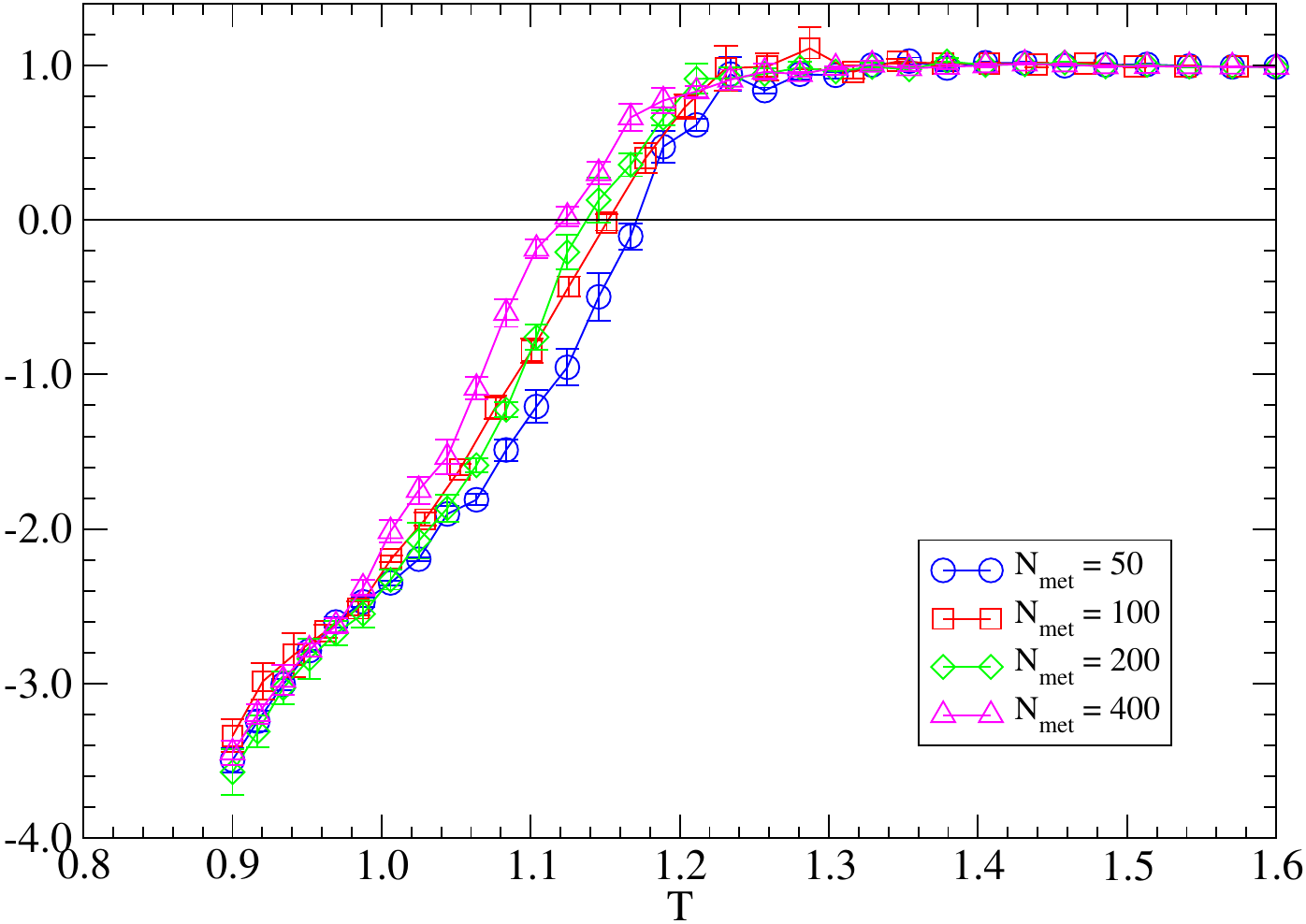}
    \caption{The comparison of $B_3$ from MC simulations with $N_\mathrm{met} = 50$, $100$, $200$ and $400$ for $p = 4\%$ (top) and $p = 6\%$ (bottom) on $24^3$ lattice volume for the symmetric depolarizing noise RCPGM Monte Carlo simulation. MC simulation results are discussed in \ref{Depolar}}
    \label{fig:p0.04_p0.06_Nmet_dep}
\end{figure}

\begin{table}
\begin{tabular}{|c|c|c|c|c|c|c|}
\hline
 $p$  & $L$ & $N_{\rm sweep}$ & $N_{\rm met}$ & $T$ step & $T_{\rm min}$ & $T_{\rm max}$\\
 \hline\hline
$5.11 \times 10^{-3}$ & 8 & 1000 & 100 & 32 & 0.800 & 1.600 \\ & 8 & 10000 & 100 & 32 & 0.950 & 1.350\\
                      & 12 & 5000 & 100 & 32 & 0.800 & 1.200\\
                      & 12 & 1000 & 100 & 32 & 0.800 & 1.600\\
                      & 12 & 10000 & 100 & 32 & 0.950 & 1.350\\
                      & 16 & 5000 & 100 & 32 & 0.800 & 1.200\\
                      & 16 & 1000 & 100 & 32 & 0.800 & 1.600\\
                      & 16 & 10000 & 100 & 32 & 0.950 & 1.350\\
                      & 20 & 5000 & 100 & 32 & 0.800 & 1.200\\
                      & 20 & 1000 & 100 & 32 & 0.800 & 1.600\\
                      & 20 & 5000 & 100 & 32 & 0.950 & 1.350\\
                      & 24 & 5000 & 100 & 32 & 0.800 & 1.200\\
                      & 24 & 1000 & 100 & 32 & 0.800 & 1.600\\
\hline
$6.82 \times 10^{-3}$ & 8 & 1000 & 100 & 32 & 0.400 & 1.600 \\ & 8 & 10000 & 100 & 32 & 0.650 & 1.350\\
                      & 12 & 5000 & 100 & 32 & 0.400 & 1.200\\
                      & 12 & 1000 & 100 & 32 & 0.400 & 1.600\\
                      & 12 & 10000 & 100 & 32 & 0.650 & 1.350\\
                      & 16 & 5000 & 100 & 32 & 0.400 & 1.200\\
                      & 16 & 1000 & 100 & 32 & 0.400 & 1.600\\
                      & 16 & 10000 & 100 & 32 & 0.650 & 1.350\\
                      & 20 & 5000 & 100 & 32 & 0.400 & 1.200\\
                      & 20 & 1000 & 100 & 32 & 0.400 & 1.600\\
                      & 20 & 5000 & 100 & 32 & 0.650 & 1.350\\
                      & 24 & 5000 & 100 & 32 & 0.400 & 1.200\\
                      & 24 & 1000 & 100 & 32 & 0.400 & 1.600\\
\hline 
$8.52 \times 10^{-3}$ & 8 & 1000 & 100 & 32 & 0.300 & 1.200\\ & 8 & 1000 & 100 & 64 & 0.400 & 1.300\\
                      & 8 & 1000 & 100 & 32 & 0.400 & 1.600\\
                      & 8 & 1000 & 100 & 64 & 0.400 & 1.600\\
                      & 12 & 1000 & 100 & 32 & 0.300 & 1.200\\ 
                      & 12 & 1000 & 100 & 64 & 0.400 & 1.300\\
                      & 12 & 1000 & 100 & 32 & 0.400 & 1.600\\
                      & 12 & 1000 & 100 & 64 & 0.400 & 1.600\\
                      & 16 & 1000 & 100 & 32 & 0.300 & 1.200\\ 
                      & 16 & 1000 & 100 & 32 & 0.400 & 1.600\\
                      & 20 & 1000 & 100 & 32 & 0.300 & 1.200\\ 
                      & 20 & 1000 & 100 & 32 & 0.400 & 1.600\\
                      & 24 & 1000 & 100 & 32 & 0.300 & 1.200\\ 
                      & 24 & 1000 & 100 & 32 & 0.400 & 1.600\\
\hline\hline
\end{tabular}\\
\caption{List of run parameters for anisotropic, independent $XZ$ noise RPGM Monte Carlo Simulations (continued from Table. \ref{tab:aniso_RPGM}). MC results are discussed in \ref{indepXZ}}
\label{tab:aniso_RPGM2}
\end{table}

\begin{table}
\begin{tabular}{|c|c|c|c|c|c|c|}
\hline
 $p$  & $L$ & $N_{\rm sweep}$ & $N_{\rm met}$ & $T$ step & $T_{\rm min}$ & $T_{\rm max}$\\
 \hline\hline
0.00      & 8 & 1000 & 100 & 32 & 1.750 & 2.150\\
          & 8 & 1000 & 100 & 32 & 1.900 & 2.050\\
          & 8 & 1000 & 100 & 32 & 1.900 & 2.100\\
          & 8 & 1000 & 100 & 32 & 1.940 & 2.050\\
          & 12 & 1000 & 100 & 32 & 1.750 & 2.150\\
          & 12 & 1000 & 100 & 32 & 1.900 & 2.000\\
          & 12 & 1000 & 100 & 32 & 1.900 & 2.050\\
          & 12 & 1000 & 100 & 32 & 1.900 & 2.100\\
          & 12 & 1000 & 100 & 32 & 1.930 & 1.960\\
          & 12 & 1000 & 100 & 32 & 1.930 & 1.970\\
          & 16 & 1000 & 100 & 32 & 1.750 & 2.150\\
          & 16 & 1000 & 100 & 32 & 1.900 & 2.000\\
          & 16 & 1000 & 100 & 32 & 1.900 & 2.050\\
          & 16 & 1000 & 100 & 32 & 1.930 & 1.970\\
          & 16 & 1000 & 100 & 32 & 1.940 & 1.960\\
          & 20 & 1000 & 100 & 32 & 1.940 & 1.950\\
          & 24 & 1000 & 100 & 32 & 1.750 & 2.150\\
          & 24 & 1000 & 100 & 32 & 1.900 & 2.000\\
          & 24 & 1000 & 100 & 32 & 1.900 & 2.050\\
          & 24 & 1000 & 100 & 32 & 1.930 & 1.970\\
          & 24 & 1000 & 100 & 32 & 1.940 & 1.950\\
\hline
0.02      & 8 & 1000 & 100 & 32 & 1.650 & 2.200\\
          & 8 & 1000 & 100 & 32 & 1.700 & 1.950\\
          & 8 & 1000 & 100 & 32 & 1.700 & 2.000\\
          & 12 & 1000 & 100 & 32 & 1.650 & 2.200\\
          & 12 & 1000 & 100 & 32 & 1.700 & 1.950\\
          & 12 & 1000 & 100 & 32 & 1.700 & 2.000\\
          & 12 & 1000 & 100 & 32 & 1.720 & 1.900\\
          & 16 & 1000 & 100 & 32 & 1.650 & 2.200\\
          & 16 & 1000 & 100 & 32 & 1.700 & 1.950\\
          & 16 & 1000 & 100 & 32 & 1.720 & 1.900\\
          & 20 & 1000 & 100 & 32 & 1.720 & 1.900\\
          & 24 & 1000 & 100 & 32 & 1.650 & 2.200\\
          & 24 & 1000 & 100 & 32 & 1.700 & 1.950\\
          & 24 & 1000 & 100 & 32 & 1.720 & 1.900\\
\hline
0.04      & 8 & 1000 & 100 & 32 & 1.000 & 2.500\\
          & 8 & 1000 & 100 & 32 & 1.000 & 2.100\\
          & 8 & 1000 & 100 & 32 & 1.450 & 2.000\\
          & 12 & 1000 & 100 & 32 & 1.000 & 2.500\\
          & 12 & 1000 & 100 & 32 & 1.200 & 2.100\\
          & 12 & 1000 & 100 & 32 & 1.450 & 1.850\\
          & 12 & 1000 & 100 & 32 & 1.450 & 2.000\\
          & 16 & 1000 & 100 & 32 & 1.000 & 2.500\\
          & 16 & 1000 & 100 & 32 & 1.450 & 1.850\\
          & 16 & 1000 & 100 & 32 & 1.450 & 2.000\\
          & 20 & 1000 & 100 & 32 & 1.450 & 1.850\\
          & 24 & 1000 & 100 & 32 & 1.000 & 2.500\\
          & 24 & 1000 & 100 & 32 & 1.450 & 1.850\\
          & 24 & 1000 & 200 & 32 & 1.450 & 1.850\\
          & 24 & 1000 & 400 & 32 & 1.450 & 1.850\\
          & 24 & 1000 & 50 & 32 & 1.450 & 1.850\\
          & 24 & 1000 & 100 & 32 & 1.450 & 2.000\\
\hline
\end{tabular}\\
\caption{List of run parameters for the symmetric depolarizing noise RCPGM Monte Carlo Simulations. MC results are discussed in \ref{Depolar}.}
\label{tab:depolar_RCPGM}
\end{table}

\begin{table}
\begin{tabular}{|c|c|c|c|c|c|c|}
\hline
 $p$  & $L$ & $N_{\rm sweep}$ & $N_{\rm met}$ & $T$ step & $T_{\rm min}$ & $T_{\rm max}$\\
 \hline\hline
0.05      & 8 & 1000 & 100 & 32 & 0.350 & 2.000\\
          & 8 & 1000 & 1000 & 32 & 0.350 & 2.000\\
          & 12 & 1000 & 100 & 32 & 0.350 & 2.000\\
          & 12 & 1000 & 1000 & 32 & 0.350 & 2.000\\
          & 16 & 1000 & 100 & 32 & 0.350 & 2.000\\
          & 16 & 1000 & 1000 & 32 & 0.350 & 2.000\\
          & 20 & 1000 & 1000 & 32 & 0.350 & 2.000\\
          & 24 & 1000 & 100 & 32 & 0.350 & 2.000\\
\hline
0.06      & 8 & 1000 & 100 & 32 & 0.350 & 2.000\\
          & 8 & 1000 & 1000 & 32 & 0.350 & 2.000\\
          & 8 & 1000 & 1000 & 32 & 0.900 & 1.800\\
          & 8 & 1000 & 1000 & 32 & 1.200 & 2.000\\
          & 8 & 1000 & 100 & 32 & 1.200 & 2.100\\
          & 12 & 1000 & 100 & 32 & 0.350 & 2.000\\
          & 12 & 1000 & 1000 & 32 & 0.350 & 2.000\\
          & 12 & 1000 & 200 & 32 & 0.900 & 1.600\\
          & 12 & 1000 & 400 & 32 & 0.900 & 1.600\\
          & 12 & 1000 & 50 & 32 & 0.900 & 1.600\\
          & 12 & 1000 & 100 & 32 & 0.900 & 1.800\\
          & 12 & 1000 & 800 & 32 & 1.200 & 1.600\\
          & 12 & 1000 & 100 & 32 & 1.200 & 2.000\\
          & 12 & 1000 & 100 & 32 & 1.200 & 2.100\\
          & 16 & 1000 & 100 & 32 & 0.350 & 2.000\\
          & 16 & 1000 & 1000 & 32 & 0.350 & 2.000\\
          & 16 & 1000 & 200 & 32 & 0.900 & 1.600\\
          & 16 & 1000 & 400 & 32 & 0.900 & 1.600\\
          & 16 & 1000 & 50 & 32 & 0.900 & 1.600\\
          & 16 & 1000 & 100 & 32 & 0.900 & 1.800\\
          & 16 & 1000 & 100 & 32 & 1.200 & 2.100\\
          & 20 & 1000 & 100 & 32 & 0.350 & 2.000\\
          & 20 & 1000 & 1000 & 32 & 0.350 & 2.000\\
          & 20 & 1000 & 100 & 32 & 0.900 & 1.600\\
          & 24 & 1000 & 100 & 32 & 0.350 & 2.000\\
          & 24 & 1000 & 200 & 32 & 0.900 & 1.600\\
          & 24 & 1000 & 400 & 32 & 0.900 & 1.600\\
          & 24 & 1000 & 50 & 32 & 0.900 & 1.600\\
          & 24 & 1000 & 100 & 32 & 0.900 & 1.800\\
          & 24 & 1000 & 100 & 32 & 1.200 & 2.100\\
\hline
0.07      & 8 & 1000 & 100 & 32 & 0.100 & 1.800\\
          & 8 & 1000 & 1000 & 32 & 0.100 & 1.800\\
          & 8 & 1000 & 100 & 32 & 0.350 & 2.000\\
          & 12 & 1000 & 100 & 32 & 0.100 & 1.600\\
          & 12 & 1000 & 1000 & 32 & 0.100 & 1.800\\
          & 12 & 1000 & 100 & 32 & 0.350 & 2.000\\
          & 16 & 1000 & 100 & 32 & 0.100 & 1.600\\
          & 16 & 1000 & 1000 & 32 & 0.100 & 1.800\\
          & 16 & 1000 & 100 & 32 & 0.350 & 2.000\\
          & 20 & 1000 & 100 & 32 & 0.100 & 1.600\\
          & 20 & 1000 & 1000 & 32 & 0.100 & 1.800\\
          & 24 & 1000 & 100 & 32 & 0.100 & 1.600\\
          & 24 & 1000 & 100 & 32 & 0.350 & 2.000\\
\hline
0.09      & 8 & 1000 & 100 & 32 & 0.150 & 2.000\\
          & 8 & 1000 & 1000 & 32 & 0.350 & 2.000\\
          & 12 & 1000 & 100 & 32 & 0.350 & 2.000\\
          & 16 & 1000 & 100 & 32 & 0.350 & 2.000\\
          & 20 & 1000 & 100 & 32 & 0.350 & 2.000\\
          & 24 & 1000 & 100 & 32 & 0.350 & 2.000\\
\hline\hline
\end{tabular}\\
\caption{List of run parameters for the symmetric depolarizing noise RCPGM Monte Carlo Simulations (continued from Table \ref{tab:depolar_RCPGM}). MC results are discussed in \ref{Depolar}.}
\label{tab:depolar_RCPGM2}
\end{table}

\begin{table}
\begin{tabular}{|c|c|c|c|c|c|c|}
\hline
 $p$  & $L$ & $N_{\rm sweep}$ & $N_{\rm met}$ & $T$ step & $T_{\rm min}$ & $T_{\rm max}$\\
 \hline\hline
$2.88 \times 10^{-5}$ & 8 & 1000 & 100 & 32 & 2.100 & 2.145\\ & 8 & 1000 & 100 & 32 & 2.150 & 2.750\\
          & 8 & 1000 & 100 & 32 & 2.240 & 2.340\\
          & 12 & 1000 & 100 & 32 & 1.900 & 2.450\\ 
          & 12 & 1000 & 100 & 32 & 1.950 & 2.500\\
          & 12 & 1000 & 100 & 32 & 2.240 & 2.340\\
          & 12 & 1000 & 100 & 32 & 2.245 & 2.300\\ 
          & 16 & 1000 & 100 & 32 & 1.800 & 2.450\\ 
          & 16 & 1000 & 100 & 32 & 1.950 & 2.500\\
          & 16 & 1000 & 100 & 32 & 2.050 & 2.450\\
          & 16 & 1000 & 100 & 32 & 2.200 & 2.320\\ 
          & 16 & 1000 & 100 & 32 & 2.250 & 2.290\\
          & 16 & 1000 & 100 & 32 & 2.255 & 2.750\\ 
          & 20 & 1000 & 100 & 32 & 2.255 & 2.275\\
          & 24 & 1000 & 100 & 32 & 2.255 & 2.275\\
\hline
$5.77 \times 10^{-3}$ & 8 & 1000 & 100 & 32 & 0.350 & 2.000 \\ & 8 & 1000 & 100 & 32 & 1.500 & 1.950\\
          & 8 & 1000 & 100 & 32 & 1.500 & 2.200\\
          & 12 & 1000 & 100 & 32 & 1.300 & 1.950\\
          & 12 & 1000 & 100 & 32 & 1.350 & 2.000\\
          & 12 & 1000 & 100 & 32 & 1.600 & 1.850\\
          & 16 & 1000 & 100 & 32 & 1.200 & 1.950\\
          & 16 & 1000 & 100 & 32 & 1.350 & 2.000\\
          & 16 & 1000 & 100 & 32 & 1.650 & 1.800\\
          & 20 & 1000 & 100 & 32 & 1.650 & 1.800\\
          & 24 & 1000 & 100 & 32 & 1.650 & 1.800\\
\hline
0.00115 & 8 & 1000 & 100 & 32 & 0.450 & 1.600 \\ & 8 & 1000 & 100 & 32 & 0.450 & 2.000\\
          & 8 & 1000 & 100 & 32 & 1.000 & 1.950\\
          & 8 & 1000 & 100 & 32 & 1.500 & 2.100\\
          & 12 & 1000 & 100 & 32 & 0.350 & 1.600\\
          & 12 & 1000 & 100 & 32 & 0.950 & 1.700\\
          & 12 & 1000 & 100 & 32 & 0.950 & 1.950\\
          & 16 & 1000 & 100 & 32 & 0.250 & 1.600\\
          & 16 & 1000 & 100 & 32 & 0.700 & 1.950\\
          & 16 & 1000 & 100 & 32 & 1.000 & 1.600\\
          & 20 & 1000 & 100 & 32 & 1.000 & 1.500\\
          & 24 & 1000 & 100 & 32 & 1.000 & 1.500\\
\hline          
0.00124 & 8 & 1000 & 100 & 32 & 1.000 & 1.950 \\  & 12 & 1000 & 100 & 32 & 0.950 & 1.700 \\       
        & 12 & 1000 & 100 & 32 & 1.000 & 1.950 \\       
        & 16 & 1000 & 100 & 32 & 0.950 & 1.700 \\       
        & 16 & 1000 & 100 & 32 & 1.000 & 1.950 \\       
        & 20 & 1000 & 100 & 32 & 0.950 & 1.700 \\       
        & 20 & 1000 & 100 & 32 & 1.000 & 1.950 \\       
        & 24 & 1000 & 100 & 32 & 0.950 & 1.700 \\       
        & 24 & 1000 & 100 & 32 & 1.000 & 1.950 \\       
\hline          
0.00130 & 8 & 1000 & 100 & 32 & 1.000 & 1.950 \\  & 12 & 1000 & 100 & 32 & 0.950 & 1.700 \\          
        & 12 & 1000 & 100 & 32 & 1.000 & 1.950 \\          
        & 16 & 1000 & 100 & 32 & 0.950 & 1.700 \\          
        & 16 & 1000 & 100 & 32 & 1.000 & 1.950 \\          
        & 20 & 1000 & 100 & 32 & 0.950 & 1.700 \\          
        & 20 & 1000 & 100 & 32 & 1.000 & 1.950 \\          
        & 24 & 1000 & 100 & 32 & 0.950 & 1.700 \\          
        & 24 & 1000 & 100 & 32 & 1.000 & 1.950 \\          
\hline\hline
\end{tabular}\\
\caption{List of run parameters for the circuit-level noise RCPGM Monte Carlo Simulations. MC results are discussed in \ref{Depolar}.}
\label{tab:aniso_RCPGM}
\end{table}

\begin{table}
\begin{tabular}{|c|c|c|c|c|c|c|}
\hline
 $p$  & $L$ & $N_{\rm sweep}$ & $N_{\rm met}$ & $T$ step & $T_{\rm min}$ & $T_{\rm max}$\\
 \hline\hline
0.00144 & 8 & 1000 & 100 & 32 & 0.450 & 2.000 \\ & 8 & 1000 & 1000 & 32 & 0.450 & 2.000 \\ 
        & 12 & 1000 & 100 & 32 & 0.350 & 2.000 \\ 
        & 12 & 1000 & 1000 & 32 & 0.450 & 2.000 \\ 
        & 12 & 1000 & 100 & 32 & 0.500 & 1.200 \\ 
        & 16 & 1000 & 1000 & 32 & 0.450 & 2.000 \\ 
        & 16 & 1000 & 100 & 32 & 0.700 & 1.950 \\ 
        & 20 & 1000 & 1000 & 32 & 0.450 & 2.000 \\ 
        & 20 & 1000 & 100 & 32 & 0.700 & 1.950 \\ 
\hline
0.00159 & 8 & 1000 & 100 & 32 & 0.450 & 2.000 \\ & 8 & 1000 & 1000 & 32 & 0.450 & 2.000 \\ 
        & 12 & 1000 & 100 & 32 & 0.450 & 2.000 \\ 
        & 12 & 1000 & 1000 & 32 & 0.450 & 2.000 \\ 
        & 16 & 1000 & 100 & 32 & 0.450 & 2.000 \\ 
        & 16 & 1000 & 1000 & 32 & 0.450 & 2.000 \\ 
        & 20 & 1000 & 100 & 32 & 0.450 & 2.000 \\ 
        & 20 & 1000 & 1000 & 32 & 0.450 & 2.000 \\ 
\hline
0.00173 & 8 & 1000 & 100 & 32 & 0.300 & 2.000 \\ & 8 & 1000 & 1000 & 32 & 0.300 & 2.000 \\ 
        & 8 & 1000 & 100 & 32 & 0.450 & 1.600 \\ 
        & 8 & 1000 & 100 & 32 & 0.500 & 1.950 \\ 
        & 12 & 1000 & 100 & 32 & 0.100 & 1.400 \\ 
        & 12 & 1000 & 100 & 32 & 0.300 & 1.600 \\ 
        & 12 & 1000 & 1000 & 32 & 0.300 & 2.000 \\ 
        & 12 & 1000 & 100 & 32 & 0.500 & 1.200 \\ 
        & 16 & 1000 & 100 & 32 & 0.100 & 1.400 \\ 
        & 16 & 1000 & 100 & 32 & 0.100 & 1.600 \\ 
        & 16 & 1000 & 100 & 32 & 0.200 & 1.400 \\ 
        & 16 & 1000 & 1000 & 32 & 0.300 & 2.000 \\ 
        & 16 & 1000 & 100 & 32 & 0.350 & 1.000 \\ 
        & 20 & 1000 & 100 & 32 & 0.200 & 1.400 \\ 
        & 20 & 1000 & 1000 & 32 & 0.300 & 2.000 \\ 
        & 20 & 1000 & 100 & 32 & 0.350 & 1.000 \\ 
        & 24 & 1000 & 100 & 32 & 0.200 & 1.400 \\ 
\hline
0.00187 & 8 & 1000 & 100 & 32 & 0.050 & 1.400 \\ & 12 & 1000 & 100 & 32 & 0.050 & 1.400 \\ 
        & 16 & 1000 & 100 & 32 & 0.050 & 1.400 \\ 
        & 16 & 1000 & 100 & 32 & 0.050 & 1.400 \\ 
\hline
0.00202 & 8 & 1000 & 100 & 32 & 0.050 & 1.400 \\ & 8 & 1000 & 100 & 32 & 0.300 & 1.400\\
        & 12 & 1000 & 100 & 32 & 0.050 & 1.400 \\ 
        & 12 & 1000 & 100 & 32 & 0.100 & 1.400\\
        & 16 & 1000 & 100 & 32 & 0.050 & 1.400 \\ 
        & 16 & 1000 & 100 & 32 & 0.100 & 1.400\\
        & 20 & 1000 & 100 & 32 & 0.050 & 1.400 \\ 
        & 20 & 1000 & 100 & 32 & 0.100 & 1.400\\
        & 24 & 1000 & 100 & 32 & 0.100 & 1.400\\
\hline
0.00231 & 8 & 1000 & 100 & 32 & 0.050 & 0.950 \\ & 8 & 1000 & 100 & 32 & 0.100 & 0.950\\
          & 8 & 1000 & 100 & 32 & 0.100 & 1.950\\
          & 12 & 1000 & 100 & 32 & 0.050 & 0.950 \\ 
          & 12 & 1000 & 100 & 32 & 0.100 & 0.950\\
          & 16 & 1000 & 100 & 32 & 0.050 & 0.950 \\ 
          & 16 & 1000 & 100 & 32 & 0.100 & 0.950\\
          & 20 & 1000 & 100 & 32 & 0.050 & 0.950 \\ 
          & 24 & 1000 & 100 & 32 & 0.050 & 0.950 \\ 
\hline\hline
\end{tabular}\\
\caption{List of run parameters for the circuit-level noise RCPGM Monte Carlo Simulations (continued from Table \ref{tab:aniso_RCPGM}. MC results are discussed in \ref{Depolar}.}
\label{tab:aniso_RCPGM2}
\end{table}

\subsection{Systematics of simulation results}

Numerical results from Monte Carlo simulations are influenced by whether the MC simulations reached thermal equilibrium, and whether the averages are taken over sufficiently large number of {\it independent} Monte Carlo samples. Taking large volume limit of MC simulation results using the simulations with finite number of degrees of freedom requires a finite volume scaling study. Studying various statistical physics models in this work is much harder due to random distribution of wrong-sign bond or wrong-sign plaquette in the models: there may be many local minima and MC simulations may be trapped in one of such minima despite the parallel tempering steps. 

At least for a small probability (i.e., near $p = 0$) of wrong-sign bonds or wrong-sign plaquettes, thermal fluctuations will dominate in MC simulations and varying $N_\mathrm{sweep}, N_\mathrm{met}$ and $L$ in MC simulations will suggest how large their effects on the MC results are. As the probability for wrong-sign bond or wrong-sign plaquette increases, the disorder of wrong-signs will become important and spin-configuration swapping in parallel tempering step through $(T_\mathrm{min}, T_\mathrm{max})$ and $T_\mathrm{step}$ will play an important role. 

Also, we found that the large volume limit of transition temperatures obtained from $B_3$ and $\chi$ below the threshold probability exists but above the threshold probability either the large volume limit does not exist or it is highly non-linear in $L$ (see Figure \ref{fig:aniso_RPM_fit_vs_V}, \ref{fig:depolarizing_RCPGM_fit_vs_V} and \ref{fig:anisotropic_RCPGM_fit_vs_V}).

In the following, we examine these aspects separately by comparing MC simulations with different parameters. Based on these investigations, we present MC results in the main text which are obtained with $N_\mathrm{thermalization} = 10000$, $N_\mathrm{sweep} = 1000 \sim 10000$, $N_\mathrm{met} = 100$, $T_\mathrm{step} = 32$. 
 
\subsubsection{Metropolis steps}

In Figure .\ref{fig:p0.04_p0.06_Nmet_dep} and \ref{fig:p0.04_p0.06_Nmet_dep}, we compare the third order cumulant, $B_3$, from 4 different number of Metropolis updates between parallel tempering steps for MC simulation of symmetric depolarizing noise random coupled-plaquette gauge model on $24^3$ lattice volume (MC simulation results are discussed in \ref{Depolar}). The left figure is from $p = 0.04$ random wrong-sign plaquette probability and the right from $p = 0.06$. $p = 0.04$ is well below the threshold probability and there is no significant dependence in $T_c$ as we change the number of Metropolis steps between parallel tempering. Near the threshold probability ($p = 0.06$), the transition temperature from $B_3$ does depend on $N_\mathrm{met}$ although the change in the transition temperature is numerically small. Thus, we think that below and near the threshold probability, MC simulations with $N_\mathrm{met} = 100$ between parallel tempering steps is sufficient for the study of phase diagram.

\subsubsection{Run lengths}

\begin{figure}
    \centering
    \includegraphics[width=0.85\columnwidth]{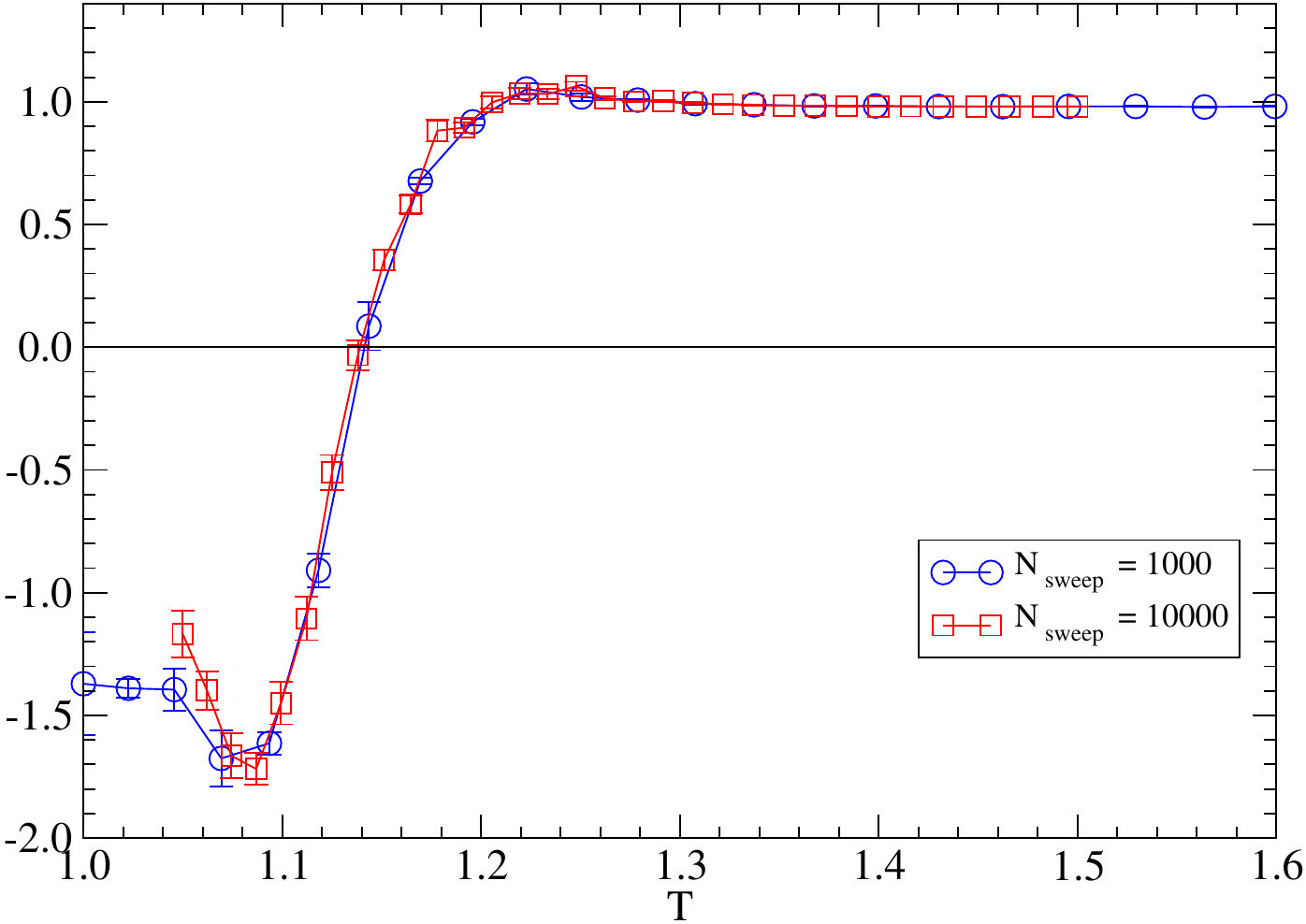}
    \caption{The comparison of $B_3$ from MC simulations with $N_\mathrm{sweep} = 1000$ (blue symbol) and $10000$ (red symbol) for $p = 3.41 \times 10^{-3}$ on $16^3$ lattice volume for the anisotropic, independent $XZ$ noise RPGM Monte Carlo simulation. MC simulation results are discussed in \ref{indepXZ}.}
    \label{fig:p0.02_vs_Nsweep}
\end{figure}

In Figure .\ref{fig:p0.02_vs_Nsweep}, $B_3$ from MC simulation (MC simulation results are discussed in \ref{indepXZ}) with $N_\mathrm{sweep} = 1000$ (blue symbol) is compared with $N_\mathrm{sweep} =  10000$ (red symbol) for $p = 3.41 \times 10^{-3}$ on $16^3$ lattice volume for the anisotropic, independent $XZ$ noise RPGM Monte Carlo simulation (the temperature range is slightly different ($(T_\mathrm{min}, T_\mathrm{max}) = (1.000, 2.000)$ (blue) and $(1.050,1.500)$(red)). There is no noticeable difference between these two simulation results other than the size of statistical error. The determined transition temperature is similar to each other and thus MC sampling of $\sim 10^5$ appears to be sufficient for the phase diagram study in this statistical physics model.

\subsubsection{Parallel tempering temperature steps}

Parallel tempering step exchanges the whole spin configurations between two neighboring temperature related by Eq. \ref{tempering_ratio} with the probability, 
\begin{equation}
{\rm Pr} (k \rightarrow k + 1) = \min \left[1, e^{(\beta_k - \beta_{k+1}) (E_k - E_{k+1})} \right] , 
\end{equation}
where $E_k$ is the energy (i.e., $\frac{H}{k_B T} = - \beta_k E_k$ schematically, where $k_B$ is the Boltzmann constant) and $\beta_k = \frac{J}{k_B T}$ schematically. Thus, the efficiency of parallel tempering depends on how small the difference in neighboring temperature is, and on how small the energy difference is between the neighboring temperature. For a given range of the temperature $(T_\mathrm{min}, T_\mathrm{max})$, increasing $T_\mathrm{step}$ facilitates the exchange of spin configurations between neighboring temperature because the difference in $\beta$'s is small.

\begin{figure}
    \centering
    \includegraphics[width=0.85\columnwidth]{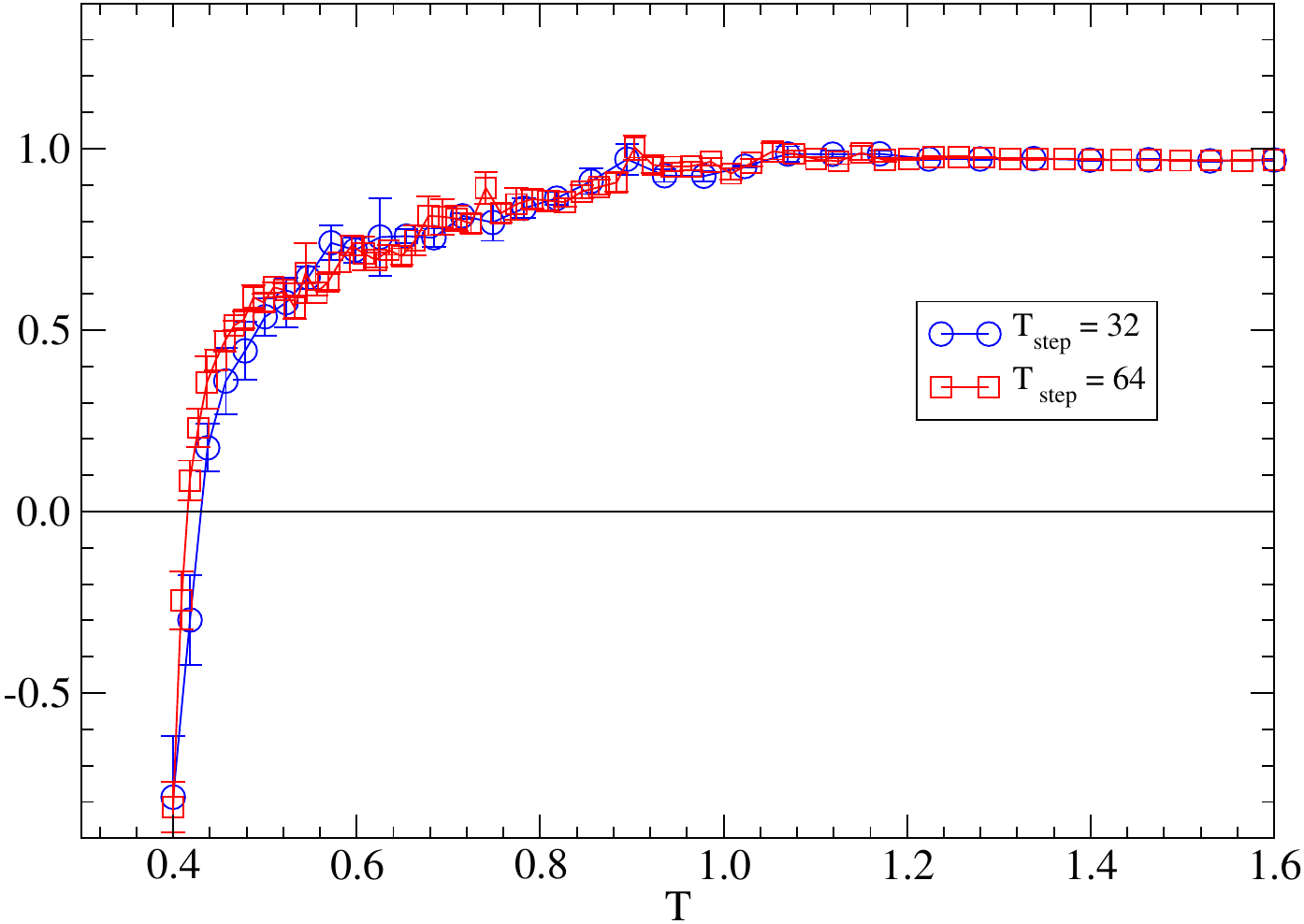}
    \caption{The comparison of $B_3$ from MC simulations with $T_\mathrm{step} = 32$ (blue symbol) and $64$ (red symbol) for $p = 8.52 \times 10^{-3}$ on $12^3$ lattice volume for the anisotropic, independent $XZ$ noise RPGM Monte Carlo simulation. MC simulation results are discussed in \ref{indepXZ}.}
    \label{fig:p0.05_vs_nbeta}
\end{figure}

Figure \ref{fig:p0.05_vs_nbeta} shows $B_3$ from MC simulations with $T_\mathrm{step} = 32$ (blue symbol) and $64$ (red symbol) for $p = 8.52 \times 10^{-3}$ on $12^3$ lattice volume for the anisotropic, independent $XZ$ noise RPGM Monte Carlo simulation (MC simulation results are discussed in \ref{indepXZ}. In this model, large volume limit of MC simulations with $p = 8.52 \times 10^{-3}$ does not show a transition. See Fig. \ref{fig:B3_RPM}). In this figure, the difference in finite volume transition temperature from two different $T_\mathrm{step}$ is small, which suggest that scanning with $T_\mathrm{step} = 32$ in this temperature range may be sufficient for the phase diagram study.

\clearpage 
\bibliography{thebibliography}{}

\end{document}